\documentclass[11pt,fleqn]{article}
\usepackage{amsmath,amssymb,amsthm,epsfig,graphics}
\newcommand{\Der}{\mathop{\rm Der}\nolimits}
\newcommand{\tr}{\mathop{\rm tr}\nolimits}
\newcommand{\rank}{\mathop{\rm rank}\nolimits}
\newcommand{\ad}{\mathop{\rm ad}\nolimits}
\newcommand{\diag}{\mathop{\rm diag}\nolimits}
\newcommand{\rsemioplus}{\mathbin{\mbox{$\lefteqn{\hspace{.65ex}\rule{.4pt}{1.2ex}}{\ni}$}}}

\newcommand{\CC}{{\mathbb C}}
\newcommand{\R}{{\mathbb R}}
\newcommand{\N}{{\mathbb N}}
\newcommand{\Z}{{\mathbb Z}}

\newtheorem{theorem}{Theorem}
\newtheorem{lemma}{Lemma}
\newtheorem*{lemma*}{Lemma}
\newtheorem{corollary}{Corollary}

\newtheorem*{conjecture*}{Conjecture}

{\theoremstyle{definition}
\newtheorem{definition}{Definition}
\newtheorem{example}{Example}
\newtheorem{remark}{Remark}

\newtheorem*{note*}{Note}
}

\allowdisplaybreaks

\begin{document}

\begin{center}
\LARGE \bf
Contractions of Low-Dimensional Lie Algebras
\end{center}

\begin{center} \bf
Maryna NESTERENKO~$^\dag$ and Roman POPOVYCH~$^{\dag\S}$
\end{center}

\noindent $^\dag$~Institute of Mathematics of NAS of Ukraine, 3
Tereshchenkivs'ka Str., Kyiv-4, 01601 Ukraine\\
$\phantom{^\dag}$~E-mail: maryna@imath.kiev.ua, rop@imath.kiev.ua

\noindent $^\S$~Fakult\"at f\"ur Mathematik, Universit\"at Wien, Nordbergstra{\ss}e 15, A-1090 Wien, Austria

\begin{abstract}
\noindent
Theoretical background of continuous contractions
of finite-dimensional Lie algebras is rigorously formulated and developed.
In~particular, known necessary criteria of contractions are collected and new criteria are proposed.
A number of requisite invariant and semiinvariant quantities are calculated for wide classes of Lie algebras
including all low-dimensional Lie algebras.

An algorithm that allows one to handle one-parametric contractions
is presented and applied to low-dimensional Lie algebras.
As a result, all one-parametric continuous contractions for
both the complex and real Lie algebras of dimensions not greater than four
are constructed with intensive usage of necessary criteria of contractions and
with studying correspondence between real and complex cases.

Levels and colevels of low-dimensional Lie algebras are discussed in detail.
Properties of multi-parametric and repeated contractions are also investigated.
\end{abstract}

\section{Introduction}\label{Introduction}

Limiting processes between Lie algebras were first investigated by Segal~\cite{Segal1951}.
The most known example concerning these processes is given by connection between relativistic and classical mechanics
with their underling Poincar\'e and Galilean symmetry groups.
If the velocity of light is assumed to go to infinity, relativistic mechanics `transforms' into classical mechanics.
This also induces a singular transition from the Poincar\'e algebra to the Galilean one.
The other well-known example is a limit process from quantum mechanics to classical mechanics under $\hbar\to 0$,
which corresponds to the contraction of the Heisenberg algebras to the Abelian ones of the same dimensions.

Existing works on contractions can be conditionally divided into two main streams which are scarcely connected with each other.
One of them is more `physical' and is mainly oriented to applications of contractions.
The other one is more `algebraic' and usually have better mathematical background.
Let us simultaneously survey works on the main types of contractions existing in the frameworks of both approaches. 

After Segal, the concept of limiting processes between physical theories in terms of contractions of 
the underling symmetry groups was also formulated by In\"on\"u and Wigner~\cite{Inonu&Wigner1953, Inonu&Wigner1954}.
They introduced so-called \emph{In\"on\"u--Wigner contractions} (\emph{IW-contractions})
which, in spite of their simplicity, were effectively applied to a wide range of physical and mathematical problems.
Later Saletan~\cite{Saletan1961} studied the most general class of one-parametric contractions 
for which the elements of the corresponding matrices are first-order polynomials with respect to the contraction parameter.
In\"on\"u--Wigner contractions obviously form a special subclass in the class of \emph{Saletan contractions}.

Another extension of the class of In\"on\"u--Wigner contractions is given by \emph{generalized In\"on\"u--Wigner contractions}. 
They are generated by matrices which become diagonal after suitable choices of bases of initial and contracted algebras, 
and, moreover, diagonal elements should be integer powers of the contraction parameters. 
Contractions of this kind were introduced by Doebner and Melsheimer~\cite{Doebner&Melsheimer1967}.
At the best of our knowledge, the name `generalized In\"on\"u--Wigner contractions' first appears in~\cite{Hegerfeldt1967}. 
The other names (\emph{p-contractions}, \emph{Doebner--Melsheimer contractions} and \emph{singular IW-contractions}~\cite{Lyhmus1969}) 
are also used. 
Similar contractions are applied in the `purely mathematical' framework and called \emph{one-parametric subgroup degenerations}
\cite{Burde&Steinhoff1999,Burde1999,Burde2005,Grunewald&Halloran1988,Steinhoff1997}.  
The last name came from the algebraic invariant theory~\cite{Kraft}. 
Generalized IW-contractions are very useful for applications and were revisited many times.  
In particular, it was incorrectly conjectured that any continuous one-parametric
contraction is equivalent to a generalized In\"on\"u--Wigner contraction.

A general definition of contractions was first formulated by Segal~\cite{Segal1951} in terms of limiting processes of bases. 
It is used as an operational definition for calculations up to now. 
Saletan~\cite{Saletan1961} gave a more rigorous general definition of contractions, which is based on limiting processes of Lie brackets  
and allows one to avoid a confusion with limit state of bases, existing in the Segal approach.    
Saletan's definition was generalized for the case of arbitrary field in terms of Lie algebra orbit closures 
with respect to the Zariski topology. 
The generalization is a basis of modern investigation on contractions and was used by a number of authors, 
e.g.,~\cite{Burde1999,Burde2005,Burde&Steinhoff1999,Carles1979,Grunewald&Halloran1988,Kirillov&Neretin1984-1987,Lauret2003,Neretin1987,Seeley1990,Steinhoff1997}.
The name `\emph{degeneration}' is often used instead of the name `\emph{contraction}' in the generalized context.

A still more general notion of degenerations, which works in case of algebras of different dimensions, was proposed 
in~\cite{Gorbatsevich1991,Gorbatsevich1994,Gorbatsevich1998}. 
The algebra $\mathfrak g$ degenerates to the algebra $\mathfrak g_0$ according to Gorbatsevich if 
$\mathfrak g\oplus pA_1$ is contracted to $\mathfrak g_0\oplus qA_1$  in the usual sense for some $p,q\in\N\cup\{0\}$, 
where $pA_1$ and $qA_1$ are the $p$- and $q$-dimensional Abelian algebras.

The other type of contractions is given by the purely algebraic notion of \emph{graded contractions}
\cite{CouturePateraSharpWinternitz1991,HavlicekPateraPelantovaTolar2004,Hrivnak&Novotny&Patera&Tolar2006,Montigny&Patera2000,Patera1992,Weimar-Woods1995}.
The graded contraction procedure is the following. 
Structure constants of a graded Lie algebra are multiplied by numbers which 
are chosen in such a way that the multiplied structure constants give a Lie algebra with the same grading.
Graded contractions include discrete contractions as a subcase but do not cover all continuous~ones.

Different kinds of contractions and their properties were reviewed and compared in~\cite{Lyhmus1969}.
The interrelations between contractions and deformations or expansions were widely investigated 
\cite{Gerstenhaber1964,Levy-Nahas1967,Lyhmus1969,Fialowski&O'Halloran1990,Grunewald&Halloran1993}.
The related but principally different problem is given by contractions of Lie groups, which are also widely studied and applied. 
Notions of such contractions were introduced in~\cite{Brennich1974,Hermann1966,Mickelsson&Niederle1972,Saletan1961} with different levels of generality.

Problems concerning contractions of Lie algebra (or group) representations and 
simultaneous contractions of Lie algebras or Lie groups and their representations 
are also important and demand a special technique which differs from the techniques  
associated with pure contractions of Lie algebras and Lie groups.
In spite of existing works on the subject and a range of applications, 
these problems are not studied enough although a number of interesting results have been obtained. 
For example, the contractions of representations of de Sitter groups were described in~\cite{Mickelsson&Niederle1972}.
Contractions of matrix representations of concrete physically significant Lie algebras were investigated, e.g.,\ 
in~\cite{Montigny&Niederle&Nikitin2006, Nikitin2006, Weimar-Woods1991a}. 
Related theoretical inventions and different examples of application can also be found 
in~\cite{Celeghini&Tarlini1981, Inonu&Wigner1953, 
Inonu1964, Heredero&Levi&Rodriguez&Winternitz2000, Leng&Patera1994, Leng&Patera1995, Lyhmus1969, Moody&Patera1991, Patera1992} 
and in the references therein.

Intensive investigation of real and complex low-dimensional Lie algebras in last decades 
is motivated by a number of causes.
As subalgebras of important higher-dimensional Lie algebras, 
these algebras are widely applied in the theory of induced representations
(representations of subalgebras/subgroups are used to construct representations of the whole algebra/group),
in the representation theory
(chains of subalgebras can provide sets of commuting operators, 
eigenfunctions of which form bases of representation spaces for the corresponding Lie group)
and in study of broken symmetries.
Low-dimensional Lie algebras are also interesting per se and supply theoretical consideration with substantial examples.
In this connection classifications, subalgebras, realizations, invariants, contractions, deformations
and other objects concerning low-dimensional Lie algebras were
studied~\cite{Boyko&Patera&Popovych2006,Fialowski&Montigny2006,Fialowski&Penkava2005,
Levy-Nahas1967,Mubarakzyanov1963a,Patera&Winternitz1977,Popovych&Boyko&Nesterenko&Lutfullin2003b}.

Contractions of low-dimensional Lie algebras naturally appeared as illustrating examples in a number of papers. 
Thus, in the~pioneer paper on contractions~\cite{Segal1951} Segal adduced two such contractions, namely, 
the contractions from $so(3)$ and $sl(2,\R)$ to the Weyl--Heisenberg algebra $\mathfrak h_3=A_{3.1}$.
Some examples are contained also in the known paper by Saletan~\cite{Saletan1961}. 
Afterwards contractions of low-dimensional Lie algebras became independent subject of investigation. 
In\"on\"u--Wigner contractions of real three-dimensional Lie algebras were considered~\cite{Sharp1960} but some cases were missed. 
These results were partially amended in~\cite{Lyhmus1969}.
First In\"on\"u--Wigner contractions of real three-dimensional Lie algebras were exhaustively described 
by Conatser~\cite{Conatser1972}. 
Using the known classification of subalgebras of real low-dimensional Lie algebras~\cite{Patera&Winternitz1977},
Huddleston~\cite{Huddleston1978} constructed In\"on\"u--Wigner contractions of the four-dimensional real Lie algebras.
All~inequivalent continuous one-parametric contractions of real three-dimensional Lie algebras were obtained 
in~\cite{Weimar-Woods1991} but contractions inside parameterized series of algebras were not discussed.
The~same problem was nicely solved by Lauret~\cite{Lauret2003} in terms of orbit closures 
using a non-evident connection between algebraic characterization of Lie groups having metrics with special 
curvature properties and existence of degenerations for Lie algebras. 
Orbit closures of complex three- and four-dimensional Lie algebras were studied 
in~\cite{Burde&Steinhoff1999,Burde2005,Steinhoff1997}. 
It is the works from which we adopted the fruitful idea on usage of a wide set of necessary contraction criteria. 
The same subject was also investigated in~\cite{Agaoka1999,Agaoka2002}. 
In these papers obtained results were presented in a very simple and clear form due to special 
improvement of classification of complex three- and four-dimensional Lie algebras.

Complexity of description of algebra orbit closures is exponentially increased under growing dimension of the underlying vector space. 
A possible ways of simplification is to consider a closed subclass of Lie algebras (e.g.,\ nilpotent algebras)
instead of the whole class of Lie algebras of a fixed dimension. 
Degenerations of nilpotent algebras were studied 
in~\cite{Grunewald&Halloran1988}, \cite{Seeley1990} and \cite{Burde1999,Burde2005} in the case of 
dimensions five, six and seven correspondingly. 

Deformations of low-dimensional Lie algebras are also treated intensively.
Thus, deformations of three-dimensional real Lie algebras were described in~\cite{Levy-Nahas1967}.
The four-dimensional case was completely studied over the complex field~\cite{Fialowski&Penkava2005}.
There also exist a number of papers on contractions and deformations
of  higher- or even infinite-dimensional Lie algebras (see, e.g.,~\cite{Montigny1996}).
Since this subject is out of the scope of our paper, we do not review it here in detail.

Investigation of contractions is motivated by numerous applications in different fields of physics and mathematics, 
e.g.,\ in study of representations, invariants and special functions~\cite{Moody&Patera1991,Montigny&Niederle&Nikitin2006,Campoamor-Stursberg2002}.
It is one of the tools to recognize structure of Lie algebra varieties~\cite{Burde2005}. 
The Wigner coefficients of the Euclidean group $E(3)$
were constructed with contracting the Wigner coefficients of the special orthogonal
group $SO(4)$~\cite{Holman1969}.
Contractions were used to establish connection between various kinematical groups and
to shed a light on their physical meaning.
In this way relationship between the conformal and Schr\"odinger groups was elucidated~\cite{Barut1973}
and various Lie algebras including a relativistic position operator were interrelated.
Under dynamical group description of interacting systems,
contractions corresponding to the coupling constant going to zero give noninteracting systems~\cite{Doebner&Melsheimer1968}.
Application of contractions allows to derive interesting results in the special function theory and on the variable separation method 
\cite{Izmestev&Pogosyan&Sissakian&Winternitz1996-1999-2001, Pogosyan&Sissakian&Winternitz2002, Heredero&Levi&Rodriguez&Winternitz2000}.

Contractions of low-dimensional Lie algebras also play an important role from the physical point of view. 
It is illustrated by the following simple examples which are related to physics.
We use the standard physical notations and numeration by Mubarakzyanov~\cite{Mubarakzyanov1963a} simultaneously. 
Hereafter, describing a Lie algebra, we adduce only the nonzero commutators of fixed basis elements. 
See Section~\ref{SectionOnOne-parContractionsOfRealLow-DimLieAlgebras} for notations and more examples.

\looseness=1
The four-dimensional Lie algebra $u(2)=sl(2,\R)\oplus A_1$ 
has the nonzero commutation relations $[e_1,e_2]=e_1$, $[e_2,e_3]=e_3$ and $[e_1,e_3]=2e_2$.
The matrix $U_1(\varepsilon)=I_{10}\diag(\varepsilon,\varepsilon,1,1)$ provides a contraction 
of $u(2)$ to the algebra $e(2)\oplus A_1=A_{3.5}^0\oplus A_1$ 
($[e_1,e_3]=-e_2$, $[e_2,e_3]=e_1$),
i.e.,\ to the direct sum of the three-dimensional Euclidean algebra and the one-dimensional Abelian algebra.

The other example is the contraction of $u(2)$ to the harmonic oscillator algebra $\mathfrak h_4=A_{4.8}^{-1}$ 
($[e_2,e_3]=e_1$, $[e_2,e_4]=e_2$, $[e_3,e_4]=-e_3$)
which frequently occurs in physics. 
The `physical' name of $\mathfrak h_4$ is justified since 
the set consisting of the creation ($a^+$), annihilation ($a^-$), identity ($I$) and single-mode photon number ($N=a^+a^-$) 
operators is closed under commutation and generates a Lie algebra isomorphic to $\mathfrak h_4$ 
with $e_1=I$, $e_2=a^-$, $e_3=a^+$, $e_4=N$. 
The algebra $u(2)$ is contracted to $\mathfrak h_4$ with the matrix
$U_2(\varepsilon)=I_{19}\diag(\varepsilon,1,\varepsilon,1)$

The subalgebra $\mathfrak h_3=\langle e_1,\, e_2,\, e_3\rangle$ ($[e_2,e_3]=e_1$) 
of $\mathfrak h_4$ is also widely applied since 
it is isomorphic to the algebra formed by the quantum mechanical position operator $Q$, the momentum operator $P$
and the identity operator $I$ via designation
\[
e_1=I,\quad e_2=\frac{Q+iP}{\sqrt{2\hslash}},\quad e_3=\frac{Q-iP}{\sqrt{2\hslash}}.
\]

The main purpose of our paper is to classify contractions of the real and complex Lie algebras of dimensions
not greater than four. We rigorously formulate and develop a theoretical background to do this.
Effectiveness of the applied algorithm for handling of contractions is based on using a wide set of necessary contraction criteria. 
A number of known necessary contraction criteria are collected and new criteria are proposed.
Requisite invariant and semiinvariant quantities are calculated for classes of Lie algebras including all low-dimensional Lie algebras.
Multi-parametric and repeated contractions are also investigated since they give a tool for finding contraction matrices in 
complicated cases.
An important by-effect of the present investigation is that 
the contractions under consideration supply with a number of model examples and contrary instances 
for statements and conjectures of the contraction theory. 
Availability of exhaustive information about them also allows us to describe levels and colevels of low-dimensional Lie algebras completely.

This paper is arranged in the following way.

In Section~\ref{SectionOnDefsOfContractionsAndTheirEquvalence} different definitions 
of general contractions of Lie algebras 
and contraction equivalence are given and discussed. 
We also construct a contrary instance on a conjecture on equivalence of contractions. 
Simplest types of contractions (In\"on\"u--Wigner contractions, Saletan contractions and 
generalized In\"on\"u--Wigner contractions) 
are described in Section~\ref{SectionOnSimplestTypesOfContractions}. 
Necessary contraction criteria are listed and proved in Section~\ref{SectionOnNecessaryContractionCriteria}.
Calculation of invariant quantities for wide classes of Lie algebras is adduced in Section~\ref{SectionOnCalculationOfInvariantQuantities}.
Section~\ref{SectionOnLow-DimRealLieAlgebras} collects algebraic quantities and objects concerning real three- and
four-dimensional Lie algebras. 
These quantities are used in Section~\ref{SectionOnOne-parContractionsOfRealLow-DimLieAlgebras} 
as a base for application of necessary contraction criteria 
in order to conclude whether there is a contraction in an arbitrary pair of the algebras of the same dimension.
An algorithm for handling of contractions of low-dimensional Lie algebras is precisely formulated
in Section~\ref{SectionOnContractionIdentification} and illustrated by examples.
All inequivalent one-parametric contractions of the real low-dimensional Lie algebras are arranged 
in Section~\ref{SectionOnOne-parContractionsOfRealLow-DimLieAlgebras}
and supplied with diagrams and explicit forms of the contraction matrices. 
All cases where contractions are equivalent to simple or generalized In\"on\"u--Wigner contractions are separated. 
Levels and colevels of low-dimensional Lie algebras are also investigated.
Using known correspondence between lists of nonisomorphic real and complex low-dimensional Lie algebras, 
we construct all inequivalent contractions over the complex field in Section~\ref{SectionOnOne-parContractionsOfComplexLow-DimLieAlgebras}. 
They are compared with the degenerations of four-dimensional Lie algebras, which were found in~\cite{Agaoka2002,Burde&Steinhoff1999}.
Multi-parametric and repeated contractions are studied in Section~\ref{SectionOnMulti-parametricDecomposableAndRepeatedContractions} 
and used for construction of contraction matrices in the most complicated cases which are not covered by generalized In\"on\"u--Wigner contractions.
Some problems arising under analysis of obtained results are formulated in the Conclusion.

\section{Definitions of contractions and their equivalence}\label{SectionOnDefsOfContractionsAndTheirEquvalence}

Consider an $n$-dimensional Lie algebra ${\mathfrak g}=(V,[\cdot,\cdot])$
with an underlying $n$-dimensional vector space $V$
over $\R$ or $\CC$ and a Lie bracket~$[\cdot,\cdot]$.
Usually the Lie algebra ${\mathfrak g}$ is defined by means of
commutation relations in a~fixed basis~$\{e_1, \ldots, e_n\}$ of $V$.
More precisely, it is sufficient to write down only the nonzero commutators $[e_i,e_j]=c^{k}_{ij}e_k$,
where $c^{k}_{ij}$ are components of the structure constant tensor of ${\mathfrak g}$.
Hereafter the indices $i$, $j$, $k$, $i'\!$, $j'\!$, $k'\!$, $i''\!$, $j''\!$ and $k''$ 
run from 1 to $n$ and the summation over the repeated indices is implied.

Consider a \textit{continuous} function $U\colon (0,\varepsilon_1]\to GL(V)$, where $\varepsilon_1>0$.
In other words, $U_\varepsilon=U(\varepsilon)$ is a nonsingular linear operator on $V$
for all $\varepsilon \in (0,\varepsilon_1]$. Without loss of generality we can put $\varepsilon_1=1$.
A parameterized family of new Lie brackets on~$V$ is determined via the old one by the following way:
\[
\forall\; \varepsilon \in (0,1],\ \forall \; x, y\in V\colon\quad
[x,y]_{\varepsilon}=U_\varepsilon{}^{-1}[U_\varepsilon x,U_\varepsilon y].
\]
It is reasonable that for any $\varepsilon \in (0,1]$ the Lie algebra ${\mathfrak g}_{\varepsilon}=(V,[\cdot,\cdot]_{\varepsilon})$
is isomorphic to ${\mathfrak g}$.

\begin{definition}\label{DefOfContractions1}
If the limit
$
\lim\limits_{\varepsilon \to +0}[x,y]_\varepsilon=
\lim\limits_{\varepsilon \to +0}U_\varepsilon{}^{-1}[U_\varepsilon x,U_\varepsilon y]=:[x,y]_0
$
exists for any $x, y\in V$ then $[\cdot,\cdot]_0$ is a well-defined Lie bracket.
The Lie algebra ${\mathfrak g}_0=(V,[\cdot,\cdot]_0)$ is called a \emph{one-parametric continuous contraction}
(or simply a \emph{contraction}) of the Lie algebra~${\mathfrak g}$.
\end{definition}

If a basis of~$V$ is fixed, the operator $U_\varepsilon$ is defined by the corresponding matrix.
Definition~\ref{DefOfContractions1} can be reformulated in terms of structure constants.

{\addtocounter{definition}{-1}\renewcommand{\thedefinition}{\arabic{definition}$'$}
\begin{definition}\label{DefOfContractions2}
Let ${c}^k_{ij}$ be the structure constants of the algebra~${\mathfrak g}$ in the fixed basis~$\{e_1, \ldots, e_n\}$.
If the limit
\[\lim\limits_{\varepsilon\to+0}(U_\varepsilon)_{i'}^i(U_\varepsilon)_{j'}^j(U_\varepsilon{}^{-1})_k^{k'}c^{k}_{ij}=:\tilde c^{k'}_{i'\!j'}\]
exists for all values of $i'$, $j'$ and $k'$ then
$\tilde c^{k'}_{i'\!j'}$ are components of the well-defined structure constant tensor of a Lie algebra~${\mathfrak g}_0$.
In this case the Lie algebra~${\mathfrak g}_0$ is called
a \emph{one-parametric continuous contraction} (or simply \emph{contraction}) of the Lie algebra ${\mathfrak g}$.
The parameter $\varepsilon$ and the matrix-function $U=U(\varepsilon)$ are called a \emph{contraction parameter} and a \emph{contraction matrix}
correspondingly. The procedure that provides the Lie algebra~${\mathfrak g}_0$ from the
algebra~${\mathfrak g}$ is also called a \emph{contraction}.
\end{definition}}

Definitions~\ref{DefOfContractions1} and~\ref{DefOfContractions2} are equivalent.
The first definition is basis-free and convenient for theoretical consideration.
The second one is more usable for calculations of concrete contractions.
In this paper we mainly use Definition~\ref{DefOfContractions2}.

The well-known In\"on\"u--Wigner~\cite{Inonu&Wigner1953}, Saletan~\cite{Saletan1961} and generalized In\"on\"u--Wigner~\cite{Doebner&Melsheimer1967} 
contractions are particular cases of the above one-parametric continuous contractions.

\begin{definition}
We call a contraction from the Lie algebra ${\mathfrak g}$ to the Lie algebra ${\mathfrak g}_0$
\emph{trivial} if ${\mathfrak g}_0$ is Abelian and \emph{improper} if ${\mathfrak g}_0$ is isomorphic to ${\mathfrak g}$.
\end{definition}

If there exists a componentwise limit $\lim\limits_{\varepsilon\to+0} U_\varepsilon=:U_0$
and $U_0\in GL(V)$ then it is obvious that the contraction is improper.
Therefore, in order to generate a proper contraction, the matrix-function~$U$ have to satisfy one of the conditions:
1) there is no limit of~$U$ at $\varepsilon\to+0$,
i.e.,\ at least one of the elements of~$U$ is singular under $\varepsilon\to+0$,
or
2) there exists $\lim\limits_{\varepsilon\to+0} U_\varepsilon=:U_0$ but the matrix~$U_0$ is singular.
Both the conditions are not sufficient for the contraction to be proper.

The trivial and improper contractions exist for any Lie algebra.
The trivial contraction is easily provided, e.g.,\ by the matrix
$U_\varepsilon={\rm diag}(\varepsilon, \varepsilon, \dots, \varepsilon)$.
As a contraction matrix of the improper contraction, the identity matrix $U_\varepsilon={\rm diag}(1, 1, \dots, 1)$ can be always used.
Sometimes the trivial and improper contractions are united in the common class
of trivial contractions~\cite{Weimar-Woods2000}.

The Abelian algebra is contracted only to itself. 
It is a special case when the contraction is trivial and improper at the same time.

\begin{definition}\label{equivalent contr}
Let the Lie algebras ${\mathfrak g}$ and $\tilde{\mathfrak g}$ be contracted
to the algebras ${\mathfrak g}_0$ and $\tilde{\mathfrak g}_0$, correspondingly.
If $\tilde{\mathfrak g}$ is isomorphic to ${\mathfrak g}$ and
$\tilde{\mathfrak g}_0$ is isomorphic to ${\mathfrak g}_0$ then the contractions are called \emph{weakly equivalent}.
\end{definition}

Roughly speaking, all contractions in the same pairs of Lie algebras are weakly equivalent.
Under usage of weak equivalence, attention is concentrated on possibility and results of contractions.
Difference in ways of contractions is neglected by this approach.
For parametric contractions we can also introduce different notions of stronger equivalence,
which take into account ways of contractions.
Hereafter $\mathop{\rm Aut}(\mathfrak g)$ denotes the automorphism group of the Lie algebra~$\mathfrak g$ and
$\mathop{\rm Iso}(\mathfrak g,\tilde{\mathfrak g})$ denotes the set of isomorphisms from the Lie algebra~$\mathfrak g$
to the Lie algebra~$\tilde{\mathfrak g}$. Additionally we identify isomorphisms with the corresponding matrices in a fixed basis.

\begin{definition}\label{DefStrongEquivOfContractions}
Two one-parametric contractions in the same pair of Lie algebras $(\mathfrak g,\mathfrak g_0)$
with the contraction matrices $U(\varepsilon)$ and $\tilde U(\varepsilon)$ are called \emph{strictly equivalent} if
there exists $\delta\in(0,1]$,
there exist functions
$\hat U\colon(0,\delta]\to\mathop{\rm Aut}(\mathfrak g)$ and
$\check U\colon(0,\delta]\to\mathop{\rm Aut}(\mathfrak g_0)$
and a continuous monotonic function $\varphi\colon(0,\delta]\to(0,1]$,
$\displaystyle\lim_{\varepsilon\to +0}\varphi(\varepsilon)=0$, such that
\[
\tilde U_\varepsilon=\hat U_\varepsilon U_{\varphi(\varepsilon)}\check U_\varepsilon,\quad \varepsilon\in(0,\delta].
\]
\end{definition}

The latter definition can be reformulated for different pairs of algebras, which are term-by-term isomorphic.

{\addtocounter{definition}{-1}\renewcommand{\thedefinition}{\arabic{definition}$'$}
\begin{definition}\label{DefStrongEquivOfContractions2}
Let the isomorphic Lie algebras $\mathfrak g$ and $\tilde{\mathfrak g}$ be contracted
to the isomorphic algebras $\mathfrak g_0$ and $\tilde{\mathfrak g}_0$
with the contraction matrices $U(\varepsilon)$ and $\tilde U(\varepsilon)$ correspondingly.
These contractions are called \emph{strictly equivalent} if there exists $\delta\in(0,1]$,
there exist functions
$\hat U\colon(0,\delta]\to\mathop{\rm Iso}(\mathfrak g,\tilde{\mathfrak g})$ and
$\check U\colon(0,\delta]\to\mathop{\rm Iso}(\mathfrak g_0,\tilde{\mathfrak g}_0)$ and
a continuous monotonic function $\varphi\colon(0,\delta]\to(0,1]$,
$\displaystyle\lim_{\varepsilon\to +0}\varphi(\varepsilon)=0$, such that
\[
\tilde U_\varepsilon=\hat U_\varepsilon^{-1}U_{\varphi(\varepsilon)}\check U_\varepsilon,\quad \varepsilon\in(0,\delta].
\]
\end{definition}}

Strictly equivalent contractions obviously are weakly equivalent.
In our consideration we use only the notion of weak equivalence hence weakly equivalent contractions
will be called equivalent ones for simplicity.

\begin{remark}
The restriction that $\hat U$ and $\check U$ should be isomorphism matrices cannot be omitted with preserving correctness.

The contractions of a Lie algebra, which are defined by 
the matrices $U_\varepsilon$ and $W_0 U_\varepsilon\tilde W_0$, where
$U\colon (0,1]\to GL(V)$, $W_0,\tilde W_0\in GL(V)$, are weakly inequivalent in the general case.
For example, the algebra $sl(2,{\mathbb R})$ ($[e_1,e_2]=e_1$, $[e_2,e_3]=e_3$, $[e_1,e_3]=2e_2$) is contracted
to the Heisenberg algebra $A_{3.1}$ ($[e_2,e_3]=e_1$) with the matrix $I_3\diag(\varepsilon,\varepsilon,1)$ and
to the algebra $A_{3.5}^0$ (\mbox{$[e_1,e_3]=-e_2$}, $[e_2,e_3]=e_1$) with the matrix $I_5\diag(\varepsilon,\varepsilon,1)$.
Here $I_3$ and $I_5$ are nonsingular matrices defined in Section~\ref{Contractions_lists_3}.

Moreover, let $W,U,\tilde W\colon (0,1]\to GL(V)$ and
\[
\exists\lim_{\varepsilon\to +0}W_\varepsilon=:W_0\in GL(V),\quad
\exists\lim_{\varepsilon\to +0}\tilde W_\varepsilon=:\tilde W_0\in GL(V).
\]
Generally speaking, the matrices $W_\varepsilon U_\varepsilon\tilde W_\varepsilon$ and
$W_0 U_\varepsilon\tilde W_0$ can also give weakly inequivalent contractions.
This statement is illustrated by the below example. Therefore, Lemma~2.2 of~\cite{Weimar-Woods2000} is incorrect.
\end{remark}

\begin{example}
Consider the one-parametric continuous contraction of the four-dimensional real Lie algebras
$so(3)\oplus A_1\to A_{4.1}$ given by the matrix
\begin{gather*}
U_\varepsilon=
\left(
\begin{array}{cccc}
0 & 0 & \varepsilon^2 & 0\\
0 & -\varepsilon^3 & 0 & 0\\
0 & 0 & 0 & \varepsilon\\
-\varepsilon^2 & 0 & -1 & 0
\end{array}
\right)\quad
\text{with}\quad
U_\varepsilon^{-1}=
\left(
\begin{array}{cccc}
-\varepsilon^{-4} & 0 & 0 & -\varepsilon^{-2}\\
0 & -\varepsilon^{-3} & 0 & 0\\
\varepsilon^{-2} & 0 & 0 & 0\\
0 & 0 & \varepsilon^{-1} & 0
\end{array}
\right).
\end{gather*}
Taking the canonical commutation relations $[e_1,e_2]=e_3$, $[e_2,e_3]=e_1$, $[e_3,e_1]=e_2$
of the algebra $so(3)\oplus A_1$ (the commutators with~$e_4$ vanish),
we calculate the transformed commutators up to antisymmetry:
\begin{gather*}
[e_1,e_2]_{\varepsilon}=0,
\quad
[e_1,e_3]_{\varepsilon}=0,
\quad
[e_1,e_4]_{\varepsilon}=0,
\quad
[e_2,e_3]_{\varepsilon}=\varepsilon^4e_4,
\\
[e_2,e_4]_{\varepsilon}=e_1-\varepsilon^2e_3,
\quad
[e_3,e_4]_{\varepsilon}=e_2.
\end{gather*}
After the limiting process $\varepsilon\to+0$ we obtain the canonical commutation relations
$[e_2,e_4]_0=e_1$, $[e_3,e_4]_0=e_3$ of the algebra $A_{4.1}$.

Let us fix an arbitrary $\varepsilon\in (0,1]$.
Since the matrix $U_\varepsilon$ is nonsingular,
its polar decomposition has the form $U_\varepsilon=P_\varepsilon T_\varepsilon$, where
$P_\varepsilon :=(U_\varepsilon U_\varepsilon{}^\text{T})^{1/2}$ is
a real symmetric matrix with positive eigenvalues and
$T_\varepsilon :=P_\varepsilon{}^{-1}U_\varepsilon$ is a real orthogonal matrix.
Denote a real orthogonal matrix which reduces $P_\varepsilon$ to a diagonal matrix $D_\varepsilon$ by $W_\varepsilon $,
i.e.,\ $P_\varepsilon=W_\varepsilon D_\varepsilon W_\varepsilon{}^\text{T}$.
As a result, we derive the representation $U_\varepsilon=W_\varepsilon D_\varepsilon\tilde W_\varepsilon$,
where $\tilde W_\varepsilon=W_\varepsilon{}^\text{T}T_\varepsilon=D_\varepsilon{}^{-1} W_\varepsilon{}^\text{T}U_\varepsilon$
is an orthogonal matrix.
The~explicit form of the matrices $W_\varepsilon$, $D_\varepsilon$ and $\tilde W_\varepsilon$ is
\begin{gather*}
W_\varepsilon =
\left(
\begin{array}{cccc}
-\theta_- & 0 & 0 & \theta_+\\
0 & 1 & 0 & 0\\
0 & 0 & 1 & 0\\
\theta_+ & 0 & 0 & \theta_-
\end{array}
\right),\quad
\tilde W_\varepsilon =
\left(
\begin{array}{cccc}
-\theta_- & 0 & -\theta_+ & 0\\
0 & -1 & 0 & 0\\
0 & 0 & 0 & 1\\
-\theta_+ & 0 & \theta_- & 0
\end{array}
\right),
\\[.5ex]
D_\varepsilon =\diag\Bigl(K+\frac12,0,0,K-\frac12\Bigr),
\end{gather*}
where
\[
K=\frac12\sqrt{4\varepsilon^4+1},\quad
\theta_+=\sqrt{\frac{2K+1}{4K}},\quad
\theta_-=\sqrt{\frac{2K-1}{4K}}.
\]

The matrices $W_\varepsilon$ and $\tilde W_\varepsilon$ converge under $\varepsilon\to+0$
to the constant nonsingular matrices
\begin{gather*}
W_0=
\left(
\begin{array}{cccc}
0 & 0 & 0 & 1\\
0 & 1 & 0 & 0\\
0 & 0 & 1 & 0\\
1 & 0 & 0 & 0
\end{array}
\right)
\quad\mbox{and}\quad
\tilde W_0=
\left(
\begin{array}{cccc}
0 & 0 & -1 & 0\\
0 & -1 & 0 & 0\\
0 & 0 & 0 & 1\\
-1 & 0 & 0 & 0
\end{array}
\right).
\end{gather*}

Consider the matrix $\tilde U_\varepsilon=W_0D_\varepsilon\tilde W_0$ constructed from the representation
$U_\varepsilon=W_\varepsilon D_\varepsilon\tilde W_\varepsilon$
with replacement of the matrices $W_\varepsilon$ and $\tilde W_\varepsilon$ by their regular limits.
We transform the canonical commutation relations of the algebra $so(3)\oplus A_1$
with the matrix $\tilde U_\varepsilon$ and limit $\varepsilon\to+0$:
\begin{gather*}
[e_1,e_2]_{\varepsilon}=\frac12(\sqrt{4\varepsilon^4+1}-1)e_4\to 0,
\quad
[e_1,e_3]_{\varepsilon}=0,
\\
[e_1,e_4]_{\varepsilon}=-\frac{\sqrt{4\varepsilon^4+1}-1}{2\varepsilon^2}e_2\to 0,
\quad
[e_2,e_3]_{\varepsilon}=0,
\\
[e_2,e_4]_{\varepsilon}=\frac{2\varepsilon^4}{\sqrt{4\varepsilon^4+1}-1}e_1\to e_1,
\quad
[e_3,e_4]_{\varepsilon}=0.
\end{gather*}
As a result, we obtain commutation relations of the algebra $A_{3.1}\oplus A_1$. 
Therefore, the matrices $U_\varepsilon$ and $\tilde U_\varepsilon$ lead to weakly inequivalent contractions.
\end{example}

The notion of sequential contractions is introduced similar to continuous contractions.  
See, e.g.,~\cite{Huddleston1978,Weimar-Woods2000}.

Consider a sequence of $U_p\in GL(V)$, $p\in\N$.
The corresponding sequence of new Lie brackets on~$V$ is determined via the old one by the condition
$[x,y]_p=U_p{}^{-1}[U_p x,U_p y]$ $\forall\; p\in\N$, $\forall \; x, y\in V$.
For any $p\in\N$ the Lie algebra ${\mathfrak g}_p=(V,[\cdot,\cdot]_p)$ is isomorphic to ${\mathfrak g}$.

\begin{definition}\label{DefOfSequetialContractions1}
If the limit
$
\lim\limits_{p\to\infty}[x,y]_p=
\lim\limits_{p\to\infty}U_p{}^{-1}[U_p x,U_p y]=:[x,y]_0
$
exists for any $x, y\in V$ then the Lie bracket~$[\cdot,\cdot]_0$ is well-defined.
The Lie algebra ${\mathfrak g}_0=(V,[\cdot,\cdot]_0)$ is called a \emph{sequential contraction} of the Lie algebra~${\mathfrak g}$.
\end{definition}

Any continuous contraction from $\mathfrak g$ to $\mathfrak g_0$ gives an infinite family of matrix sequences resulting in
the sequential contraction from $\mathfrak g$ to $\mathfrak g_0$.
More precisely, if $U_\varepsilon$ is the matrix of the continuous contraction and
the sequence $\{\varepsilon_p,\, p\in\N\}$ satisfies the conditions $\varepsilon_p\in(0,1]$, $\varepsilon_p\to+0$, $p\to\infty$,
then $\{U_{\varepsilon_p},\, p\in\N\}$ is a suitable matrix sequence.

The notion of contraction is generalized to arbitrary fields in terms of orbit closures in the variety of Lie algebras
\cite{Burde&Steinhoff1999,Burde1999,Burde2005,Gorbatsevich1991,Gorbatsevich1998,Grunewald&Halloran1988,Lauret2003}.

Let $V$ be an $n$-dimensional vector space over a field~$\mathbb K$ and 
$\mathcal L_n=\mathcal L_n(\mathbb K)$ denote the set of all possible Lie brackets on~$V$. 
We identify $\mu\in\mathcal L_n$ with the corresponding Lie algebra $\mathfrak g=(V,\mu)$.
$\mathcal L_n$ is an algebraic subset of the variety $V^*\otimes V^*\otimes V$ of bilinear maps from $V\times V$ to $V$.
Indeed, under setting a basis $\{e_1,\dots,e_n\}$ of~$V$ 
there is the one-to-one correspondence between $\mathcal L_n$ and 
\[
\mathcal C_n=\{(c_{ij}^k)\in\mathbb K^{n^3}\mid c_{ij}^k+c_{ji}^k=0,\,
c_{ij}^{i'\!}c_{i'\!k}^{k'\!}+c_{ki}^{i'\!}c_{i'\!j}^{k'\!}+c_{jk}^{i'\!}c_{i'\!i}^{k'\!}=0\},
\]
which is determined for any Lie bracket $\mu\in\mathcal L_n$ and 
any structure constant tuple $(c_{ij}^k)\in\mathcal C_n$ by the formula $\mu(e_i,e_j)=c_{ij}^ke_k$.
$\mathcal L_n$ is called the \emph{variety of $n$-dimensional Lie algebras (over the field~$\mathbb K$)}
or, more precisely, the variety of possible Lie brackets on~$V$.
The group~$GL(V)$ acts on $\mathcal L_n$ in the following way: 
\[
(U\cdot\mu)(x,y)=U\bigl(\mu(U^{-1}x,U^{-1}y)\bigr)\quad \forall U\in GL(V),\forall \mu\in\mathcal L_n,\forall x,y\in V.
\]
(It is a left action in contrast to the right action which is more usual for the `physical' contraction tradition 
and defined by the formula $(U\cdot\mu)(x,y)=U^{-1}\bigl(\mu(Ux,Uy)\bigr)$ that is not of fundamental importance.
We use the right action all over the paper except this paragraph.) 
Denote the orbit of $\mu\in\mathcal L_n$ under the action of~$GL(V)$ by $\mathcal O(\mu)$ and 
the closure of it with respect to the Zariski topology on~$\mathcal L_n$ by $\overline{\mathcal O(\mu)}$.

\begin{definition}\label{DefOfContractionsViaOrbitClosure}
The Lie algebra ${\mathfrak g}_0=(V,\mu_0)$ is called a \emph{contraction} (or \emph{degeneration}) 
of the Lie algebra~${\mathfrak g}=(V,\mu)$ if $\mu_0\in\overline{\mathcal O(\mu)}$. 
The contraction is \emph{proper} if $\mu_0\in\overline{\mathcal O(\mu)}\backslash\mathcal O(\mu)$.
The contraction is \emph{nontrivial} if $\mu_0\not\equiv0$.
\end{definition}

In the case of $\mathbb K=\CC$ or $\mathbb K=\R$ the orbit closures with respect to the Zariski topology coincide 
with the orbit closures with respect to the Euclidean topology and Definition~\ref{DefOfContractionsViaOrbitClosure} is 
reduced to the usual definition of contractions.

\section{Simplest types of contractions}\label{SectionOnSimplestTypesOfContractions}

In\"on\"u--Wigner contractions present limit processes between Lie algebras with contraction matrices of simplest types. 
Most of contractions of low-dimensional Lie algebras are equivalent to such contractions.
We discuss their properties which are essential for further consideration.

\emph{Simple In\"on\"u--Wigner contractions} or shortly IW-contractions first proposed in~\cite{Inonu&Wigner1953} 
are generated by matrices of the form $U_\varepsilon=U_0+\varepsilon U'_0$, where $U_0$ and $U'_0$ are constant $n\times n$ matrices.
The matrix $U_\varepsilon$ is additionally assumed to be transformable
to the special diagonal form 
$\hat WU_\varepsilon \check W^{-1}=\diag(1+\varepsilon v,\dots,1+\varepsilon v,\varepsilon,\dots,\varepsilon)
=:D_\varepsilon$
by means of the regular constant matrices $\hat W$ and $\check W$.
The assumption was investigated by In\"on\"u and Wigner themselves~\cite{Inonu&Wigner1954}. 
Without loss of generality we can put $v=0$. 
The matrix~$D_\varepsilon$ provides the contractions from~$\tilde{\mathfrak g}$ to~$\tilde{\mathfrak g}_0$. 
Here~$\tilde{\mathfrak g}$ and~$\tilde{\mathfrak g}_0$ are Lie algebras with the Lie brackets 
$[x,y]\,\tilde{}=\hat W[\hat W^{-1}x,\hat W^{-1}y]$ and 
$[x,y]_0\!\tilde{}=\check W[\check W^{-1}x,\check W^{-1}y]_0$, 
which are obviously isomorphic to~$\mathfrak g$ and~$\mathfrak g_0$. 
Therefore, it can be assumed at once that $U_\varepsilon=D_\varepsilon$, i.e.
\[
U_\varepsilon=\diag(1,\dots,1,\varepsilon,\dots,\varepsilon).
\]
Denote the number of diagonal elements equal to $1$ by~$s$. Then, the dimension of $\varepsilon$-block is $n-s$.
It is convenient to divide the set of basis elements $\{e_1,\dots, e_n\}$ of~$V$ 
into two subsets $\{e_1,\dots,e_s\}$ and $\{e_{s+1},\dots,e_n\}$ according to the values of diagonal elements.
Since 
\[
[e_{i_1},e_{j_1}]_\varepsilon=c_{i_1j_1}^{k_1}e_{k_1}+\frac{1}{\varepsilon}c_{i_1j_1}^{k_2}e_{k_2}+O(\varepsilon)
\to \tilde c_{i_1j_1}^{k_1}e_{k_1}+\tilde c_{i_1j_1}^{k_2}e_{k_2}, \quad \varepsilon\to+0,
\]
where the indices $i_1$, $j_1$ and $k_1$ run from 1 to~$s$ and the indices $i_2$, $j_2$ and $k_2$ run from $s+1$ to~$n$, 
then $c_{i_1j_1}^{k_2}=0$.
Therefore, the basis elements $e_1,\dots,e_s$ generate a subalgebra $\mathfrak h$ of the initial algebra $\mathfrak g$.
It is the unique condition for the contraction to exist.
All structure constants of the resulting algebra $\mathfrak g_0$ are easily calculated:
\begin{gather*}
\tilde c_{i_1j_1}^{k_1}=c_{i_1j_1}^{k_1},\quad
\tilde c_{i_1j_1}^{k_2}=c_{i_1j_1}^{k_2}=0,\quad
\tilde c_{i_1j_2}^{k_1}=0,\quad
\tilde c_{i_1j_2}^{k_2}=c_{i_1j_2}^{k_2},\quad
\tilde c_{i_2j_2}^{k_1}=\tilde c_{i_2j_2}^{k_2}=0.
\end{gather*}

Let us make a summary of properties of the IW-contractions (see, e.g.,~\cite{Saletan1961,Lyhmus1969} for some properties). 
Each subalgebra $\mathfrak h$ of the Lie algebra~$\mathfrak g$ can be used to obtain an IW-contraction of~$\mathfrak g$. 
Improper subalgebras correspond to improper ($\mathfrak h=\mathfrak g$) or trivial ($\mathfrak h=\{0\}$) IW-contractions.
Different choices of basis complement to a basis of~$\mathfrak h$ or replacement of~$\mathfrak h$ by an equivalent subalgebra of~$\mathfrak g$ 
give the same contracted algebra up to isomorphism. 
The contracted algebra $\mathfrak g_0$ has the structure of semidirect sum $\mathfrak h\rsemioplus\mathfrak a$, where 
$\mathfrak a$ is the Abelian ideal spanned on the chosen basis complement to a basis of~$\mathfrak h$.
The subalgebra $\mathfrak h$ is isomorphic to the quotient algebra $\mathfrak g_0/\mathfrak a$. 
And vice versa, the Lie algebra~$\mathfrak g_0$ is an IW-contraction of the algebra $\mathfrak g$ with the subalgebra $\mathfrak h$ iff 
there exists an Abelian ideal $\mathfrak a\subset\mathfrak g_0$ 
for which the quotient algebra $\mathfrak g_0/\mathfrak a$ is isomorphic to  $\mathfrak h$.
The repeated IW-contraction with the same subalgebra $\mathfrak h$ again results in the algebra~$\mathfrak g_0$.

Any IW-contraction satisfies two assumptions: 1) the contraction matrix is linear with respect to the contraction parameter;
2) there exist constant regular matrices $\hat W$ and $\check W$ diagonalizing the contraction matrix.
It is well known that IW-contractions do not exhaust all possible contractions even in the case of three-dimensional Lie algebras. 
IW-contractions of the three-dimensional rotation algebra $so(3)$
result in only one nontrivial and proper contraction to the Lie algebra $A^0_{3.5}$.
At the same time, there also exists the proper contraction from $so(3)$ to the Heisenberg algebra $\mathfrak h_3=A_{3.1}$
and it is not provided by IW-contractions (see Sections~\ref{SectionOnLow-DimRealLieAlgebras} and \ref{SectionOnOne-parContractionsOfRealLow-DimLieAlgebras} for details).

\looseness=1
Saletan~\cite{Saletan1961} studied the whole class of contractions linear with respect to the contraction parameter, 
i.e.,\ contractions generated by the matrices of the form $U(\varepsilon)=U_0+\varepsilon U'_0$, 
where $U_0$ and $U'_0$ are constant matrices.
Now such contractions are called \emph{Saletan contractions} or, shortly, S-contractions.
The assumption~$U(1)=E$, where $E$ is the unit matrix, can be imposed with basis change 
and reparameterization without loss of generality. 
Then the contraction matrix takes the form 
$U(\varepsilon)=\varepsilon E+(1-\varepsilon)\tilde U$, where $\tilde U$~is a constant matrix.
Conditions on the matrix~$\tilde U$ of a well-defined S-contraction were formulated in~\cite{Saletan1961}.
Iterations of S-contractions with the same contraction matrix result in a finite chain of nonisomorphic algebras.
The repeated contraction from the first algebra to the last algebra of the chain is an 
IW-contraction~\cite{Saletan1961,Lyhmus1969}.
Any IW-contraction obviously is an S-contraction and there exist S-contractions which are inequivalent to IW-contractions.    
Thus, Saletan proved that the contraction $so(3)\oplus A_1\to A_{4.9}^0$ is realized via an S-contraction 
and is equivalent to no IW-contraction.
At the same time, S-contractions do not also exhaust all possible contractions of Lie algebras. 
An illustrative example is again given by the contraction $so(3)\to\mathfrak h_3=A_{3.1}$, 
which is not provided even by S-contractions~\cite{Saletan1961}.

Another generalization of the class of IW-contractions is given by 
\emph{generalized IW-contractions} (or \emph{Doebner--Melsheimer contractions}) \cite{Doebner&Melsheimer1967,Hegerfeldt1967,Lyhmus1969} 
for which the linearity condition is replaced by the condition that the elements of the diagonalized contraction matrix are (integer) powers 
of the contraction parameter. Namely, the contraction matrix of a generalized IW-contraction has the form 
$
U(\varepsilon)=\hat W^{-1}\diag(\varepsilon^{\alpha_1},\varepsilon^{\alpha_2},\dots,\varepsilon^{\alpha_n})\check W,
$
where $\hat W$ and $\check W$ are nonsingular constant matrices and $\alpha_1,\alpha_2,\dots,\alpha_n\in\Z$. 
As in the case of simple IW-contractions, due to possibility of replacement of Lie algebras by isomorphic ones 
we can assume that $\hat W=\check W=E$, i.e. 
\[
U(\varepsilon)=\diag(\varepsilon^{\alpha_1},\varepsilon^{\alpha_2},\dots,\varepsilon^{\alpha_n}). 
\] 
Then the structure constants of the resulting algebra $\mathfrak g_0$ are calculated by the formula 
\[
\tilde c_{ij}^{k}=\lim_{\varepsilon\to +0}\varepsilon^{\alpha_i+\alpha_j-\alpha_k}c_{ij}^{k}
\]
with no summation with respect to the repeated indices.
Therefore, the constraints 
\[
\alpha_i+\alpha_j\geqslant \alpha_k,\quad i,j,k=1,\dots,n\quad \mbox{if} \quad c_{ij}^k\ne 0
\]
are necessary and sufficient
for the existence of the well-defined generalized IW-contraction with the contraction matrix $U(\varepsilon)$ and 
\[
\tilde c_{ij}^{k}=c_{ij}^{k}\quad \mbox{if} \quad\alpha_i+\alpha_j=\alpha_k \qquad\mbox{and}\qquad
\tilde c_{ij}^{k}=0\quad \mbox{otherwise}.
\]
The conditions on contraction existence and the resulting algebra can be reformulated in the
basis-independent terms of filtrations on the initial algebra and of associated graded 
Lie algebras~\cite{Grunewald&Halloran1988}. 

IW-contractions clearly form subclass of generalized IW-contractions with $\alpha_i\in\{0,1\}$.
A~natural question is whether any generalized IW-contraction can be decomposed to a sequence 
of successive IW-contractions. 
The unique nontrivial generalized IW-contraction between three-dimensional algebras 
is given by two successive IW-contractions $so(3)\to A_{3.5}^0\to A_{3.1}$.
In\"on\"u~\cite{Inonu1964} and Sharp~\cite{Sharp1960} formulated the proposition 
that the decomposition is not always possible.
It was shown~\cite{Lyhmus1969} that decomposability of a generalized IW-contraction 
implies additional constraints on structure constants of the initial Lie algebra. 
We construct a number of generalized IW-contractions of four-dimensional algebras, which are not decomposed to a 
sequence of simple IW-contractions. 
For example, the algebra $A_{4.4}$ having the nonzero commutation relations 
$[e_1,e_4]=e_1$, $[e_2,e_4]=e_1+e_2$, $[e_3,e_4]=e_2+e_3$ 
is contracted to the algebra $A_{4.1}$ ($[e_2,e_4]=e_1$, $[e_3,e_4]=e_2$) 
by the generalized IW-contraction with the matrix $\diag(\varepsilon^2,\varepsilon,1,\varepsilon)$. 
This contraction obviously illustrates the above statement since it is direct, i.e.,\ 
there are no Lie algebra~$\mathfrak g$ 
such that the contractions $A_{4.4}\to\mathfrak g$ and $\mathfrak g\to A_{4.1}$ are proper.
See Remark~\ref{RemarkOnIWcontractionsOf4DimAlgeabras} for more examples.

\begin{remark}
If some of the powers $\alpha_1,\alpha_2,\dots,\alpha_n$ in the contraction matrix $U(\varepsilon)$
are negative, the limit of $ U(\varepsilon)$ under $\varepsilon\to+0$ does not exist.
It is not precisely known up to now in what situations it is sufficient to consider only nonnegative powers of $\varepsilon$.
Results of this paper imply that all contractions of the three- and four-dimensional
Lie algebras are weakly equivalent to the ones for which the limit of the contraction matrices exists.
\end{remark}

\section{Necessary contraction criteria}\label{SectionOnNecessaryContractionCriteria}

An optimal way of exhaustive investigation of contractions in a set of Lie algebras includes
intensive usage of necessary criteria based on quantities which are \emph{invariant} or \emph{semiinvariant}
under contractions.
The invariant quantities are preserved under contractions.
Semiinvariance means existence of inequalities between the corresponding quantities of initial and contracted algebras.
Since contractions are limit processes, the terms of \emph{continuity} and \emph{semicontinuity} can be used
instead of invariance and semiinvariance.

For convenience we collect the relations between invariant or semiinvariant quantities
as necessary criteria of contractions in Theorem~\ref{TheoremOnNecessaryContractionCriteria1}.

Below we use the following notations of quantities and objects connected with the algebra~${\mathfrak g}$:
the differentiation algebra $\Der {\mathfrak g}$,
the orbit $\mathcal O({\mathfrak g})$ under action of $GL(V)$ in the variety~$\mathcal L_n$ of $n$-dimensional Lie algebras, 
the center $Z({\mathfrak g})$,
the radical $R({\mathfrak g})$,
the nilradical $N({\mathfrak g})$,
the maximal dimension~$n_{\rm A}({\mathfrak g})$ of Abelian subalgebras,
the maximal dimension~$n_{\rm Ai}({\mathfrak g})$ of Abelian ideals,
the Killing form $\kappa$,
the rank~$r_{\mathfrak g}$, i.e.,\ the dimension of the Cartan subalgebras,
the adjoint and coadjoint representations $\ad \mathfrak g$ and $\ad^* \mathfrak g$,
the adjoint representation $\ad_x$ of the element $x\in\mathfrak g$ and 
the ranks of adjoint and coadjoint representations which are calculated in a fixed basis by the formulas
\[
\rank(\ad \mathfrak g)=\max\limits_{x \in V} \rank (c_{ij}^k x^j)
\quad\mbox{and}\quad
\rank(\ad^{*} \mathfrak g)=\max\limits_{u \in V^*}\rank (c_{ij}^k u_k).
\]

Let us also define three standard series of characteristic ideals of~${\mathfrak g}$, namely,
\begin{gather*}
\mbox{the lower central series:}\quad
\mathfrak g^0\supset \mathfrak g^1\supset \dots \supset \mathfrak g^l\supset \cdots,
\\
\mbox{the derived series:}\quad
\mathfrak g^{(0)}\supset \mathfrak g^{(1)}\supset \dots \supset \mathfrak g^{(l)}\supset \cdots,
\\
\mbox{the upper central series:}\quad
\mathfrak g_{(0)}^{\phantom{1}}\subset \mathfrak g_{(1)}\subset \dots \subset \mathfrak g_{(l)}\subset \cdots,
\end{gather*}
where
$\mathfrak g^0=\mathfrak g$, $\mathfrak g^l=[\mathfrak g, \mathfrak g^{l-1}]$,
$\mathfrak g^{(0)}=\mathfrak g$, $\mathfrak g^{(l)}=[\mathfrak g^{(l-1)}, \mathfrak g^{(l-1)}]$,
$\mathfrak g_{(0)}=\{0\}$, $\mathfrak g_{(l)}/\mathfrak g_{(l-1)}$ is the center of $\mathfrak g/\mathfrak g_{(l-1)}$,
$l\in \N$.
In particular, $\mathfrak g^1=\mathfrak g^{(1)}=[\mathfrak g, \mathfrak g]$, $\mathfrak g_{(1)}=Z(\mathfrak g)$.
If ${\mathfrak g}$ is a solvable (nilpotent) Lie algebra,
$r_{\rm s}=r_{\rm s}(\mathfrak g)$ ($r_{\rm n}=r_{\rm n}(\mathfrak g)$) 
denotes the solvability (nilpotency) rank of~${\mathfrak g}$,
i.e.,\ the minimal number $l$ such that ${\mathfrak g}^{(l)}=\{0\}$ (${\mathfrak g}^l=\{0\}$).

Suppose that $\tr(\ad_u\!{}^p)\not=0$, $\tr(\ad_u\!{}^q)\not=0$ and $\tr(\ad_u\!{}^p\ad_v\!{}^q)\not=0$ for
some $p,q\in \N$ and $u,v\in\mathfrak g$ and the value
\begin{gather*}
\mathfrak C_{pq}=\frac{\tr(\ad_u\!{}^p)\tr(\ad_v\!{}^q)}{\tr(\ad_u\!{}^p\ad_v\!{}^q)},\quad p,q\in \N
\end{gather*}
does not depend on $u$ and $v$.
Then $\mathfrak C_{pq}=\mathfrak C_{pq}(\mathfrak g)$ is a well-defined invariant characteristic of the algebra~$\mathfrak g$,
i.e.,\ it is a constant on the orbit~$\mathcal O(\mathfrak g)$.

Denote the rank of positive (negative) part of the Killing form~$\kappa_\mathfrak g$,
i.e.,\ the number of positive (negative) diagonal elements of a diagonal form of its matrix, by
$\rank_+\kappa_\mathfrak g$ ($\rank_-\kappa_\mathfrak g$).
In view of the law of inertia of quadratic forms, $\rank_+\kappa_\mathfrak g$ and $\rank_-\kappa_\mathfrak g$ are invariant
under basis transformations over~$\R$.
For any $\alpha\in\R$ we introduce the modified Killing form
\[
\tilde\kappa^\alpha_{\mathfrak g}=\tr(\ad_u \ad_v)+\alpha\tr(\ad_u)\tr(\ad_v)
\]
and the corresponding values $\rank_+\tilde\kappa^\alpha_\mathfrak g$ and $\rank_-\tilde\kappa^\alpha_\mathfrak g$.
The Killing form is the special case of the modified Killing form with $\alpha=0$.

The following technical lemma is very useful for further considerations.

\begin{lemma}\label{LemmaOnRankOfMatrixLimit}
Let $A_p$, $p\in\N$, be a sequence of real or complex matrices of the same dimensions
and there exists componentwise limit of $A_p$, $p\to\infty$, denoted by $A_0$.
If $\rank A_p=r$ $\forall p\in\N$ then $\rank A_0\leqslant r$.
\end{lemma}

\begin{theorem}\label{TheoremOnNecessaryContractionCriteria1}
If the Lie algebra ${\mathfrak g}_0$ is a proper (continuous or sequential) contraction of the Lie algebra ${\mathfrak g}$,
then the following set of relations holds true:
\renewcommand{\labelenumi}{{\rm \theenumi)}}
\begin{enumerate}\itemsep=0ex
\item\label{criterion_dim_Der}
$\dim \Der\mathfrak g_0>\dim \Der\mathfrak g$ (and $\dim \mathcal O({\mathfrak g}_0)<\dim \mathcal O({\mathfrak g}))$;
\item\label{criterion_dim_Ab}
$n_{\rm A}(\mathfrak g_0)\geqslant n_{\rm A}(\mathfrak g)$;
\item\label{criterion_dim_Z}
$\dim Z(\mathfrak g_0)\geqslant \dim Z(\mathfrak g)$;
moreover, $\dim{\mathfrak g}_{0(l)}\geqslant \dim{\mathfrak g}_{(l)}$, $l\in \N$;
\item\label{criterion_dim_derived_series_components}
$\dim{\mathfrak g}_0^{(l)}\leqslant \dim{\mathfrak g}^{(l)},$ $l\in \N$;
\item\label{criterion_dim_lower_central_series_components}
$\dim{\mathfrak g}_0^{l}\leqslant \dim{\mathfrak g}^{l},$ $l\in \N$;
\item\label{criterion_dim_R}
$\dim R(\mathfrak g_0)\geqslant \dim R(\mathfrak g)$;
\item\label{criterion_dim_N}
$\dim N(\mathfrak g_0)\geqslant \dim N(\mathfrak g)$;
\item\label{criterion_dim_AbI}
$n_{\rm Ai}(\mathfrak g_0)\geqslant n_{\rm Ai}(\mathfrak g)$;
\item\label{criterion_rank}
$r_{\mathfrak g_0}\geqslant r_{\mathfrak g}$;
\item\label{criterion_rank_ad}
$\rank\ad\mathfrak g_0 \leqslant \rank\ad\mathfrak g$,
$\rank\ad^*\mathfrak g_0 \leqslant \rank\ad^*\mathfrak g$;
\item\label{criterion_Killing_formRank}
$\mathop{\rm rank} \kappa_{\mathfrak g_0}\leqslant \mathop{\rm rank} \kappa_{\mathfrak g}$;
\item\label{criterion_unimodular_property}
${\mathfrak g}_0$ is unimodular if $\mathfrak g$ is unimodular, i.e.,\
$\tr(\ad_u)=0$ for any $u$ in $\mathfrak g$ implies the same condition in $\mathfrak g_0$;
{\addtocounter{enumi}{-1}\renewcommand{\theenumi}{\arabic{enumi}$'$}
\item\label{criterion_l-unimodular_property}
moreover, for any fixed $l\in\N$ ${\mathfrak g}_0$ is $l$-unimodular if $\mathfrak g$ is $l$-unimodular, i.e.,\
$\tr(\ad_u{}^l)=0$ for any $u$ in $\mathfrak g$ implies the same condition in $\mathfrak g_0$;}
\item\label{criterion_solvability}
if ${\mathfrak g}$ is solvable Lie algebra then ${\mathfrak g}_0$ is also solvable and
$r_{\rm s}({\mathfrak g}_0)\leqslant r_{\rm s}({\mathfrak g})$;
\item\label{criterion_nilpotency}
if ${\mathfrak g}$ is nilpotent Lie algebra then ${\mathfrak g}_0$ is also nilpotent and
$r_{\rm n}({\mathfrak g}_0)\leqslant r_{\rm n}({\mathfrak g})$;
\item\label{criterion_Cpq}
$\mathfrak C_{pq}(\mathfrak g_0)=\mathfrak C_{pq}(\mathfrak g)$ for all values $p,q\in\N$,
where the invariants $\mathfrak C_{pq}(\mathfrak g_0)$ and $\mathfrak C_{pq}(\mathfrak g)$ are well-defined;
\item\label{criterion_Killing_formPNRanks}
(only over~$\R$!)
$\rank_+\kappa_{\mathfrak g_0}\leqslant \rank_+\kappa_{\mathfrak g}$ and
$\rank_-\kappa_{\mathfrak g_0}\leqslant \rank_-\kappa_{\mathfrak g}$; moreover for any $\alpha\in\R$
$\rank_+\tilde\kappa^\alpha_{\mathfrak g_0}\leqslant \rank_+\tilde\kappa^\alpha_{\mathfrak g}$ and
$\rank_-\tilde\kappa^\alpha_{\mathfrak g_0}\leqslant \rank_-\tilde\kappa^\alpha_{\mathfrak g}$;
\item\label{criterion_versus_deformation}
if the algebra ${\mathfrak g}_0$ is rigid then it is not a contraction of any ${\mathfrak g}$ and
if there is no deformation from ${\mathfrak g}_0$ to ${\mathfrak g}$,
then there is no contraction from ${\mathfrak g}$ to ${\mathfrak g}_0$.
\end{enumerate}
\end{theorem}

\begin{proof}
It is sufficient to prove the theorem in the case of sequential contractions.
The statement on continuous contractions directly follows from the one on  sequential contractions.
We use the notations introduced at the beginning of the section.

At first, Criteria~\ref{criterion_dim_derived_series_components} and~\ref{criterion_dim_lower_central_series_components}
are proved in detail.
The statements are true if $\dim [\mathfrak g,\mathfrak g]=\dim \mathfrak g=:n$. 
Indeed, in this case $\dim {\mathfrak g}^{(l)}=\dim{\mathfrak g}^{l}=n$ for all $l\in \N$
that results in Criteria~\ref{criterion_dim_derived_series_components} and~\ref{criterion_dim_lower_central_series_components}
in view of the obvious conditions
$\dim {\mathfrak g}_0^{(l)}\leqslant\dim \mathfrak g_0$, $\dim{\mathfrak g}_0^{l}\leqslant\dim \mathfrak g_0$ and
$\dim \mathfrak g_0=\dim \mathfrak g=n$.

Suppose that $\dim [\mathfrak g,\mathfrak g]<n$.
Let $\{e^1, \ldots, e^n\}$ be the basis in the dual space $V^*$, which is dual to the basis
$\{e_1, \ldots, e_n\}$, i.e.,\ $\langle e^i,e_j \rangle=\delta^i_j$.
Here $\delta^i_j$ is the Kronecker delta.
We define $A$ as the $n\times n^2$ matrix consisting of the elements $c^k_{ij}=\langle e^k,[e_i,e_j] \rangle$,
where the index $k$ runs the row range and the index pair~$(i,j)$ runs the column range:
\[
A=
\left(\begin{array}{ccccccccccc}
c_{11}^1 & \cdots & c_{1n}^1 & c_{21}^1 & \cdots & c_{2n}^1 & \cdots & c_{n1}^1 & \cdots & c_{nn}^1\\[1ex]
c_{11}^2 & \cdots & c_{1n}^2 & c_{21}^2 & \cdots & c_{2n}^2 & \cdots & c_{n1}^2 & \cdots & c_{nn}^2\\
\vdots & \ddots & \vdots & \vdots & \ddots & \vdots & \ddots & \vdots & \ddots & \vdots \\
c_{11}^n & \cdots & c_{1n}^n & c_{21}^3 & \cdots & c_{2n}^n & \cdots & c_{n1}^n & \cdots & c_{nn}^n
\end{array}\right).
\]
Due to antisymmetry of $c^k_{ij}$ in subscripts, we can take only columns with $i<j$ into account.
Analogously we introduce the matrices $A_p$ and $A_0$ for the algebras
$\mathfrak g_p$ and $\mathfrak g_0$.

The dimensions of $[\mathfrak g,\mathfrak g]$ and $[\mathfrak g,\mathfrak g]_p$ coincide.
Denote the common value of dimensions as $n_1$.
These statements are reformulated in terms of the introduced matrices $A$ and $A_p$:
\begin{gather*}
\rank A=\rank A_p=n_1.
\end{gather*}
Therefore, all $(n_1+1)$-dimensional minors of any matrix $A_p,\, p\in\N$ equal to zero.
Moreover, we have
\begin{gather*}
c_{p,ij}^k=\langle e^k,[e_i,e_j]_p \rangle \to c_{0,ij}^k=\langle e^k,[e_i,e_j]_0 \rangle,\quad p\to\infty,
\end{gather*}
i.e.,\ the elements of the matrix $A_p$ go to the corresponding elements of the matrix $A_0$.
It leads to the convergency of minors.
Consequently, any $(n_1+1)$-dimensional minor of the matrix $A_0$ vanishes.
It implies that $\rank A_0\leqslant n_1$, i.e.,\ $[\mathfrak g_0,\mathfrak g_0]\leqslant n_1$,
Criteria~\ref{criterion_dim_derived_series_components} and~\ref{criterion_dim_lower_central_series_components} for $l=1$
have been proved.  

Criteria~\ref{criterion_dim_derived_series_components} and~\ref{criterion_dim_lower_central_series_components}
for the other values of $l$ are proved analogously.
The requisite matrices are defined in the similar way as the matrices $A$, $A_p$ and $A_0$ in the case $l=1$
with replacement of the usual commutators $[e_i, e_j]$ by the corresponding repeated commutators of basis elements.

Criteria~\ref{criterion_rank},~\ref{criterion_rank_ad} and~\ref{criterion_Killing_formRank} are proved in a similar 
and simpler way via the limit process $p\to\infty$ in the formulas
\begin{gather*}
r_{\mathfrak g}=n-\max\limits_{x \in V} \rank (c_{ij}^k x^j)^n=r_{\mathfrak g_p}=n-\max\limits_{x \in V} \rank (c_{p,ij}^k x^j)^n,
\\
\rank(\ad \mathfrak g)=\max\limits_{x \in V} \rank (c_{ij}^k x^j)=\rank(\ad \mathfrak g_p)=\max\limits_{x \in V} \rank (c_{p,ij}^k x^j),
\\
\rank(\ad^{*} \mathfrak g)=\max\limits_{u \in V^*}\rank (c_{ij}^k u_k)=\rank(\ad^{*} \mathfrak g_p)=\max\limits_{u \in V^*}\rank (c_{p,ij}^k u_k),
\\
\rank(\kappa_\mathfrak g)=\rank (c_{ij}^k c_{i'\!k}^j)=\rank(\kappa_{\mathfrak g_p})=\rank (c_{p,ij}^k c_{p,i'\!k}^j), \quad p\in\N.
\end{gather*}

Criteria~\ref{criterion_unimodular_property} and~\ref{criterion_l-unimodular_property} 
are obvious since $\tr(\ad_u{\!}^l)=0$ for any $u$ in $\mathfrak g$ implies the same condition in $\mathfrak g_p$
and $\tr(\ad_{\mathfrak g_p,u}{\!}^l)\to\tr(\ad_{\mathfrak g_0,u}{\!}^l)$, $p\to\infty$.

Criteria~\ref{criterion_solvability} and~\ref{criterion_nilpotency} directly follow from
Criteria~\ref{criterion_dim_derived_series_components} and~\ref{criterion_dim_lower_central_series_components}.

Since the radical $R(\mathfrak g)$ is the orthogonal complement of the derivative $[\mathfrak g,\mathfrak g]$
with respect to the Killing form~$\kappa_\mathfrak g$~\cite{Jacobson} then $\dim R(\mathfrak g)=\dim R(\mathfrak g_p)$ coincides with the value
\[
n-\rank (c_{ij}^k c_{i'\!k}^jc_{i''\!j''\!}^{i'\!})=n-\rank (c_{p,ij}^k c_{p,i'\!k}^jc_{p,i''\!j''\!}^{i'\!}), \quad p\in\N.
\]
In the matrices the index pair~$(i'',j'')$ runs the row range and the index $i$ runs the column range.
The limit process $p\to\infty$ in the latter formula results in Criterion~\ref{criterion_dim_R}.

The center $Z(\mathfrak g)$ coincides with the set of solutions of the system $[e_i,x]=0$, or $c_{ij}^k x^j=0$ in the coordinate form.
Therefore, $\dim Z(\mathfrak g)=\dim Z(\mathfrak g_p)$ equals to
\[
n-\rank (c_{ij}^k)=n-\rank (c_{p,ij}^k), \quad p\in\N,
\]
where the index pair~$(k,j)$ runs the row range and the index $i$ runs the column range.
The limit process $p\to\infty$ in the latter formula implies Criterion~\ref{criterion_dim_Z} for $l=1$.
Proof for the other values of~$l$ is similar.
Instead of $(c^k_{ij})=(\langle e^k,[e_i,e_j] \rangle)$,
the matrix $(\langle e^k,[\dots[e_i,e_{j_1}],\dots,e_{j_l}] \rangle)$ should be used,
where the index tuple~$(k,j_1,\dots,j_l)$ runs the row range and the index $i$ runs the column range.

Criterion~\ref{criterion_Cpq} is true in view of invariance property of~$\mathfrak C_{pq}$.

Proof of Criterion~\ref{criterion_dim_Ab} is also adduced in detail since 
it presents another typical trick which is used in deriving necessary contraction criteria.
Let $n_{\rm A}(\mathfrak g)=l$. Then $n_{\rm A}(\mathfrak g_p)=l$ too. 
We change the basis of~$\mathfrak g_p$ with a nonsingular matrix~$W_p$ that 
\[
\tilde c_{p,ij}^k=0 \quad\mbox{if}\quad i,j\leqslant l.
\]
Here $\tilde c_{p,ij}^k=(W_p)_{i'}^i(W_p)_{j'}^j(W_p{}^{-1})_k^{k'}c_{p,i'\!j'}^{k'\!}$ are 
the structure constants of~$\mathfrak g_p$ in the new basis. 
Due to possibility of `orthogonalization' of the basis, we can assume without loss generality that 
$W_p$ is an orthogonal (unitary) matrix in the case of the real (complex) field. 
The set of orthogonal (or unitary) matrices is compact in the induced `Euclidean' matrix norm. 
Therefore, there exists a convergent subsequence $\{W_{p_q},\, q\in\N\}$. 
Denote the orthogonal (unitary) matrix being the limit of this subsequence by $W_0$. Then 
\[
\tilde c_{p_q,ij}^k=(W_{p_q})_{i'}^i(W_{p_q})_{j'}^j(W_{p_q}{}^{-1})_k^{k'}c_{p_q,i'\!j'}^{k'\!}\to 
(W_0)_{i'}^i(W_0)_{j'}^j(W_0{}^{-1})_k^{k'}c_{0,i'\!j'}^{k'\!}=\tilde c_{0,ij}^k, \quad q\to\infty.
\]   
Hence $\tilde c_{p,ij}^k=0 \quad\mbox{if}\quad i,j\leqslant l$ in view of the same condition for $c_{p,ij}^k$, 
i.e.,\ $\mathfrak g_0$ contains an $l$-dimensional Abelian subalgebra that implies Criterion~\ref{criterion_dim_Ab}.

Criteria~\ref{criterion_dim_R}, \ref{criterion_dim_N} and~\ref{criterion_dim_AbI} are proved in a similar way 
with replacement of the Abelian subalgebra condition by the ideal condition 
\[
\tilde c_{p,ij}^k=0 \quad\mbox{if}\quad (i\leqslant l\ \mbox{or}\ j\leqslant l)\ \mbox{and}\ k> l 
\]
completed with the conditions  
\begin{gather*}
\tilde c_{p,ij}^k=0 \quad\mbox{if}\quad i,j\leqslant l\ \mbox{and}\ k>\max(i,j), \quad l= \dim R(\mathfrak g), 
\\
\tilde c_{p,ij}^k=0 \quad\mbox{if}\quad i,j\leqslant l\ \mbox{and}\ k\geqslant\max(i,j), \quad l= \dim N(\mathfrak g), 
\\
\tilde c_{p,ij}^k=0 \quad\mbox{if}\quad i,j\leqslant l, \quad l= n_{\rm Ai}(\mathfrak g) 
\end{gather*}
of solvability for the radical, nilpotency for the nilradical and commutativity for an Abelian ideal of 
the maximal dimension correspondingly. Other conditions of solvability and nilpotency can also be used. 
Let us note that we derive the second proof of Criterion~\ref{criterion_dim_R}.

Similar technique based on compactness of the set of orthogonal matrices is used in proof of Criterion~\ref{criterion_Killing_formPNRanks}. 
Denote the number of positive (negative) diagonal elements of a diagonal form of a symmetric matrix~$K$ by $\rank_+K$ ($\rank_-K$).

It is sufficient to prove that for any convergent sequence of symmetric matrices $K_p\to K_0$, $p\to\infty$, with 
$\rank_+K_p=r_+$ and $\rank_-K_p=r_-$, $p\in\N$, 
the inequalities $\rank_+K_0\leqslant r_+$ and $\rank_-K_0\leqslant r_-$ are true. 
Then this statement is applied to the sequence of the matrices of the (modified) Killing forms of the algebras $\mathfrak g_p$, 
which have the same values of $\rank_+$ and $\rank_-$ in view of the inertia law of quadratic forms and converge to 
the matrix of the (modified) Killing form of the resulting algebra $\mathfrak g_0$. 

Let $W_p$ be the orthogonal matrix which reduces $K_p$ to the matrix $D_p=\diag(d_{p,1},\ldots,d_{p,n})$, 
where 
$d_{p,i_1}>0$ for $i_1=1,\ldots,r_+$, 
$d_{p,i_2}<0$ for $i_2=r_++1,\ldots,r_++r_-$ and 
$d_{p,i_3}=0$ for $i_3=r_++r_-+1,\ldots,n$ 
($r_++r_-\leqslant n$). So, $K_p=W_pD_pW_p{}^\text{T}$.
We choose a convergent subsequence $\{W_{p_q},\, q\in\N\}$ and denote the orthogonal matrix being the limit of this subsequence by $W_0$. 
Then 
\[
D_{p_q}=W_{p_q}{}^\text{T}K_{p_q}W_{p_q}\to W_0{}^\text{T}K_0W_0=:D_0, \quad q\to\infty.
\]
$D_0$ is a diagonal matrix $\diag(d_{0,1},\ldots,d_{0,n})$ as the limit of the sequence of the diagonal matrices $D_{p_q}$, $q\in\N$. 
Moreover, 
$d_{0,i_1}\geqslant0$ for $i_1=1,\ldots,r_+$, 
$d_{0,i_2}\leqslant0$ for $i_2=r_++1,\ldots,r_++r_-$ and 
$d_{0,i_3}=0$ for $i_3=r_++r_-+1,\ldots,n$ that implies the requisite statement. 

Criteria~\ref{criterion_dim_Der} and~\ref{criterion_versus_deformation}
were proved, e.g.,\ in~\cite{Borel1969,Grunewald&Halloran1988,Steinhoff1997}.
\end{proof}

\begin{remark}
The criteria can be reformulated in terms of closed subsets of the variety~$\mathcal{A}_n$ of
$n$-dimensional Lie algebras.
Thus, the sets of nilpotent, solvable and unimodular algebras are closed.
The sets $\{\mathfrak g\in\mathcal{A}_n\mid \dim \mathfrak g^l\leqslant r\}$,
$\{\mathfrak g\in\mathcal{A}_n\mid \dim \mathfrak g^{(l)}\leqslant r\}$,
$\{\mathfrak g\in\mathcal{A}_n\mid \dim \mathfrak g_{(l)}\geqslant r\}$
and similar ones are closed for each $l$ and $r=0,\dots,n$.
\end{remark}

\begin{remark}
Necessary criteria already appeared in early papers on contractions of Lie algebras. 
Thus, in his pioneer paper~\cite{Segal1951} Segal used the criterion based on the law of inertia of the Killing forms 
(the first part of Criterion~\ref{criterion_Killing_formPNRanks}) in the case of compact semisimple Lie algebras. 
The inequality on dimensions of the derived algebras 
$\dim[\mathfrak g_0,\mathfrak g_0]\leqslant \dim[\mathfrak g,\mathfrak g]$  
was proved in~\cite{Saletan1961} in the practically same way as above.
Criteria~\ref{criterion_dim_Z} and~\ref{criterion_dim_lower_central_series_components} 
on upper and lower central series arise under studying varieties of nilpotent algebras~\cite{Vergne1966b}.  
Very important Criterion~\ref{criterion_dim_Der} is a direct consequence of the lemma on orbit closures,
which is adduced, e.g.,\ in~\cite{Borel1969}. 
Criterion~\ref{criterion_dim_Ab} was proved in~\cite{Grunewald&Halloran1988} with usage of the Iwasawa decomposition.
It was pointed out in~\cite{Seeley1990} that the same technique can be used to prove other necessary contraction criteria. 
Restricting ourself with real and complex cases, 
we simplify this technique and use it also to prove Criteria~\ref{criterion_dim_R},~\ref{criterion_dim_N} and~\ref{criterion_dim_AbI}.
Criteria~\ref{criterion_dim_Z},~\ref{criterion_dim_derived_series_components} and~\ref{criterion_dim_lower_central_series_components} 
were applied simultaneously in~\cite{Kirillov&Neretin1984-1987}, see also discussion in~\cite{Gorbatsevich1998}.  
All the above criteria are based on semiinvariant values. 
Criterion~\ref{criterion_Cpq} on the invariant algebra characteristic $\mathfrak C_{pq}$ was first proposed and 
effectively used in~\cite{Burde&Steinhoff1999,Steinhoff1997}.
A number of criteria are collected ibid and key ideas on proof of them were also formulated. 
In particular, Criterion~\ref{criterion_rank} was adopted by us from there. 
Let us note that $\mathfrak C_{pq}$ is a generalization of the invariant $J=\tr(\ad_u\!{}^2)/(\tr\ad_u)^2$ 
introduced in~\cite{Kirillov&Neretin1984-1987}.
The part of Criterion~\ref{criterion_rank_ad} on rank of coadjoint representations arises 
in investigation of connections between invariants of initial and contracted algebras~\cite{Campoamor-Stursberg2002,Campoamor-Stursberg2003}.    
Criterion~\ref{criterion_versus_deformation} was discussed, e.g.,\ in~\cite{Lyhmus1969}.
There exist also other criteria, e.g.,\ ones connected with cohomologies of Lie algebras~\cite{Burde2005}.
\end{remark}

\begin{remark}
The list of criteria can be extended with other quantities which concern algebras
and are semiinvariant under contractions~\cite{Burde&Steinhoff1999}.
The criteria used in this paper are simple from the computing point of view. 

The set (or even a subset) of the adduced criteria is complete for the
three- and four-dimensional Lie algebras in the sense that
they precisely separate all pairs of algebras, which do not admit contractions.
The question on completeness of the adduced criteria
in the case of Lie algebras of higher dimensions is still open.

The set of criteria is not minimal.
Some criteria are induced by others. For example, Criteria~\ref{criterion_dim_derived_series_components} 
and~\ref{criterion_dim_lower_central_series_components}
imply Criteria~\ref{criterion_solvability} and \ref{criterion_nilpotency}.

Criteria differ from each other in effectiveness.
Criteria~\ref{criterion_dim_Der} and~\ref{criterion_unimodular_property} are most powerful since they exclude possibility of contractions in
most pairs of low-dimensional Lie algebras.
This fact is illustrated by examples of Section~\ref{SectionOnContractionIdentification}.

Criterion~\ref{criterion_Killing_formPNRanks} is the unique criterion which is special for the real field.
Only it works for pairs of algebras having a contraction over~$\CC$ and no contractions over~$\R$. 
See Remark~\ref{RemarkOnRealCriteriaIn4Dim} additionally. 
\end{remark}

\begin{remark}
Criterion~\ref{criterion_dim_Der} is singular and particularly powerful due to appearance of strict inequality in it.
It is the unique criterion which enables investigation of contractions in series of Lie algebras in a simple way.
Since the dimensions of the differentiation algebras
for the nonsingular values of the parameters in series of Lie algebras coincide,
Criterion~\ref{criterion_dim_Der} implies the absence of contractions between these cases.

The weakened version of Criterion~\ref{criterion_dim_Der} with unstrict inequality is proved analogous 
to a number of other criteria. 
Indeed, in a fixed basis of~$V$ the coefficients $d^i_j$ of the matrix of any operator from $\Der\mathfrak g$ 
satisfy the homogenous system of linear equations
\[
c_{ij}^{k'\!}d^k_{k'\!}=c_{i'\!j}^kd^{i'\!}_i+c_{ij'\!}^kd^{j'\!}_j.
\]
Let $A$ be the matrix of this system and $A_p$ and $A_0$ be the similar matrices for the algebras~$\mathfrak g_p$ 
and~$\mathfrak g_0$. 
Then $\dim\Der\mathfrak g=n^2-\rank A=\dim\Der\mathfrak g_p=n^2-\rank A_p$ and $\dim\Der\mathfrak g_0=n^2-\rank A_0$. 
Therefore, the inequality $\dim\Der\mathfrak g_0\geqslant \dim\Der\mathfrak g$ obviously follows from 
Lemma~\ref{LemmaOnRankOfMatrixLimit}.
(Note that $\dim\mathcal O(\mathfrak g)=\rank A$.)

Can other criteria or their combinations be strengthened with replacement of unstrict inequalities by strict ones?
The answer to this question is unknown up to now. 
There existed the conjecture that dimension of an element of the upper or lower central series or the derived series 
should vary under a proper contraction. 
A contrary instance on the conjecture was adduced in~\cite{Gorbatsevich1998}.
It is given by the contraction of the three-dimensional algebras 
\[A_{3.2}\, ([e_1,e_3]=e_1,[e_2,e_3]=e_1+e_2)\to A_{3.3}\, ([e_1,e_3]=e_1,[e_2,e_3]=e_1),\] 
which is realized via the simple IW-cintraction associated with the subalgebra~$\langle e_1, e_2+e_3\rangle$. 
Let us note additionally that all invariant and semiinvariant quantities adduced in this section 
excluding only $\dim\Der$ coincide for the above algebras. 
Therefore, only Criterion~\ref{criterion_dim_Der} is effectual for this pair of algebras.
\end{remark}

\section{Calculation of invariant quantities}\label{SectionOnCalculationOfInvariantQuantities}

There are two simple classes of Lie algebras, which cover most low-dimensional algebras.
The first one is formed by \emph{almost Abelian algebras} having Abelian ideals of codimension one.
The algebras from the second class have WH+A ideals of codimension one, which are isomorphic to
the direct sum of the Weyl--Heisenberg algebra $\mathfrak h_3=A_{3.1}$ and the Abelian algebra of codimension four.
Characteristics of the above algebras are found in a uniform way.
The other low-dimensional algebras should be investigated separately.
Below we adduce only calculations of the invariants~$\mathfrak C_{pq}$ which were recently proposed in~\cite{Burde&Steinhoff1999,Steinhoff1997}
and the ranks of some algebras.

\subsection{Almost Abelian algebras}\label{SectionOnAlmostAbelianAlgebras}

Consider an $n$-dimensional Lie algebra
over $\CC$ or $\R$ which has an $(n-1)$-dimensional Abelian ideal.
It is a solvable and, moreover, metabelian algebra.
Let $e_1$, \ldots, $e_{n-1}$ form a basis of the ideal and $e_n$ completes it to a basis of the algebra.
The nonzero commutation relations between elements of the constructed bases are
\begin{gather*}
[e_j,e_n]=\sum\limits_{k=1}^{n-1}a^k_je_k,\quad j=1,\dots, n-1.
\end{gather*}
The $(n-1)\times(n-1)$ matrix $A=(a^k_j)$ defines the algebra completely hence we will denote this algebra by $\mathfrak a_A$,
i.e.,\ $\mathfrak a_A:=A_1\oplus_A(n-1)A_1$.

The algebras $\mathfrak a_A$ and $\mathfrak a_{A'}$ are isomorphic iff the matrices $A$ and $A'$ are similar up to scalar multiplier.
The isomorphisms are established via changes of bases in the Abelian ideals and scaling of the complementary elements of bases.
Up to the algebra isomorphisms the matrix $A$ can be assumed reduced to the Jordan normal form, and its eigenvalues
can be additionally normalized with a nonvanishing multiplier.

For any algebra ${\mathfrak g}$ the matrix $\hat\ad_u$ of the adjoint representation $\ad_u$
of an arbitrary element $u\in \mathfrak g$ is found
by the formula $(\hat\ad_u)^j_k=c^k_{ij}u_i$,
where $c^k_{ij}$ are the structure constants of ${\mathfrak g}$ in the fixed basis.
Since for $\mathfrak a_A$ $\smash{c^k_{ij}=0}$ if neither $i$ nor $j$ equals to $n$,
the matrix $\smash{\hat\ad_u}$ can be easily calculated:
\begin{gather*}
\hat\ad_u=
\sum\limits_{i=1}^{n-1}u_i\left(
\begin{array}{cccc}
0 & \cdots & 0 & a^i_1\\
\vdots & \ddots & \vdots & \vdots\\
0 & \cdots & 0 & a^i_{n-1}\\
0 & \cdots & 0 & 0
\end{array}
\right)
-u_n\left(
\begin{array}{cccc}
a^1_1& \cdots & a^{n-1}_1 & 0\\
\vdots&\ddots&\vdots&\vdots\\
a^1_{n-1}&\dots&a^{n-1}_{n-1}&0\\
0&\cdots&0&0
\end{array}
\right),
\end{gather*}
or shortly $\hat\ad_u$ have the form
\begin{gather*}
\hat\ad_u=
\left(\begin{array}{cc}
u_nA & -A\bar{u}
\\[1.5ex]
\underline{0} & 0
\end{array}\right),
\quad\text{where}\quad
\bar{u}=(u_1,\dots,u_{n-1})^{\text{T}}\quad \text{and}\quad \underline{0}=(0,\dots,0).
\end{gather*}

To calculate the invariant characteristic $\mathfrak C_{pq}$ of  $\mathfrak a_A$,
we find powers of $\hat\ad_u$ and their traces
\begin{gather*}
\hat\ad_u{}^p=
\left(\begin{array}{cc}
u_n{}^pA^p & -u_n{}^{p-1}A^p\bar{u}
\\[1.5ex]
\underline{0} & 0
\end{array}\right),
\quad
\tr(\ad_u\!{}^p)=u_n{}^p\tr(A^p).
\end{gather*}

Matrix trace is not affected by matrix similarity transformations.
If $\lambda_1,$ \dots, $\lambda_{n-1}$ are the roots of the characteristic polynomial $\chi_A(\lambda)$ of the matrix $A$ in~$\CC$
then $\tr(A^p)=\lambda_1^p+\cdots+\lambda_{n-1}^p$ for any $p\in\N$.
Consequently, the invariant value $\mathfrak C_{pq}$ can be calculated explicitly.

The rank of $\mathfrak a_A$ can be easily calculated as by-product.
Indeed, the characteristic polynomial $\chi_{\hat\ad_u}(\lambda)$ of $\hat\ad_u$ equals to $\lambda\chi_{u_nA}(\lambda)$,
i.e.,\ any element $u\in\mathfrak a_A$ with $u_n\not=0$ is regular and the rank of $\mathfrak a_A$ coincides with the number of zero roots
of the polynomial $\lambda\chi_A(\lambda)$.

As a result, we obtain the following statement.

\begin{lemma}\label{LemmaCpqOfaA}
Let $\mathfrak a_A$ be an $n$-dimensional Lie algebra with an $(n-1)$-dimensional Abelian ideal and with
commutation relations which are defined via the matrix $A$ and
$\lambda_1,\dots,\lambda_{n-1}$ be roots of the characteristic polynomial of~$A$ over~$\CC$. If
\begin{gather*}
\tr(A^p)=\lambda_1^p+\cdots+\lambda_{n-1}^p\not=0,\quad
\tr(A^q)=\lambda_1^q+\cdots+\lambda_{n-1}^q\not=0,\\
\tr(A^{p+q})=\lambda_1^{p+q}+\cdots+\lambda_{n-1}^{p+q}\not=0,\quad
\end{gather*}
then the value $\mathfrak C_{pq}$ is well-defined invariant characteristics of
the algebra $\mathfrak a_A$ and is given by the formula
\begin{gather*}
\mathfrak C_{pq}=\frac{\tr(A^p)\tr(A^q)}{\tr(A^{p+q})}=
\frac{(\lambda_1^p+\cdots+\lambda_{n-1}^p)(\lambda_1^q+\cdots+\lambda_{n-1}^q)}
{(\lambda_1^{p+q}+\cdots+\lambda_{n-1}^{p+q})}.
\end{gather*}
The rank of the algebra $\mathfrak a_A$ (i.e.,\ the dimension of its Cartan subalgebra)
equals to the order of zero root of the characteristic polynomial of the matrix $A$ plus one.
\end{lemma}

\subsection{Lie algebras with WH+A ideals of codimension one}\label{SectionOnAlgebrasWithWHAIdeals}

Consider an $n$-dimensional complex or real Lie algebra
with an $(n-1)$-dimensional ideal which is isomorphic to
the direct sum of the Weyl--Heisenberg algebra $\mathfrak h_3=A_{3.1}$ and the $(n-4)$-dimensional Abelian algebra.
Let $e_1$, $e_2$ and $e_3$ form the canonical basis of a $\mathfrak h_3$-isomorphic component,
$e_4$, \ldots, $e_{n-1}$ give a basis of the Abelian component of the ideal and $e_n$ completes the basis of the ideal to a basis of the whole algebra.
The nonzero commutation relations between elements of the constructed bases are
\begin{gather*}
[e_2,e_3]=e_1,\quad
[e_j,e_n]=\sum\limits_{k=1}^{n-1}a^k_je_k,\quad j=1,\dots, n-1.
\end{gather*}
The $(n-1)\times(n-1)$ matrix $A=(a^k_j)$ defines the algebra completely hence 
we will denote this algebra by $\mathfrak w_A$, i.e.,\ $\mathfrak w_A:=A_1\oplus_A(\mathfrak h_3\oplus(n-4)A_1)$.
The Jacobi identity implies the following constraints on elements of~$A$:
\[
a^1_1=a^2_2+a^3_3, \quad a^k_1=0, \ k=2,\dots,n-1, \quad a^2_i=a^3_i=0, \ i=4,\dots,n-1.
\]

The matrix $\smash{\hat\ad_u}$ of the adjoint representation of an arbitrary element $u\in \mathfrak w_A$
is calculated in a way which is analogous to the previous case and has the form
\begin{gather*}
\hat\ad_u=
\left(\begin{array}{cc}
u_nA+u_3E_2^1-u_2E_3^1 & -A\bar{u}
\\[1.5ex]
\underline{0} & 0
\end{array}\right),
\end{gather*}
where
$\bar{u}=(u_1,\dots,u_{n-1})^{\text{T}}$, $\underline{0}=(0,\dots,0)$,
$E^i_j$ is the $(m-1)\times(m-1)$ matrix with unit in $^i_j$-entry and zero otherwise.
In view of the restrictions on the matrix~$A$ we again have
\begin{gather*}
\tr(\ad_u\!{}^p)=u_n{}^p\tr(A^p), \quad \chi_{\ad_u}(\lambda)=-\lambda\chi_{u_nA}(\lambda).
\end{gather*}

Therefore, Lemma~\ref{LemmaCpqOfaA} can be completely reformulated for the algebra~$\mathfrak w_A$.

\begin{lemma}\label{LemmaCpqOfwA}
Let $\mathfrak w_A=A_1\oplus_A(\mathfrak h_3\oplus(n-4)A_1)$ and
$\lambda_1,\dots,\lambda_{n-1}$ be roots of the characteristic polynomial of~$A$ over~$\CC$. If
\begin{gather*}
\tr(A^p)=\lambda_1^p+\cdots+\lambda_{n-1}^p\not=0,\quad
\tr(A^q)=\lambda_1^q+\cdots+\lambda_{n-1}^q\not=0,\\
\tr(A^{p+q})=\lambda_1^{p+q}+\cdots+\lambda_{n-1}^{p+q}\not=0,\quad
\end{gather*}
then $\mathfrak C_{pq}$ is well-defined invariant characteristics of
the algebra $\mathfrak a_A$ and is given by the formula
\begin{gather*}
\mathfrak C_{pq}=\frac{\tr(A^p)\tr(A^q)}{\tr(A^{p+q})}=
\frac{(\lambda_1^p+\cdots+\lambda_{n-1}^p)(\lambda_1^q+\cdots+\lambda_{n-1}^q)}
{(\lambda_1^{p+q}+\cdots+\lambda_{n-1}^{p+q})}.
\end{gather*}
The rank of the algebra $\mathfrak w_A$ (i.e.,\ the dimension of its Cartan subalgebra)
equals to the order of zero root of the characteristic polynomial of the matrix $A$ plus one.
\end{lemma}

\subsection{Special cases}

{\samepage The adjoint action of any element $u\in so(3)$ is presented in the canonical basis in the form $\hat\ad_u \hat v=\hat u\times\hat v$.
Hereafter $\hat u$ and $\hat v$ are the coordinate columns of~$u$ and~$v$ treated as elements of~$\mathbb R^3$,
`$\cdot$' and `$\times$' denote the usual scalar and vector products in~$\mathbb R^3$. By induction,
\[
\ad_u\!\!{}^{2p'-1} v=(-|\hat u|^2)^{p'-1}\hat u\times\hat v, \quad
\ad_u\!\!{}^{2p'} v=(-|\hat u|^2)^{p'-1}((\hat u\cdot\hat v)\hat u-|\hat u|^2\hat v), \quad p'\in\N,
\]
i.e.,\ $\tr(\ad_u\!\!{}^{2p'-1})=0$, $\tr(\ad_u\!\!{}^{2p'})=(-|\hat u|^2)^{p'}$, $p'\in\N$.
Therefore, $\mathfrak C_{2p',2q'}=2$. For the other pairs of the indices~$p$ and~$q$ the invariant~$\mathfrak C_{pq}$ is undefined.

}

The same statement is true for the algebras $sl(2,\R)$, $so(3)\oplus A_1$ and $sl(2,\R)\oplus A_1$. The arguments are that
$sl(2,\R)$ is equivalent to~$so(3)$ over~$\CC$ and the algebras~$\mathfrak g$ and $\mathfrak g\oplus kA_1$ have the same invariants~$\mathfrak C_{pq}$.

In the canonical basis of $2A_{2.1}$, where the commutation relations are
$[e_1,e_2]=e_1$, $[e_3,e_4]=e_3$, the matrices of $\ad_u\!{}^p$, $p\in\N$, have the form
\[
\hat\ad_u\!{}^p=\left(\begin{array}{cccc}u_2^p&-u_2^{p-1}u_1&0&0\\0&0&0&0\\0&0&u_4^p&-u_4^{p-1}u_3\\0&0&0&0\end{array}\right),
\]
i.e.,\ $\tr(\ad_u\!{}^p)=u_2^p+u_4^p$.
Since the fraction with traces from the definition of $\mathfrak C_{pq}$ explicitly depends on $u$ and~$v$ in the case of $2A_{2.1}$,
the value $\mathfrak C_{pq}$ is undefined  for any $p,q\in\N$.

The same statement is true for the algebra $A_{4.10}$ being isomorphic to~$2A_{2.1}$ over~$\CC$.

\section{Low-dimensional real Lie algebras}\label{SectionOnLow-DimRealLieAlgebras}

We use the complete lists of nonisomorphic classes of real three- and four-dimensional Lie algebras,
which were constructed by Mubarakzyanov~\cite{Mubarakzyanov1963a} and slightly enhanced
in~\cite{Popovych&Boyko&Nesterenko&Lutfullin2003a,Popovych&Boyko&Nesterenko&Lutfullin2003b}.
Enumeration of algebras by Mubarakzyanov is followed in general.

A number of algebraic characteristics and quantities are adduced for each Lie algebra~$\mathfrak g$ from the lists.
More precisely we deal with
the type of the algebra (such as decomposable, solvable, nilpotent, etc.);
the dimension $n_{\rm D}$ of the differentiation algebra $\Der {\mathfrak g}$;
the dimension $n_{\rm Z}$ of the center;
the maximal dimension $n_{\rm A}$ of the Abelian subalgebras;
the Killing form $\kappa$;
the rank $r_{\mathfrak g}$ (equal to the dimension of the Cartan subalgebras);
the rank of solvability $r_{\rm s}$ (if $\mathfrak g$ is solvable);
the rank of nilpotency $r_{\rm n}$ (if $\mathfrak g$ is nilpotent);
the tuple of dimensions of the components of derived series
$\rm{DS}=[\dim {\mathfrak g}^{(1)},\dim {\mathfrak g}^{(2)},\dots,\dim {\mathfrak g}^{(k)}]$,
where $k$ is the minimal number with $\dim\mathfrak g^{(k)}=\dim\mathfrak g^{(i)}$ $\forall\ i>k$;
the tuple of dimensions of the components of lower central series
$\rm{CS}=[\dim {\mathfrak g}^{1},\dim\mathfrak g^{2},\dots,\dim \mathfrak g^k]$,
where $k$ is the minimal number with $\dim \mathfrak g^k=\dim \mathfrak g^i$ $\forall\ i>k$;
the trace $\tr(\ad_v)$ of the adjoint representations of arbitrary element $v\in V$ and
the invariant~$\mathfrak C_{pq}$ for the values $p,q\in \N$ when this invariant is well defined.

These characteristics and quantities are used in Sections~\ref{SectionOnContractionIdentification} and~\ref{SectionOnOne-parContractionsOfRealLow-DimLieAlgebras}
as a basis for application of necessary contraction criteria.

\subsection{Three-dimensional algebras}\label{3_dim_algebras}

\noindent                
{\mathversion{bold}$3A_{1}\colon$} (Abelian, unimodular);
\\[1ex]
$n_{\rm D}=9$,\ \
$n_{\rm Z}=3$,\ \
$n_{\rm A}=3$,\ \
$\kappa=0$,\ \
$r_{\mathfrak g}=3$,\ \
$r_{\rm n}=r_{\rm s}=1$,\ \
$\rm{DS}=[0]$,\ \
$\rm{CS}=[0]$,\ \
$\tr(\ad_v)=0$.

\vspace{1.7ex}
\noindent                
{\mathversion{bold}$A_{2.1}\oplus A_1\colon$}\; $[e_1,e_2]=e_1$ (decomposable, solvable);
\\[1ex]
$n_{\rm D}=4$,\ \
$n_{\rm Z}=1$,\ \
$n_{\rm A}=2$,\ \
$\kappa=x_2y_2$,\ \
$r_{\mathfrak g}=2$,\ \
$r_{\rm s}=2$,\ \
$\rm{DS}=[1,0]$,\ \
$\rm{CS}=[1]$,\ \
$\tr(\ad_v)=-v_2$,\ \
$\mathfrak{C}_{pq}=1$.

\vspace{1.7ex}
\noindent                
{\mathversion{bold}$A_{3.1}\colon$}\;$[e_2,e_3]=e_1$ (Heisenberg, indecomposable, nilpotent, unimodular);
\\[1ex]
$n_{\rm D}=6$,\ \
$n_{\rm Z}=1$,\ \
$n_{\rm A}=2$,\ \
$\kappa=0$,\ \
$r_{\mathfrak g}=3$,\ \
$r_{\rm n}=r_{\rm s}=2$,\ \
$\rm{DS}=[1,0]$,\ \
$\rm{CS}=[1,0]$,\ \
$\tr(\ad_v)=0$.

\vspace{1.7ex}
\noindent                
{\mathversion{bold}$A_{3.2}\colon$}\;$[e_1,e_3]=e_1,\ [e_2,e_3]=e_1+e_2$ (indecomposable, solvable);
\\[1ex]
$n_{\rm D}=4$,\ \
$n_{\rm Z}=0$,\ \
$n_{\rm A}=2$,\ \
$\kappa=2x_3y_3$,\ \
$r_{\mathfrak g}=1$,\ \
$r_{\rm s}=2$,\ \
$\rm{DS}=[2,0]$,\ \
$\rm{CS}=[2]$,\ \
$\tr(\ad_v)=-2v_3$,\ \
$\mathfrak{C}_{pq}=2$.

\vspace{1.7ex}
\noindent                
{\mathversion{bold}$A_{3.3}\colon$}\;$[e_1,e_3]=e_1,\ [e_2,e_3]=e_2$ (indecomposable, solvable);
\\[1ex]
$n_{\rm D}=6$,\ \
$n_{\rm Z}=0$,\ \
$n_{\rm A}=2$,\ \
$\kappa=2x_3y_3$,\ \
$r_{\mathfrak g}=1$,\ \
$r_{\rm s}=2$,\ \
$\rm{DS}=[2,0]$,\ \
$\rm{CS}=[2]$,\ \
$\tr(\ad_v)=-2v_3$,\ \
$\mathfrak{C}_{pq}=2$.

\vspace{1.7ex}
\noindent                
{\mathversion{bold}$A_{3.4}^{-1}\colon$}\;$[e_1,e_3]=e_1,\ [e_2,e_3]=-e_2$
(indecomposable, solvable, unimodular);
\\[1ex]
$n_{\rm D}=4$,\ \
$n_{\rm Z}=0$,\ \
$n_{\rm A}=2$,\ \
$\kappa=2x_3y_3$,\ \
$r_{\mathfrak g}=1$,\ \
$r_{\rm s}=2$,\ \
$\rm{DS}=[2,0]$,\ \
$\rm{CS}=[2]$,\ \
$\tr(\ad_v)=0$,\ \
$\mathfrak{C}_{2p,2q}=2$.

\vspace{1.7ex}
\noindent                
{\mathversion{bold}$A_{3.4}^a\colon$}\;$[e_1,e_3]=e_1,\ [e_2,e_3]=ae_2,\ 0<|a|<1$
(indecomposable, solvable);
\\[1ex]
$n_{\rm D}=4$,\ \
$n_{\rm Z}=0$,\ \
$n_{\rm A}=2$,\ \
$\kappa=(1+a^2)x_3y_3$,\ \
$r_{\mathfrak g}=1$,\ \
$r_{\rm s}=2$,\ \
$\rm{DS}=[2,0]$,\ \
$\rm{CS}=[2]$,\ \
$\tr(\ad_v)=-(1+a)v_3$,\ \
$\mathfrak{C}_{pq}=1+\frac{a^p+a^q}{1+a^{p+q}}$.

\vspace{1.7ex}
\noindent                
{\mathversion{bold}$A_{3.5}^0\colon$}\;$[e_1,e_3]=-e_2,\ [e_2,e_3]=e_1$
(indecomposable, solvable, unimodular);
\\[1ex]
$n_{\rm D}=4$,\ \
$n_{\rm Z}=0$,\ \
$n_{\rm A}=2$,\ \
$\kappa=-2x_3y_3$,\ \
$r_{\mathfrak g}=1$,\ \
$r_{\rm s}=2$,\ \
$\rm{DS}=[2,0]$,\ \
$\rm{CS}=[2]$,\ \
$\tr(\ad_v)=0$,\ \
$\mathfrak{C}_{2p,2q}=2$.

\vspace{1.7ex}
\noindent                
{\mathversion{bold}$A_{3.5}^b\colon$}\;$[e_1,e_3]=be_1-e_2,\ [e_2,e_3]=e_1+be_2,\ b>0$
(indecomposable, solvable);
\\[1ex]
$n_{\rm D}=4$,\ \
$n_{\rm Z}=0$,\ \
$n_{\rm A}=2$,\ \
$\kappa=2(b^2-1)x_3y_3$,\ \
$r_{\mathfrak g}=1$,\ \
$r_{\rm s}=2$,\ \
$\rm{DS}=[2,0]$,\ \
$\rm{CS}=[2]$,\ \
$\tr(\ad_v)=-2bv_3$,\ \
$\mathfrak{C}_{pq}=\frac{2\mathop{\rm Re}(b+i)^p \mathop{\rm Re}(b+i)^q}{\mathop{\rm Re}(b+i)^{p+q}}$.

\vspace{1.7ex}
\noindent                
{\mathversion{bold}$sl(2,\R)\colon$}\;$[e_1,e_2]=e_1,\ [e_2,e_3]=e_3,\ [e_1,e_3]=2e_2$ (indecomposable, simple, unimodular);
\\[1ex]
$n_{\rm D}=3$,\ \
$n_{\rm Z}=0$,\ \
$n_{\rm A}=1$,\ \
$\kappa=-2(2x_3y_1-x_2y_2+2x_1y_3)$,\ \
$r_{\mathfrak g}=1$,\ \
$\rm{DS}=[3]$,\ \
$\rm{CS}=[3]$,\ \
$\tr(\ad_v)=0$,\ \
$\mathfrak{C}_{2p,2q}=2$.

\vspace{1.7ex}
\noindent                
{\mathversion{bold}$so(3)\colon$}\;$[e_1,e_2]=e_3,\ [e_2,e_3]=e_1,\ [e_3,e_1]=e_2$ (indecomposable, simple, unimodular);
\\[1ex]
$n_{\rm D}=3$,\ \
$n_{\rm Z}=0$,\ \
$n_{\rm A}=1$,\ \
$\kappa=-2(x_1y_1+x_2y_2+x_3y_3)$,\ \
$r_{\mathfrak g}=1$,\ \
$\rm{DS}=[3]$,\ \
$\rm{CS}=[3]$,\ \
$\tr(\ad_v)=0$,\ \
$\mathfrak{C}_{2p,2q}=2$.

\begin{remark}\label{rem2}
Two terms of the above list (namely, $\{A_{3.4}^a\}$ and $\{A_{3.5}^b\}$) are, in fact,
Lie algebra series, and each of them being parameterized with one real parameter.
Some values of the parameters are singular, i.e.,\ 
algebra characteristics for them differ from the ones for regular values.
For example, the Killing form of the algebra $\{A_{3.5}^b\}$ identically vanishes if $b=1$.
The same parameter values are singular from the viewpoint of realizations, invariants, subalgebras etc.
See, e.g.,~\cite{PateraSharpWinternitzZassenhaus1976,Patera&Winternitz1977,
Popovych&Boyko&Nesterenko&Lutfullin2003a,Popovych&Boyko&Nesterenko&Lutfullin2003b}.

This fact prevents one from creation of a `canonical' list of inequivalent low-dimensional Lie algebras.
Whether is it reasonable to extract the algebras corresponding to singular parameter values from series? 
The question is answered in different ways.  
For example, in~\cite{Patera&Winternitz1977, PateraSharpWinternitzZassenhaus1976} 
all such algebras are separated from the series and have individual numbers.
In~\cite{Mubarakzyanov1963a} only the direct sums and single algebras (e.g., $A_{3.3}$) are considered separately from 
the corresponding series. 

Another barrier for canonization of the existing lists is generated by ambiguous choice of series parameters normalization 
and even by existence of essentially different approaches to such normalization.

We follow the numeration by Mubarakzyanov, explicitly point out all the singular values of series parameters and study the corresponding 
algebras separately from the regular algebras of series. Usage of this technique simplifies application of necessary contraction criteria.

Let us note that Agaoka~\cite{Agaoka1999,Agaoka2002} proposed classifications of three- and four-dimensional algebras, 
which are well adapted to investigation of contractions and deformations. 
The presented approach can be extended to the real case and algebras of greater dimensions.
\end{remark}

\subsection{Four-dimensional algebras}\label{4_dim_algebras}

\noindent                
{\mathversion{bold}$4A_{1}$} (Abelian, unimodular);
\\[1ex]
$n_{\rm D}=16$,\ \
$n_{\rm Z}=4$,\ \
$n_{\rm A}=4$,\ \
$r_{\mathfrak g}=4$,\ \
$r_{\rm n}=r_{\rm s}=1$,\ \
$\rm{DS}=[0]$,\ \
$\rm{CS}=[0]$,\ \
$\tr(\ad_v)=0$.

\vspace{2.3ex}
\noindent                
{\mathversion{bold}$A_{2.1}\oplus 2A_1\colon$}\;$[e_1,e_2]=e_1$ (decomposable, solvable);
\\[1ex]
$n_{\rm D}=8$,\ \
$n_{\rm Z}=2$,\ \
$n_{\rm A}=3$,\ \
$\kappa=x_2y_2$,\ \
$r_{\mathfrak g}=3$,\ \
$r_{\rm s}=2$,\ \
$\rm{DS}=[1,0]$,\ \
$\rm{CS}=[1]$,\ \
$\tr(\ad_v)=-v_2$,\ \
$\mathfrak{C}_{pq}=1$.

\vspace{2.3ex}
\noindent                
{\mathversion{bold}$2A_{2.1}\colon$}\;$[e_1,e_2]=e_1,\ [e_3,e_4]=e_3$ (decomposable, solvable);
\\[1ex]
$n_{\rm D}=4$,\ \
$n_{\rm Z}=0$,\ \
$n_{\rm A}=2$,\ \
$\kappa=x_2y_2+x_4y_4$,\ \
$r_{\mathfrak g}=2$,\ \
$r_{\rm s}=2$,\ \
$\rm{DS}=[2,0]$,\ \
$\rm{CS}=[2]$,\ \
$\tr(\ad_v)=-(v_2+v_4)$.

\vspace{2.3ex}
\noindent                
{\mathversion{bold}$A_{3.1}\oplus A_1\colon$}\;$[e_2,e_3]=e_1$ (decomposable, nilpotent, unimodular);
\\[1ex]
$n_{\rm D}=\!10$,\ \
$n_{\rm Z}=2$,\ \
$n_{\rm A}=3$,\ \
$\kappa=0$,\ \
$r_{\mathfrak g}=4$,\ \
$r_{\rm n}=r_{\rm s}=2$,\ \
$\rm{DS}=[1,0]$,\ \
$\rm{CS}=[1,0]$,\ \
$\tr(\ad_v)=0$.

\vspace{2.3ex}
\noindent                
{\mathversion{bold}$A_{3.2}\oplus A_1\colon$}\;$[e_1,e_3]=e_1,\ [e_2,e_3]=e_1+e_2$ (decomposable, solvable);
\\[1ex]
$n_{\rm D}=6$,\ \
$n_{\rm Z}=1$,\ \
$n_{\rm A}=3$,\ \
$\kappa=2\,x_3y_3$,\ \
$r_{\mathfrak g}=2$,\ \
$r_{\rm s}=2$,\ \
$\rm{DS}=[2,0]$,\ \
$\rm{CS}=[2]$,\ \
$\tr(\ad_v)=-2v_3$,\ \
$\mathfrak{C}_{pq}=2$.

\vspace{2.3ex}
\noindent                
{\mathversion{bold}$A_{3.3}\oplus A_1\colon$}\;$[e_1,e_3]=e_1,\ [e_2,e_3]=e_2$ (decomposable, solvable);
\\[1ex]
$n_{\rm D}=8$,\ \
$n_{\rm Z}=1$,\ \
$n_{\rm A}=3$,\ \
$\kappa=2\,x_3y_3$,\ \
$r_{\mathfrak g}=2$,\ \
$r_{\rm s}=2$,\ \
$\rm{DS}=[2,0]$,\ \
$\rm{CS}=[2]$,\ \
$\tr(\ad_v)=-2v_3$,\ \
$\mathfrak{C}_{pq}=2$.

\vspace{2.3ex}
\noindent                
{\mathversion{bold}$A_{3.4}^{-1}\oplus A_1\colon$}\;$[e_1,e_3]=e_1,\ [e_2,e_3]=-e_2$
(decomposable, solvable, unimodular);
\\[1ex]
$n_{\rm D}=6$,\ \
$n_{\rm Z}=1$,\ \
$n_{\rm A}=3$,\ \
$\kappa=2x_3y_3$,\ \
$r_{\mathfrak g}=2$,\ \
$r_{\rm s}=2$,\ \
$\rm{DS}=[2,0]$,\ \
$\rm{CS}=[2]$,\ \
$\tr(\ad_v)=0$,\ \
$\mathfrak{C}_{2p,2q}=2$.

\vspace{1.7ex}
\noindent                
{\mathversion{bold}$A_{3.4}^a\oplus A_1\colon$}\;$[e_1,e_3]=e_1,\ [e_2,e_3]=ae_2,\ 0<|a|<1$
(decomposable, solvable);
\\[1ex]
$n_{\rm D}=6$,\ \
$n_{\rm Z}=1$,\ \
$n_{\rm A}=3$,\ \
$\kappa=(1+a^2)x_3y_3$,\ \
$r_{\mathfrak g}=2$,\ \
$r_{\rm s}=2$,\ \
$\rm{DS}=[2,0]$,\ \
$\rm{CS}=[2]$,\ \
$\tr(\ad_v)=-(1+a)v_3$,\ \
$\mathfrak{C}_{pq}=1+\frac{a^p+a^q}{1+a^{p+q}}$.

\vspace{1.7ex}
\noindent                
{\mathversion{bold}$A_{3.5}^0\oplus A_1\colon$}\;$[e_1,e_3]=-e_2,\ [e_2,e_3]=e_1$
(decomposable, solvable, unimodular);
\\[1ex]
$n_{\rm D}=6$,\ \
$n_{\rm Z}=1$,\ \
$n_{\rm A}=3$,\ \
$\kappa=-2x_3y_3$,\ \
$r_{\mathfrak g}=2$,\ \
$r_{\rm s}=2$,\ \
$\rm{DS}=[2,0]$,\ \
$\rm{CS}=[2]$,\ \
$\tr(\ad_v)=0$,\ \
$\mathfrak{C}_{2p,2q}=2$.

\vspace{1.7ex}
\noindent                
{\mathversion{bold}$A_{3.5}^b\oplus A_1\colon$}\;$[e_1,e_3]=be_1-e_2,\ [e_2,e_3]=e_1+be_2,\ b>0$
(decomposable, solvable);
\\[1ex]
$n_{\rm D}=6$,\ \
$n_{\rm Z}=1$,\ \
$n_{\rm A}=3$,\ \
$\kappa=2(b^2-1)x_3y_3$,\ \
$r_{\mathfrak g}=2$,\ \
$r_{\rm s}=2$,\ \
$\rm{DS}=[2,0]$,\ \
$\rm{CS}=[2]$,\ \
$\tr(\ad_v)=-2bv_3$,\ \
$\mathfrak{C}_{pq}=\frac{2\mathop{\rm Re}(b+i)^p \mathop{\rm Re}(b+i)^q}{\mathop{\rm Re}(b+i)^{p+q}}$.

\vspace{1.7ex}
\noindent                
{\mathversion{bold}$sl(2,\R)\oplus A_1\colon$}\;$[e_1,e_2]=e_1,\ [e_2,e_3]=e_3,\ [e_1,e_3]=2e_2$ 
(decomposable, unsolvable, reductive,\linebreak unimodular);
\\[1ex]
$n_{\rm D}=4$,\ \
$n_{\rm Z}=1$,\ \
$n_{\rm A}=2$,\ \
$\kappa=-2(2x_3y_1-x_2y_2+2x_1y_3)$,\ \
$r_{\mathfrak g}=2$,\ \
$\rm{DS}=[3]$,\ \
$\rm{CS}=[3]$,\ \
$\tr(\ad_v)=0$,\ \
$\mathfrak{C}_{2p2q}=2$.

\vspace{1.7ex}
\noindent                
{\mathversion{bold}$so(3)\oplus A_1\colon$}\;$[e_1,e_2]=e_3,\ [e_2,e_3]=e_1,\ [e_3,e_1]=e_2$ 
(decomposable, unsolvable, reductive, unimodular);
\\[1ex]
$n_{\rm D}=4$,\ \
$n_{\rm Z}=1$,\ \
$n_{\rm A}=2$,\ \
$\kappa=-2(x_1y_1+x_2y_2+x_3y_3)$,\ \
$r_{\mathfrak g}=2$,\ \
$\rm{DS}=[3]$,\ \
$\rm{CS}=[3]$,\ \
$\tr(\ad_v)=0$,\ \
$\mathfrak{C}_{2p,2q}=2$.

\vspace{1.7ex}
\noindent                
{\mathversion{bold}$A_{4.1}\colon\; $} $[e_2,e_4]=e_1,\ [e_3,e_4]=e_2$ (indecomposable, solvable, nilpotent, unimodular);
\\[1.3ex plus .5ex minus .1ex]
$n_{\rm D}=7$,\ \
$n_{\rm Z}=1$,\ \
$n_{\rm A}=3$,\ \
$\kappa=0$,\ \
$r_{\mathfrak g}=4$,\ \
$r_{\rm n}=3$,\ \
$r_{\rm s}=2$,\ \
$\rm{DS}=[2,0]$,\ \
$\rm{CS}=[2,1,0]$,\ \
$\tr(\ad_v)=0$.

\vspace{1.7ex}
\noindent               
{\mathversion{bold}$A_{4.2}^1\colon\;$} $[e_1,e_4]=e_1,\ [e_2,e_4]=e_2,\ [e_3,e_4]=e_2+e_3$
(indecomposable, solvable);               
\\[1.3ex plus .5ex minus .1ex]
$n_{\rm D}=8$,\ \
$n_{\rm Z}=0$,\ \
$n_{\rm A}=3$,\ \
$\kappa=3x_4y_4$,\ \
$r_{\mathfrak g}=1$,\ \
$r_{\rm s}=2$,\ \
$\rm{DS}=[3,0]$,\ \
$\rm{CS}=[3]$,\ \
$\tr(\ad_v)=-3v_4$,\ \
$\mathfrak{C}_{pq}=3$.

\vspace{1.7ex}
\noindent                
{\mathversion{bold}$A_{4.2}^{-2}\colon\;$} $[e_1,e_4]=-2e_1,\ [e_2,e_4]=e_2,\ [e_3,e_4]=e_2+e_3$
(indecomposable, solvable, unimodular);               
\\[1.3ex plus .5ex minus .1ex]
$n_{\rm D}=6$,\ \
$n_{\rm Z}=0$,\ \
$n_{\rm A}=3$,\ \
$\kappa=(b^2+2)x_4y_4$,\ \
$r_{\mathfrak g}=1$,\ \
$r_{\rm s}=2$,\ \
$\rm{DS}=[3,0]$,\ \
$\rm{CS}=[3]$,\ \
$\tr(\ad_v)=0$,\ \
$\mathfrak{C}_{pq}=\frac{(2+(-2)^p)(2+(-2)^q)}{2+(-2)^{p+q}}$,\ \  $p,q\geqslant2$.

\vspace{1.7ex}
\noindent                
{\mathversion{bold}$A_{4.2}^b\colon\;$} $[e_1,e_4]=be_1,\ [e_2,e_4]=e_2,\ [e_3,e_4]=e_2+e_3,\ b\ne -2,0,1$
(indecomposable, solvable);               
\\[1.3ex plus .5ex minus .1ex]
$n_{\rm D}=6$,\ \
$n_{\rm Z}=0$,\ \
$n_{\rm A}=3$,\ \
$\kappa=(b^2+2)x_4y_4$,\ \
$r_{\mathfrak g}=1$,\ \
$r_{\rm s}=2$,\ \
$\rm{DS}=[3,0]$,\ \
$\rm{CS}=[3]$,\ \
$\tr(\ad_v)=-(2+b)v_4$,\ \
$\mathfrak{C}_{pq}=\frac{(2+b^p)(2+b^q)}{2+b^{p+q}}$.

\vspace{1.7ex}
\noindent                
{\mathversion{bold}$A_{4.3}\colon\; $} $[e_1,e_4]=e_1,\ [e_3,e_4]=e_2$ (indecomposable, solvable);
\\[1.3ex plus .5ex minus .1ex]
$n_{\rm D}=6$,\ \
$n_{\rm Z}=1$,\ \
$n_{\rm A}=3$,\ \
$\kappa=x_4y_4$,\ \
$r_{\mathfrak g}=3$,\ \
$r_{\rm s}=2$,\ \
$\rm{DS}=[2,0]$,\ \
$\rm{CS}=[2,1]$,\ \
$\tr(\ad_v)=-v_4$,\ \
$\mathfrak{C}_{pq}=1$.

\vspace{1.7ex}
\noindent                
{\mathversion{bold}$A_{4.4}\colon\; $} $[e_1,e_4]=e_1,\ [e_2,e_4]=e_1+e_2,\ [e_3,e_4]=e_2+e_3$
(indecomposable, solvable);
\\[1.3ex plus .5ex minus .1ex]
$n_{\rm D}=6$,\ \
$n_{\rm Z}=0$,\ \
$n_{\rm A}=3$,\ \
$\kappa=3\,x_4y_4$,\ \
$r_{\mathfrak g}=1$,\ \
$r_{\rm s}=2$,\ \
$\rm{DS}=[3,0]$,\ \
$\rm{CS}=[3]$,\ \
$\tr(\ad_v)=-3v_4$,\ \
$\mathfrak{C}_{pq}=3$.

\vspace{1.7ex}
\noindent               
{\mathversion{bold}$A_{4.5}^{111}\colon$}\;$[e_1,e_4]=e_1,\ [e_2,e_4]=e_2,\ [e_3,e_4]=e_3$
(indecomposable, solvable);
\\[1.3ex plus .5ex minus .1ex]
$n_{\rm D}=12$,\ \
$n_{\rm Z}=0$,\ \
$n_{\rm A}=3$,\ \
$\kappa=3x_4y_4$,\ \
$r_{\mathfrak g}=1$,\ \
$r_{\rm s}=2$,\ \
$\rm{DS}=[3,0]$,\ \
$\rm{CS}=[3]$,\ \
$\tr(\ad_v)=-3v_4$,\ \
$\mathfrak{C}_{pq}=3$.

\vspace{1.7ex}
\noindent               
{\mathversion{bold}$A_{4.5}^{-2,1,1}\colon$}\;$[e_1,e_4]=-2e_1,\ [e_2,e_4]=e_2,\ [e_3,e_4]=e_3$
(indecomposable, solvable, unimodular);
\\[1.3ex plus .5ex minus .1ex]
$n_{\rm D}=8$,\ \
$n_{\rm Z}=0$,\ \
$n_{\rm A}=3$,\ \
$\kappa=6x_4y_4$,\ \
$r_{\mathfrak g}=1$,\ \
$r_{\rm s}=2$,\ \
$\rm{DS}=[3,0]$,\ \
$\rm{CS}=[3]$,\ \
$\tr(\ad_v)=0$,\ \
$\mathfrak{C}_{pq}=\frac{(2+(-2)^p)(2+(-2)^q)}{2+(-2)^{p+q}}$,\ \  $p,q\geqslant2$.

\vspace{1.7ex}
\noindent               
{\mathversion{bold}$A_{4.5}^{a11}\colon$}\;$[e_1,e_4]=ae_1,\ [e_2,e_4]=e_2,\ [e_3,e_4]=e_3,\ a\ne -2,0,1$
(indecomposable, solvable);
\\[1.3ex plus .5ex minus .1ex]
$n_{\rm D}=8$,\ \
$n_{\rm Z}=0$,\ \
$n_{\rm A}=3$,\ \
$\kappa=(a^2+2)x_4y_4$,\ \
$r_{\mathfrak g}=1$,\ \
$r_{\rm s}=2$,\ \
$\rm{DS}=[3,0]$,\ \
$\rm{CS}=[3]$,\ \
$\tr(\ad_v)=-(a+2)v_4$,\ \
$\mathfrak{C}_{pq}=\frac{(2+a^p)(2+a^q)}{2+a^{p+q}}$.

\vspace{1.7ex}
\noindent               
{\mathversion{bold}$A_{4.5}^{a,-1,1}\colon$}\;$[e_1,e_4]=ae_1,\ [e_2,e_4]=-e_2,\ [e_3,e_4]=e_3$; $a>0$, $|a|\ne1$
(indecomposable, solvable);
\\[1.3ex plus .5ex minus .1ex]
$n_{\rm D}=6$,\ \
$n_{\rm Z}=0$,\ \
$n_{\rm A}=3$,\ \
$\kappa=(a^2+2)x_4y_4$,\ \
$r_{\mathfrak g}=1$,\ \
$r_{\rm s}=2$,\ \
$\rm{DS}=[3,0]$,\ \
$\rm{CS}=[3]$,\ \
$\tr(\ad_v)=-av_4$,\ \
$\mathfrak{C}_{pq}=\frac{(1+(-1)^p+a^p)(1+(-1)^q+a^q)}{1+(-1)^{p+q}+a^{p+q}}$.

\vspace{1.7ex}
\noindent               
{\mathversion{bold}$A_{4.5}^{a,-1-a,1}\colon$}\;$[e_1,e_4]=ae_1,\ [e_2,e_4]=-(1+a)e_2,\ [e_3,e_4]=e_3$
$a<0$, or $a=1$
(indecomposable, solvable, unimodular);
\\[1.3ex plus .5ex minus .1ex]
$n_{\rm D}=6$,\ \
$n_{\rm Z}=0$,\ \
$n_{\rm A}=3$,\ \
$\kappa=(a^2+(1+a)^2+1)x_4y_4$,\ \
$r_{\mathfrak g}=1$,\ \
$r_{\rm s}=2$,\ \
$\rm{DS}=[3,0]$,\ \
$\rm{CS}=[3]$,\ \
$\tr(\ad_v)=0$,\ \
$\mathfrak{C}_{pq}=\frac{(1+(-1-a)^p+a^p)(1+(-1-a)^q+a^q)}{1+(-1-a)^{p+q}+a^{p+q}}$,\ \  $p,q\geqslant2$.

\vspace{1.7ex}
\noindent               
{\mathversion{bold}$A_{4.5}^{ab1}\colon$}\;$[e_1,e_4]=ae_1,\ [e_2,e_4]=be_2,\ [e_3,e_4]=e_3$;
$ab\ne 0$, $-1<a<b<1$, $a+b\ne-1$
(indecomposable, solvable);
\\[1ex]
$n_{\rm D}=6$,\ \
$n_{\rm Z}=0$,\ \
$n_{\rm A}=3$,\ \
$\kappa=(a^2+b^2+1)x_4y_4$,\ \
$r_{\mathfrak g}=1$,\ \
$r_{\rm s}=2$,\ \
$\rm{DS}=[3,0]$,\ \
$\rm{CS}=[3]$,\ \
$\tr(\ad_v)=-(a+b+1)v_4$,\ \
$\mathfrak{C}_{pq}=\frac{(1+a^p+b^p)(1+a^q+b^q)}{1+a^{p+q}+b^{p+q}}$.

\vspace{1.7ex}
\noindent                
{\mathversion{bold}$A_{4.6}^{-2b,b}\colon$}\;$[e_1,e_4]=-2be_1,\ [e_2,e_4]=be_2-e_3,\ [e_3,e_4]=e_2+be_3,\ b<0$
(indecomposable, solvable, unimodular);
\\[1ex]
$n_{\rm D}=6$,\ \
$n_{\rm Z}=0$,\ \
$n_{\rm A}=3$,\ \
$\kappa=(6b^2-2)x_4y_4$,\ \
$r_{\mathfrak g}=1$,\ \
$r_{\rm s}=2$,\ \
$\rm{DS}=[3,0]$,\ \
$\rm{CS}=[3]$,\ \
$\tr(\ad_v)=0$,\ \
$\mathfrak{C}_{pq}=\frac{((-2b)^p+2\mathop{\rm Re}(b+i)^p)((-2b)^q+2\mathop{\rm Re}(b+i)^q)}
{(-2b)^{p+q}+2\mathop{\rm Re}(b+i)^{p+q}}$,\ \  $p,q\geqslant2$.

\vspace{1.7ex}
\noindent                
{\mathversion{bold}$A_{4.6}^{ab}\colon$}\;$[e_1,e_4]=ae_1,\ [e_2,e_4]=be_2-e_3,\ [e_3,e_4]=e_2+be_3,\ a>0,\ a\ne-2b$
(indecomposable, solvable);
\\[1ex]
$n_{\rm D}=6$,\ \
$n_{\rm Z}=0$,\ \
$n_{\rm A}=3$,\ \
$\kappa=(a^2+2b^2-2)x_4y_4$,\ \
$r_{\mathfrak g}=1$,\ \
$r_{\rm s}=2$,\ \
$\rm{DS}=[3,0]$,\ \
$\rm{CS}=[3]$,\ \
$\tr(\ad_v)=-(a+2b)v_4$,\ \
$\mathfrak{C}_{pq}=\frac{(a^p+2\mathop{\rm Re}(b+i)^p)(a^q+2\mathop{\rm Re}(b+i)^q)}
{a^{p+q}+2\mathop{\rm Re}(b+i)^{p+q}}$.

\vspace{1.7ex}
\noindent                
{\mathversion{bold}$A_{4.7}\colon\; $} $[e_2,e_3]=e_1,\ [e_1,e_4]=2e_1,\ [e_2,e_4]=e_2,\ [e_3,e_4]=e_2+e_3$ (indecomposable, solvable);
\\[1ex]
$n_{\rm D}=5$,\ \
$n_{\rm Z}=0$,\ \
$n_{\rm A}=2$,\ \
$\kappa=6\,x_4y_4$,\ \
$r_{\mathfrak g}=1$,\ \
$r_{\rm s}=3$,\ \
$\rm{DS}=[3,1,0]$,\ \
$\rm{CS}=[3]$,\ \
$\tr(\ad_v)=-4v_4$,\ \
$\mathfrak{C}_{pq}=\frac{(2+2^p)(2+2^q)}{2+2^{p+q}}$.

\vspace{1.7ex}
\noindent                
{\mathversion{bold}$A_{4.8}^{0}\colon\; $} $[e_2,e_3]=e_1,\ [e_1,e_4]=e_1,\ [e_2,e_4]=e_2$
(indecomposable, solvable);
\\[1ex]
$n_{\rm D}=5$,\ \
$n_{\rm Z}=0$,\ \
$n_{\rm A}=2$,\ \
$\kappa=2x_4y_4$,\ \
$r_{\mathfrak g}=2$,\ \
$r_{\rm s}=2$,\ \
$\rm{DS}=[2,0]$,\ \
$\rm{CS}=[2]$,\ \
$\tr(\ad_v)=-2v_4$,\ \
$\mathfrak{C}_{pq}=2$.

\vspace{1.7ex}
\noindent                
{\mathversion{bold}$A_{4.8}^{1}\colon\; $} $[e_2,e_3]=e_1,\ [e_1,e_4]=2e_1,\ [e_2,e_4]=e_2,\ [e_3,e_4]=e_3$
(indecomposable, solvable);
\\[1ex]
$n_{\rm D}=7$,\ \
$n_{\rm Z}=0$,\ \
$n_{\rm A}=2$,\ \
$\kappa=6x_4y_4$,\ \
$r_{\mathfrak g}=1$,\ \
$r_{\rm s}=3$,\ \
$\rm{DS}=[3,1,0]$,\ \
$\rm{CS}=[3]$,\ \
$\tr(\ad_v)=-4v_4$,\ \
$\mathfrak{C}_{pq}=\frac{(2+2^p)(2+2^q)}{2+2^{p+q}}$.

\vspace{1.7ex}
\noindent                
{\mathversion{bold}$A_{4.8}^{-1}\colon\; $} $[e_2,e_3]=e_1,\ [e_2,e_4]=e_2,\ [e_3,e_4]=-e_3$
(indecomposable, solvable; unimodular);
\\[1ex]
$n_{\rm D}=5$,\ \
$n_{\rm Z}=1$,\ \
$n_{\rm A}=2$,\ \
$\kappa=2x_4y_4$,\ \
$r_{\mathfrak g}=2$,\ \
$r_{\rm s}=3$,\ \
$\rm{DS}=[3,1,0]$,\ \
$\rm{CS}=[3]$,\ \
$\tr(\ad_v)=0$,\ \
$\mathfrak{C}_{2p,2q}=2$.

\vspace{1.7ex}
\noindent                
{\mathversion{bold}$A_{4.8}^{b}\colon\; $} $[e_2,e_3]=e_1,\ [e_1,e_4]=(1+b)e_1,\ [e_2,e_4]=e_2,\ [e_3,e_4]=be_3,\ 0<|b|<1$
(indecomposable, solvable);
\\[1ex]
$n_{\rm D}=5$,\ \
$n_{\rm Z}=0$,\ \
$n_{\rm A}=2$,\ \
$\kappa=2(1+b+b^2)x_4y_4$,\ \
$r_{\mathfrak g}=1$,\ \
$r_{\rm s}=3$,\ \
$\rm{DS}=[3,1,0]$,\ \
$\rm{CS}=[3]$,\ \
$\tr(\ad_v)=-2(1+b)v_4$,\ \
$\mathfrak{C}_{pq}=\frac{(1+b^p+(1+b)^p)(1+b^q+(1+b)^q)}{1+b^{p+q}+(1+b)^{p+q}}$.

\vspace{1.7ex}
\noindent               
{\mathversion{bold}$A_{4.9}^{0}\colon$}\;$[e_2,e_3]=e_1$,\ $[e_2,e_4]=-e_3$,\ $[e_3,e_4]=e_2$
(indecomposable, solvable, unimodular);
\\[1ex]
$n_{\rm D}=5$,\ \
$n_{\rm Z}=1$,\ \
$n_{\rm A}=2$,\ \
$\kappa=-2x_4y_4$,\ \
$r_{\mathfrak g}=2$,\ \
$r_{\rm s}=3$,\ \
$\rm{DS}=[3,1,0]$,\ \
$\rm{CS}=[3]$,\ \
$\tr(\ad_v)=0$,\ \
$\mathfrak{C}_{2p,2q}=2$.

\vspace{1.7ex}
\noindent               
{\mathversion{bold}$A_{4.9}^{a}\colon$}\;$[e_2,e_3]=e_1$,\ $[e_1,e_4]=2ae_1$,\
$[e_2,e_4]=ae_2-e_3$,\ $[e_3,e_4]=e_2+ae_3$,\ $a>0$
(indecomposable, solvable);
\\[1ex]
$n_{\rm D}=5$,\ \
$n_{\rm Z}=0$,\ \
$n_{\rm A}=1$,\ \
$\kappa=2(3a^2-1)x_4y_4$,\ \
$r_{\mathfrak g}=1$,\ \
$r_{\rm s}=3$,\ \
$\rm{DS}=[3,1,0]$,\ \
$\rm{CS}=[3]$,\ \
$\tr(\ad_v)=-4av_4$,\ \
$\mathfrak{C}_{pq}=\frac{((2a)^p+2\mathop{\rm Re}(a+i)^p)((2a)^q+2\mathop{\rm Re}(a+i)^q)}
{(2a)^{p+q}+2\mathop{\rm Re}(a+i)^{p+q}}$,\ \  $p,q\geqslant2$;.

\vspace{1.7ex}
\noindent               
{\mathversion{bold}$A_{4.10}\colon$}\;$[e_1,e_3]=e_1,\ [e_2,e_3]=e_2,\ [e_1,e_4]=-e_2,\ [e_2,e_4]=e_1$
 (indecomposable, solvable);
\\[1ex]
$n_{\rm D}=4$,\ \
$n_{\rm Z}=0$,\ \
$n_{\rm A}=2$,\ \
$\kappa=2(x_3y_3-x_4y_4)$,\ \
$r_{\mathfrak g}=2$,\ \
$r_{\rm s}=2$,\ \
$\rm{DS}=[2,0]$,\ \
$\rm{CS}=[2]$,\ \
$\tr(\ad_v)=-2v_3$.

\begin{remark}
Problems concerning series of algebras and singular values of parameters become more complicated in the case of dimension four.
In particular, in the Lie algebra series $\{A_{4.2}^b\}$, $\{A_{4.5}^{abc}\}$ and $\{A_{4.8}^b\}$ 
the dimension~$n_{\rm D}$ of the differentiation algebra varies depending on values of the series parameters. 
It implies obvious necessity of separation of series parameter subsets according to values of this semiinvariant quantity 
since Criterion~\ref{criterion_dim_Der} based on~$n_{\rm D}$ is most powerful. 

In the above list of algebras, we apply enhanced normalization of series parameters for four-dimensional
real Lie algebras, which were proposed in~\cite{Popovych&Boyko&Nesterenko&Lutfullin2003a,Popovych&Boyko&Nesterenko&Lutfullin2003b}.
\end{remark}

\section{Algorithm of contraction identification}\label{SectionOnContractionIdentification}
The proposed algorithm allows one to handle the continuous one-parametric contractions of
the low-dimensional Lie algebras. It consists of three steps.
\begin{enumerate}
\renewcommand{\labelenumi}{{\rm \theenumi)}}

\item
We take a complete list of nonisomorphic Lie algebras of a fixed dimension.
For each member of this list we calculate invariant and
semiinvariant quantities that concern necessary criteria of contractions.

\item
For each pair of algebras from the list we test possible existence of contractions
with the necessary criteria of contractions via comparing the calculated invariant and semiinvariant quantities.
Since it is sufficient to look only for nontrivial and proper contractions,
we do not have to study the pairs of any Lie algebra with itself and the Abelian one.

\item
Consider each from the pairs which satisfy all the necessary criteria of contractions.
Applying the direct method based on Definition~\ref{DefOfContractions2},
we either construct a contraction matrix in an explicit form or prove that no contraction is possible.

\end{enumerate}

The requisite invariant and semiinvariant quantities of the real three- and four-dimensional Lie algebras
are calculated and collected in Section~\ref{SectionOnLow-DimRealLieAlgebras}.

Most of contractions of low-dimensional Lie algebras are realized via simple In\"on\"u--Wigner contractions.
Any simple In\"on\"u--Wigner contraction corresponds to a subalgebra of the initial algebra and
therefore is easy to find.
Classification of subalgebras of three- and four-dimensional Lie algebras is well known~\cite{Patera&Winternitz1977}.
All simple In\"on\"u--Wigner contractions of these algebras are constructed in~\cite{Conatser1972,Huddleston1978}.
We only enhance presentation of the corresponding contraction matrices.

For the pairs without simple In\"on\"u--Wigner contractions we continue investigation
with generalized In\"on\"u--Wigner contractions.
Here the problem of finding contraction matrices can be divided into two subproblems:
\begin{itemize}\itemsep=0ex
\item
To construct appropriate transformations for the canonical bases of the initial and resulting algebras,
which do not depend on the contraction parameter.
The aim is for the nonzero new structure constants of the resulting algebra
to coincide with the corresponding new structure constants of the initial algebra;
\item
To find a diagonal matrix depending on the contraction parameter.
It is sufficient to assume that the diagonal elements are integer powers of the contraction parameter.
\end{itemize}
As a rule, we can manage to avoid basis change in resulting algebras in the case of dimensions three and four.
Consequently, the contraction matrix can be represented as a product of two matrices
$U_\varepsilon=IW(k_1,\dots,k_n)$, where $I$ is a constant nonsingular matrix and 
$W(k_1,\dots,k_n)={\rm diag}(\varepsilon^{k_1}, \dots, \varepsilon^{k_n})$, $k_1,\dots,k_n\in \Z$.

In complicated cases contraction matrices can be found using repeated contractions
(see Section~\ref{SectionOnMulti-parametricDecomposableAndRepeatedContractions}).

To demonstrate effectiveness of the algorithm, we discuss two typical examples in detail.

\begin{example}
Consider the series of three-dimensional Lie algebras~$A^a_{3.4}$
parameterized with one real parameter $a$, where $-1\leqslant a<1,\ a\ne0$.
Let us investigate all possible contractions of algebra $A^a_{3.4}$ for a fixed value of $a$.

$A^a_{3.4}$ is an indecomposable solvable Lie algebra with the canonical nonzero
commutation relations $[e_1,e_3]=e_1$,\ $[e_2,e_3]=ae_2$.
The tuple of considered quantities for the algebra $A^a_{3.4}$ is
\begin{gather*}
n_{\rm D}=4,\;
n_{\rm Z}=0,\;
n_{\rm A}=2,\;
\kappa=(1+a^2)x_3y_3,\;
\tr(\ad e_3)=-1-a,\;
r_{\rm s}=2,\\[.5ex]
\rm{DS}=[2,0],\;
\rm{CS}=[2].
\end{gather*}

According to the second step of the algorithm we look through all pairs of three-dimensional
algebras, where the initial algebra is $A^a_{3.4}$ and the resulting algebra runs the list
from Subsection~\ref{3_dim_algebras} and does not coincide with $3A_{1}$ and $A^a_{3.4}$.

For each pair we compare tuples of their semiinvariant quantities.
In view of Theorem~\ref{TheoremOnNecessaryContractionCriteria1} we conclude that
\begin{itemize}\itemsep=0ex
\item
contractions to the algebras $A_{2.1}\oplus A_1$, $A_{3.2}$,
$A_{3.4}^{\tilde a}$, {\scriptsize$\tilde a\ne a$}, $A_{3.5}^b$, {\scriptsize$b\geqslant0$}, $sl(2,\R)$ and $so(3)$
are impossible since Criterion~\ref{criterion_dim_Der} is not satisfied;
\item
contraction to the algebra $A_{3.3}$ is impossible according to Criterion~\ref{criterion_Cpq};
\item
contraction to the algebra $A_{3.1}$ may exist inasmuch as all the tested necessary criteria are held.
\end{itemize}

Other criteria can also be used to prove nonexistence of contractions.
For example, for the algebras $sl(2,\R)$ and $so(3)$ we can also use
Criterion~\ref{criterion_dim_Ab}, \ref{criterion_dim_lower_central_series_components}, \ref{criterion_Killing_formRank} 
or \ref{criterion_versus_deformation}.
In all cases we try to apply a minimal set of the most effective criteria such as Criterion~\ref{criterion_dim_Der}.
In particular, Criterion~\ref{criterion_dim_Der} is very important for the example under consideration,
since due to strict inequality it allows one to prove the absence of contractions inside the series $A_{3.4}^{a}$
in a very simple way.

Therefore, on the third step of the algorithm we investigate only the pair $(A_{3.4}^a, A_{3.1})$.

The canonical nonzero commutation relation of the algebra $A_{3.1}$ is $[e_2,e_3]=e_1$.
Since in the canonical basis of $A_{3.4}^a$ the structure constant $c_{23}^1$ equals to zero 
we carry out the basis change
$e'_1=(1-a)e_1,\ e'_2=e_1+e_2,\ e'_3=e_3.$
The new isomorphic commutation relations have the form
\begin{gather*}
[e_1,e_2]'=0,\quad [e_1,e_3]'=e_1,\quad [e_2,e_3]'=e_1+ae_2.
\end{gather*}
Now the desired contraction is provided by the matrix $\diag(\varepsilon,1,\varepsilon)$ and
the subsequent limit process $\varepsilon\to +0$ results in the algebra $A_{3.1}$:
\begin{gather*}
[e_1,e_2]_\varepsilon=0,\\[.5ex]
[e_1,e_3]_\varepsilon=\varepsilon e_1\to0,\ \varepsilon\to+0,\\[.5ex]
[e_2,e_3]_\varepsilon=e_1+\varepsilon a e_2\to e_1, \ \varepsilon\to+0.
\end{gather*}

Finally, all nontrivial proper contractions of the Lie algebra $A_{3.4}^a$
are exhausted by the single contraction $A_{3.4}^a\to A_{3.1}$ which is generated
by the matrix $I_5\rm{diag}(\varepsilon,1,\varepsilon)$,
where the explicit form of $I_5$ is adduced in Subsection~\ref{Contractions_lists_3}.
\end{example}

\begin{example}
Consider the decomposable, unsolvable, unimodular, reductive four-dimensional Lie algebra $sl(2,\R)\oplus A_1$,
having the canonical commutation relations $[e_1,e_2]=e_1$, $[e_2,e_3]=e_3$, $[e_1,e_3]=2e_2$.
The set of algebraic quantities which are used to study contractions of this algebra is exhausted by
\begin{gather*}
n_{\rm D}=4,\;
n_{\rm Z}=1,\;
n_{\rm A}=1,\;
n_{[\mathfrak{g},\mathfrak{g}]}=3,\;
\kappa=-2(2x_3y_1-x_2y_2+2x_1y_3),\;
\rm{DS}=[3],\;
\rm{CS}=[3].
\end{gather*}
The quantities of $sl(2,\R)\oplus A_1$ are compared with the analogous quantities of the other four-dimensional algebras.
All the requisite quantities are adduced in Subsection~\ref{4_dim_algebras}.
In view of necessary contraction criteria we conclude that
\begin{itemize}\itemsep=0ex
\item
contractions to the algebras
$A_{2.1}\oplus 2A_1$, $2A_{2.1}$, $A_{3.2}\oplus A_1$, $A_{3.3}\oplus A_1$,
$A_{3.4}^a\oplus A_1$, {\scriptsize$|a|<1,\ a\ne0,-1$}, $A_{3.5}^b\oplus A_1$, {\scriptsize$b>0$}, $A_{4.3}$,
$A_{4.8}^{b}$, {\scriptsize$|b|\leq 1,\ b\ne-1$}, and $A_{4.9}^{a}$, {\scriptsize$a>0$},
are impossible in view of Criterion~\ref{criterion_unimodular_property};
\item
contraction to the algebra
$so(3)\oplus A_1$ does not exist since Criterion~\ref{criterion_dim_Der} is not held;
\item
contractions to the algebras
$A_{4.2}^b$, {\scriptsize$b\ne 0$}, $A_{4.4}$, $A_{4.5}^{abc}$, {\scriptsize$abc\ne 0$}, $A_{4.6}^{a,b}$, {\scriptsize$a>0$}, $A_{4.7}$ and $A_{4.10}$
are impossible in view of Criterion~\ref{criterion_dim_Z};
\item
contractions to the algebras
$A_{3.1}\oplus A_1$, $A_{4.1}$, $A_{3.4}^{-1}\oplus A_1$, $A_{3.5}^0\oplus A_1$, $A_{4.8}^{-1}$ and $A_{4.9}^{0}$
may exist inasmuch as all the tested necessary criteria of contractions are satisfied.
\end{itemize}

Note that not only Criteria~\ref{criterion_dim_Der}, \ref{criterion_dim_Z} and~\ref{criterion_unimodular_property}
could be used to separate algebras for which there are no contractions from the algebra~$sl(2,\R)\oplus A_1$.
For example, Criterion~\ref{criterion_Cpq} implies impossibility of contractions from~$sl(2,\R)\oplus A_1$ to~$A_{4.4}$.

The contractions admitted by the necessary criteria can actually be executed.
Contractions to the algebras
$A_{3.1}\oplus A_1$, $A_{3.4}^{-1}\oplus A_1$, $A_{3.5}^0\oplus A_1$, $A_{4.1}$, $A_{4.8}^{-1}$ and
$A_{4.9}^{0}$
are provided by the contraction matrices
$I_8\rm{diag}(\varepsilon,\varepsilon,1,1)$, 
$I_7\rm{diag}(\varepsilon,\varepsilon,1,1)$,
$I_{10}\rm{diag}(\varepsilon,\varepsilon,1,1)$, 
$I_{23}\rm{diag}(\varepsilon,\varepsilon,\varepsilon,1)$,
$I_{19}\rm{diag}(\varepsilon,1,\varepsilon,1)$ and 
$I_{22}\rm{diag}(\varepsilon^2,\varepsilon,\varepsilon,1)$ correspondingly.
The explicit forms of the matrices $I$'s are presented in Subsection~\ref{Contractions_lists_4}.

Note that all the contractions except the last one are simple In\"on\"u--Wigner 
contractions and are constructed using a list of inequivalent subalgebras of $sl(2,\R)\oplus A_1$.
We illustrate the applied technique with the pair $(sl(2,\R)\oplus A_1,A_{3.1}\oplus A_1)$.
See also Section~\ref{SectionOnSimplestTypesOfContractions} for the theoretical background.

$\langle e_3,e_4\rangle$ is a subalgebra of $sl(2,\R)\oplus A_1$. 
The associated contraction matrix $\diag(\varepsilon,\varepsilon,1,1)$
produces a simple IW-contraction from $sl(2,\R)\oplus A_1$ to a Lie algebra isomorphic to $A_{3.1}\oplus A_1$. 
In order to obtain the canonical commutation relations ($[e_2,e_3]=e_1$) of the algebra $A_{3.1}\oplus A_1$,
we apply additional isomorphism transformation given by the matrix $I_8$ which commutes with $\diag(\varepsilon,\varepsilon,1,1)$.
The resulting contraction matrix is $I_8\rm{diag}(\varepsilon,\varepsilon,1,1)$.

Further consider the pair ($sl(2,\R)\oplus A_1,A_{4.9}^{0})$ in detail as an example on construction of generalized IW-contractions.
Our aim is to find an appropriate contraction matrix according to the above algorithm.

The canonical commutation relations of the algebra 
$A_{4.9}^{0}$ are $[e_2,e_3]=e_1$, $[e_2,e_4]=-e_3$, $[e_3,e_4]=e_2$.
In contrast to the algebra $A_{4.9}^{0}$, 
the canonical structure constants $c_{23}^1$, $c_{24}^3$ and $c_{34}^2$ of the algebra $sl(2,\R)\oplus A_1$ vanish. 
That is why we carry out the basis change 
\begin{gather*}
e'_1=-\frac 12 e_1-\frac 12 e_3,\quad
e'_2=e_2,\quad
e'_3=\frac 12 e_1-\frac 12 e_3,\quad
e'_4=\frac 12 e_1+\frac 12 e_3+e_4,
\end{gather*}
which is associated with the matrix $I_{22}$.
The obtained commutation relations (being isomorphic to the old one of $sl(2,\R)\oplus A_1$) have the form
\begin{gather*}
[e_1,e_2]'=-e_3,\ 
[e_1,e_3]'=e_2,\ 
[e_1,e_4]'=0,\ 
[e_2,e_3]'=e_1,\ 
[e_2,e_4]'=-e_3,\ 
[e_3,e_4]'=e_2.
\end{gather*}

Let us suppose that for the new Lie bracket $[\cdot,\cdot]'$ the requisite contraction is provided by the matrix
$\diag(\varepsilon^{k_1},\varepsilon^{k_2},\varepsilon^{k_3},\varepsilon^{k_4})$ and calculate the parameterized commutators:
\begin{gather*}
[e_1,e_2]_\varepsilon=-\varepsilon^{k_1+k_2-k_3} e_3,\quad
[e_1,e_3]_\varepsilon=\varepsilon^{k_1+k_3-k_2} e_2,\quad
[e_1,e_4]_\varepsilon=0,\\
[e_2,e_3]_\varepsilon=\varepsilon^{k_2+k_3-k_1} e_1,\quad
[e_2,e_4]_\varepsilon=-\varepsilon^{k_2+k_4-k_3} e_3,\quad
[e_3,e_4]_\varepsilon=\varepsilon^{k_3+k_4-k_2} e_2.
\end{gather*}
The limit of the commutators under $\varepsilon\to+0$ exists and gives the algebra $A_{4.9}^0$ iff 
the powers $k_1,$~\dots, $k_4$ are constrained by the conditions 
\[
k_2+k_3-k_1=0,\quad 
k_2+k_4-k_3=0,\quad 
k_3+k_4-k_2=0,\quad 
k_1+k_2-k_3>0,\quad 
k_1+k_3-k_2>0.
\] 
The tuple $k_1=2$, $k_2=k_3=1$, $k_4=0$ satisfies these conditions. 
The corresponding contraction indeed results in the algebra $A_{4.9}^{0}$:
\begin{gather*}
[e_1,e_2]_\varepsilon=-\varepsilon^{2} e_3\to 0, \ \varepsilon\to+0,\quad
[e_1,e_3]_\varepsilon=\varepsilon^{2} e_2\to 0, \ \varepsilon\to+0,\\
[e_1,e_4]_\varepsilon=0,\quad
[e_2,e_3]_\varepsilon=e_1,\quad
[e_2,e_4]_\varepsilon=-e_3,\quad
[e_3,e_4]_\varepsilon=e_2.
\end{gather*}
The complete contraction matrix is $I_{22}\rm{diag}(\varepsilon^2,\varepsilon,\varepsilon,1)$.

This example demonstrates that necessary criteria allow one
to handle contractions even in the cases of such complicated algebras as reductive ones.
\end{example}

\begin{remark}
Celeghini and Tarlini~\cite{Celeghini&Tarlini1981} proposed the conjecture that
all nonsemisimple Lie algebras of a fixed dimension could be obtained via contractions
from semisimple ones. Actually, the conjecture is incorrect.
There are no semisimple Lie algebras for some dimensions, e.g.,\ in the case of dimension four.
Therefore, a wider class (e.g.,\ the class of reductive algebras or
even the whole class of unsolvable algebras) should be used in the conjecture instead of semisimple algebras.
The other argument on incorrectness of the conjecture is
that all semisimple (and reductive) Lie algebras are unimodular and
any continuous contraction of a unimodular algebra necessarily results in a unimodular algebra.
Complexity of the actual state of affairs is illustrated by consideration of low-dimensional algebras.

The unsolvable three-dimensional algebras are exhausted by the simple algebras $sl(2,\R)$ and $so(3)$.
Any three-dimensional unimodular algebra ($sl(2,\R)$, $so(3)$, $A_{3.4}^{-1}$, $A_{3.4}^0$, $A_{3.1}$, $3A_1$)
belongs to the orbit closure of at least one of the simple algebras.

The reductive algebras $sl(2,\R)\oplus A_1$ and $so(3)\oplus A_1$ form the set of unsolvable four-dimensional algebras.
The union of orbit closures of these algebras consists of the unimodular algebras with the nontrivial centers
($sl(2,\R)\oplus A_1$, $so(3)\oplus A_1$,
$A_{4.8}^{-1}$, $A_{4.9}^0$, $A_{3.4}^{-1}\oplus A_1$, $A_{3.4}^0\oplus A_1$, $A_{4.1}$, $A_{3.1}\oplus A_1$, $4A_1$).
The unimodular algebras having the zero centers ($A_{4.2}^{-2}$, $A_{4.5}^{abc}$, {\scriptsize$a+b+c=0$}, $A_{4.6}^{-2b,b}$)
cannot be obtained via contractions from the unsolvable algebras.

The situation with contractions of representations is different~\cite{Nikitin2006}.
For example, matrix representations of all inequivalent classes of the real three-dimensional Lie algebras
are contractions of appropriately chosen representations (with $\varepsilon$-dependent similarity transformations)
of the simple algebras $sl(2,\R)$ and $so(3)$.
More precisely, concerning the parameterized series of Lie algebras ($A^a_{3.4}$, $A^b_{3.5}$),
only representations for single values of parameters can be obtained via contractions.
\end{remark}

\section{One-parametric contractions of real low-dimensional\\ Lie algebras}\label{SectionOnOne-parContractionsOfRealLow-DimLieAlgebras}
The objective of this section is to construct, order and analyze the contractions of real low-dimensional Lie algebras.

At first, we discuss all possible contractions of one- and two-dimensional Lie algebras.
Since there is only one inequivalent one-dimensional Lie algebra and it is Abelian,
all its contractions are trivial and improper at the same time.
The complete list of nonisomorphic two-dimensional Lie algebras is exhausted by
the Abelian algebra $2A_1$ and the non-Abelian algebra $A_{2.1}$ with the canonical commutation relation
$[e_1,e_2]=e_1$. The unique weakly inequivalent contraction of the algebra $2A_1$ is trivial and improper at the same time.
The contractions of the algebra $A_{2.1}$ are either trivial or improper.

\begin{figure}[ht]
\centerline{\includegraphics[scale=0.88]{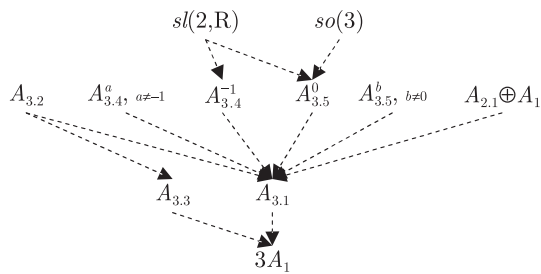}}
\caption{One-parametric contractions of real three-dimensional Lie algebras}\label{fig1}
\end{figure}

Contractions of real three- and four-dimensional Lie algebras are listed in Subsections~\ref{Contractions_lists_3} and~\ref{Contractions_lists_4}
and additionally visualized with Figures~\ref{fig1} and~\ref{fig2}.
Denote that contractions of the three-dimensional real Lie algebras were considered in~\cite{Weimar-Woods1991}.
A complete description of these contractions with proof closed to the manner of our paper was first obtained in~\cite{Lauret2003}.

Only proper direct contractions are presented on the figures. 
Let us remind that a contraction from $\mathfrak g$ to $\mathfrak g_0$
is called \emph{direct} if there is no algebra $\mathfrak g_1$ such that $\mathfrak g_1\not\sim\mathfrak g,\mathfrak g_0$,
$\mathfrak g$ is contracted to $\mathfrak g_1$ and $\mathfrak g_1$ is contracted to $\mathfrak g_0$.
Antonym to this notion is the notion of repeated contraction. See Section~\ref{SectionOnMulti-parametricDecomposableAndRepeatedContractions} for details.
The algebra $\mathfrak g$ is necessarily contracted to $\mathfrak g_0$ if
$\mathfrak g$ is contracted to $\mathfrak g_1$ and $\mathfrak g_1$ is contracted to $\mathfrak g_0$.
That is why the arrows corresponding to repeated contractions can be omitted.

In the lists of contractions we collect all the suitable pairs of Lie algebras with the same initial algebras which are adduced once.
The corresponding contraction matrices are indicated over the arrows.
In the section we use the short-cut notation for the diagonal parts of matrices of generalized In\"on\"u--Wigner contractions:
\begin{gather*}
W(k_1,k_2,\dots,k_n)={\rm diag}(\varepsilon^{k_1},\varepsilon^{k_2},\dots,\varepsilon^{k_n}),
\end{gather*}
where $k_i\in {\mathbb Z}$, $ i=\overline{1,n}$, $n$ is the dimension of the underlying vector space $V$.
The constant `left-hand' parts of matrices of generalized In\"on\"u--Wigner contractions
are denoted by numbered symbols~$I$. Their explicit forms are adduced after the lists of contractions.
The notation $\varepsilon\to+0$ is omitted everywhere.

In the case of simple In\"on\"u--Wigner contractions we additionally adduce the associated subalgebras.

\subsection{Dimension three}\label{Contractions_lists_3}

The list of all possible proper and nontrivial continuous one-parametric contractions
of real three-dimensional Lie algebras is exhausted by the following ones (see also Figure~1):
\begin{gather*}
A_{2.1}\oplus A_1\colon\quad\xrightarrow{I_1W(1,1,0)}A_{3.1},\,\langle e_1-e_3\rangle.
\\[.7ex]
A_{3.2}\colon\quad\xrightarrow{I_7W(1,0,1) \text{ or } W(2,1,1)} A_{3.1},\,\langle e_2\rangle;\quad
\xrightarrow{I_6W(0,1,0) \text{ or } W(1,2,0)} A_{3.3},\,\langle e_1, e_2+e_3\rangle.
\\[.7ex]
A_{3.4}^a\colon\quad\xrightarrow{I_2W(1,0,1)} A_{3.1},\,\langle e_1+e_2\rangle.
\\[1ex]
A_{3.5}^b\colon\quad\xrightarrow{W(1,0,1)} A_{3.1},\,\langle e_2 \rangle.
\\[.7ex]
sl(2,{\mathbb R})\colon\quad
\xrightarrow{I_3W(1,1,0)} A_{3.1},\,\langle e_3\rangle;\quad
\xrightarrow{I_4W(1,0,0)} A_{3.4}^{-1},\,\langle e_2,e_3\rangle;\quad
\xrightarrow{I_5W(1,1,0)} A_{3.5}^0,\,\langle e_1+e_3\rangle.
\\[.7ex]
so(3)\colon\quad
\xrightarrow{W(2,1,1)} A_{3.1};\quad
\xrightarrow{W(1,1,0)} A_{3.5}^0,\,\langle e_3\rangle.
\end{gather*}

\noindent
The constant parts of contraction matrices have the form

\begin{alignat*}{3}
&
I_1=
\left(\begin{array}{ccc}
1 & 0 & -1 \\
0 & 1 & 0 \\
0 & 0 & 1 \\
\end{array}\right),\quad
&&
I_2=
\left(\begin{array}{ccc}
1-a & 1 & 0 \\
0 & 1 & 0 \\
0 & 0 & 1 \\
\end{array}\right),\quad
&&
I_3=
\left(\begin{array}{ccc}
0 & 1 & 0 \\
2 & 0 & 0 \\
0 & 0 & 1 \\
\end{array}\right),\quad
\\[2ex]&
I_4=
\left(\begin{array}{ccc}
1 & 0 & 0 \\
0 & 0 & 1 \\
0 & -1 & 0 \\
\end{array}\right),
&&
I_5=
\left(\begin{array}{ccc}
0 & 0 & \frac12 \\
0 & 1 & 0 \\
1 & 0 & \frac12\\
\end{array}\right),
&&
I_6=
\left(\begin{array}{ccc}
1 & 0 & 0\\
0 & 1 & 1\\
0 & 0 & 1\\
\end{array}\right),
\\[2ex]&
I_7=
\left(\begin{array}{ccc}
-1 & 0 & 0 \\
0 & 1 & 0 \\
0 & 0 & -1 \\
\end{array}\right).
&&
\end{alignat*}

Analysis of the obtained results leads to the conclusion that
for any pair of real three-dimensional Lie algebras we have one of the two possibilities:
1) there are no contractions in view of applied necessary criteria;
2) there exists a generalized In\"on\"u--Wigner contraction.

Only the contraction $so(3)\xrightarrow{} A_{3.1}$ necessarily is a truly generalized In\"on\"u--Wigner contraction.
Nonexistence of a simple In\"on\"u--Wigner contraction in this case is implied by the following
chain of statements. Any proper and nontrivial simple In\"on\"u--Wigner contraction corresponds to a proper
subalgebra of the initial algebra. Equivalent subalgebras result in equivalent contractions.
A~complete list of inequivalent proper subalgebras of $so(3)$ is exhausted by any one-dimensional subalgebra of $so(3)$.
Any one-dimensional subalgebra generates the contraction of $so(3)$ to $A_{3.5}^0$.

All other contractions of real three-dimensional Lie algebras are equivalent to simple In\"on\"u--Wigner contractions
although sometimes generalized In\"on\"u--Wigner contraction have a simpler, pure diagonal form.
We explicitly indicate two such cases in the above list of contractions, namely,
$A_{3.2}\to A_{3.1}$ and $A_{3.2}\to A_{3.3}$.

Note additionally that all the constructed contraction matrices include only nonnegative integer powers of~$\varepsilon$, 
i.e.,\ they admit well-defined limit process under $\varepsilon\to+0$.


\begin{theorem}
Any continuous contraction of a real three-dimensional Lie algebra is equivalent to
a generalized In\"on\"u--Wigner contraction with nonnegative powers of the contraction parameter.
Moreover, only the contraction $so(3)\to A_{3.1}$ is inequivalent to a simple In\"on\"u--Wigner contraction.
\end{theorem}

\subsection{Dimension four}\label{Contractions_lists_4}

The list of all possible proper and nontrivial continuous one-parametric contractions
of real four-dimensional Lie algebras is exhausted by the following ones.

\vspace{1.7ex}

\noindent
$A_{2.1} \oplus 2 A_1\colon\quad\xrightarrow{I_{30}W(1,1,0,0)}A_{3.1}\oplus A_1$, $\langle e_3-e_1, e_4 \rangle$.
\\[1.3ex]                
$2A_{2.1}\colon\quad\xrightarrow{W(0,0,0,1)}A_{2.1}\oplus 2 A_1,\, \langle e_1, e_2, e_3\rangle;\quad
\xrightarrow{I_1W(1,1,0,1)}A_{3.1} \oplus A_1,\, \langle e_1+e_3\rangle;\quad
\\[0ex plus .2ex]
\null \qquad
\xrightarrow{U_3}A_{3.2}\oplus A_1;\quad
\xrightarrow{I_2W(0,0,0,1)}A_{3.3}\oplus A_1,\, \langle e_1, e_3, e_2+e_4\rangle;\quad
\xrightarrow{I_{27}W(1,1,0,1)}A^a_{3.4}\oplus A_1,\, \langle e_2+ae_4\rangle;
\\[0ex plus .2ex]
\null \qquad\xrightarrow{U_4}A_{4.1};\quad
\xrightarrow{I_{28}W(0,1,1,0)}A_{4.3},\, \langle e_1, e_2-e_3\rangle;\quad
\xrightarrow{I_3W(1,0,1,0)}A_{4.8}^0,\, \langle e_1+e_3, e_2+e_4\rangle$.
\\[1.3ex]               
$A_{3.2} \oplus A_1\colon\quad\xrightarrow{W(1,0,1,0)}A_{3.1} \oplus A_1,\, \langle e_2, e_4\rangle;\quad
\xrightarrow{W(0,1,0,0)}A_{3.3}\oplus A_1,\, \langle e_1, e_3, e_4\rangle;\quad
\xrightarrow{I_{29}W(2,1,0,1)}A_{4.1}$.
\\[1.3ex]               
$A_{3.3} \oplus A_1\colon\quad\xrightarrow{I_4W(1,0,1,0)}A_{3.1} \oplus A_1,\, \langle e_1, e_2+e_4\rangle$.
\\[1.3ex]                
$A_{3.4}^a \oplus A_1\colon\quad
\xrightarrow{I_5W(1,1,0,0)}A_{3.1} \oplus A_1,\, \langle e_2, e_1+e_4\rangle;\quad
\xrightarrow{I_6W(2,1,0,1)}A_{4.1}$.
\\[1.3ex]               
$A_{3.5}^b \oplus A_1\colon\quad\xrightarrow{W(1,0,1,0)}A_{3.1} \oplus A_1,\, \langle e_2, e_4\rangle;\quad
\xrightarrow{I_9W(2,1,0,1)}A_{4.1}$.
\\[1.3ex]                
$sl(2,{\mathbb R})\oplus A_1\colon\quad\xrightarrow{I_8W(1,1,0,0)}A_{3.1} \oplus A_1,\, \langle e_3, e_4\rangle;\quad
\xrightarrow{I_7W(1,1,0,0)}A_{3.4}^{-1} \oplus A_1,\, \langle e_2, e_4\rangle;\quad
\\[0ex plus .2ex]
\null\qquad
\xrightarrow{I_{10}W(1,1,0,0)}A_{3.5}^0 \oplus A_1,\, \langle e_1+e_3, e_4\rangle;\quad
\xrightarrow{I_{23}W(1,1,1,0)}A_{4.1},\, \langle e_1+e_4\rangle;
\\[0ex plus .2ex]
\null\qquad
\xrightarrow{I_{19}W(1,0,1,0)}A_{4.8}^{-1},\, \langle e_1, e_2-\frac12e_4\rangle;\quad
\xrightarrow{I_{22}W(2,1,1,0)}A_{4.9}^{0}$.
\\[1.3ex]               
$so(3)\oplus A_1\colon\quad\xrightarrow{W(2,1,1,0)}A_{3.1} \oplus A_1;\quad
\xrightarrow{W(1,1,0,0)}A_{3.5}^0\oplus A_1,\, \langle e_3,e_4\rangle;\quad
\xrightarrow{I_5W(3,2,1,1)}A_{4.1};$\quad
\\[0ex plus .2ex]
\null\qquad
$\xrightarrow{I_{11}W(2,1,1,0)}A_{4.9}^0$.
\\[1.3ex]               
$A_{4.1}\colon\quad
\xrightarrow{I_{13}(0)W(0,0,0,1)}A_{3.1} \oplus A_1,\, \langle e_1, e_2, e_4\rangle$.
\\[1.3ex]               
$A_{4.2}^b\colon\quad
\xrightarrow{I_{14}W(1,0,1,0)}A_{3.1} \oplus A_1,\, \langle e_1,e_3\rangle;\quad
\xrightarrow{b\ne1,\;I_{15}W(2,1,0,1) }A_{4.1};\quad
\xrightarrow{W(1,0,1,0)}A^{b,1,1}_{4.5},\, \langle e_2,e_4\rangle$.
\\[1.3ex]               
$A_{4.3}\colon\quad
\xrightarrow{I_{16}W(0,0,1,0)}A_{2.1}\oplus 2 A_1,\, \langle e_1,e_2,e_4\rangle;\quad
\xrightarrow{I_{14}W(1,0,1,0)}A_{3.1}\oplus A_1,\, \langle e_1,e_3\rangle;\quad
\\[0ex plus .2ex]
\null\qquad\xrightarrow{I_{17}W(2,1,0,1)}A_{4.1}$.
\\[1.3ex]                
$A_{4.4}\colon\quad
\xrightarrow{I_{13}(0)W(1,0,1,1)}A_{3.1}\oplus A_1,\, \langle e_2\rangle;\quad
\xrightarrow{W(2,1,0,1)}A_{4.1};\quad
\xrightarrow{W(0,1,1,0)}A^1_{4.2},\, \langle e_1,e_4\rangle;\quad
\\[0ex plus .2ex]
\null\qquad\xrightarrow{W(0,1,2,0)}A_{4.5}^{111}$.
\\[1.3ex]              
$A_{4.5}^{ab1}\colon\quad
\xrightarrow{a\ne b,\;I_{18}W(1,0,1,0)}A_{3.1}\oplus A_1,\, \langle \frac{1+b}{a}e_1+e_2,e_3\rangle;\quad
\xrightarrow{1\ne a\ne b\ne 1,\;I_{12}W(2,1,0,1)}A_{4.1}$.
\\[1.3ex]                
$A_{4.6}^{ab}\colon\quad\xrightarrow{I_{14}W(1,0,1,0)}A_{3.1}\oplus A_1,\, \langle e_1,e_3\rangle;\quad
\xrightarrow{I_{20}W(2,1,0,1)}A_{4.1}$.
\\[1.3ex]                
$A_{4.7}\colon\quad\xrightarrow{I_{14}W(1,0,1,0)}A_{3.1}\oplus A_1,\, \langle e_1,e_3\rangle;\quad
\xrightarrow{I_{21}W(1,1,1,0)}A_{4.1},\, \langle e_4\rangle;\quad
\xrightarrow{W(0,1,1,0)}A_{4.2}^2,\, \langle e_1,e_4\rangle;
\\[0ex plus .2ex]
\null\quad
\xrightarrow{W(0,0,1,0)}A_{4.5}^{2,1,1},\, \langle e_1,e_2,e_4\rangle;\quad
\xrightarrow{W(1,0,1,0)}A_{4.8}^1,\, \langle e_2,e_4\rangle$.
\\[1.3ex]                
$A_{4.8}^b\colon\quad\xrightarrow{W(0,0,0,1)}A_{3.1}\oplus A_1,\, \langle e_1,e_2,e_3\rangle;$\quad
$\xrightarrow{b=0,\;I_{24}W(0,0,0,1)}A_{3.2}\oplus A_1,\, \langle e_1,e_2,e_3+e_4\rangle;$
\\[0ex plus .2ex]
\null\quad
$\xrightarrow{b=0,\;I_{13}(0)W(0,0,0,1)}A_{3.3}\oplus A_1,\, \langle e_1,e_2,e_4\rangle;$\quad
$\xrightarrow{b=-1,\;I_{13}(0)W(1,1,0,1)}A^{-1}_{3.4}\oplus A_1,\, \langle e_4\rangle;$\quad
\\[0ex plus .2ex]
\null\quad
$\xrightarrow{b\ne1,\;I_{25}W(1,1,1,0)}A_{4.1},\, \langle e_2-e_3\rangle;$\quad
$\xrightarrow{-1<b<0,\;W(0,0,1,0)}A_{4.5}^{1+b,1,b},\, \langle e_1,e_2,e_4\rangle$;
\\[0ex plus .2ex]
\null\quad
$\xrightarrow{0<b\leqslant1,\;\diag(1,1,1,\frac 1{1+b})W(0,0,1,0)}A_{4.5}^{1,\frac{1}{1+b},\frac{b}{1+b}},\,
\langle e_1,e_2,e_4\rangle$.
\\[1.3ex]                
$A_{4.9}^a\colon\quad\xrightarrow{W(0,0,0,1)}A_{3.1} \oplus A_1,\, \langle e_1,e_2,e_3\rangle;\quad
\xrightarrow{a=0,\;I_{14}W(1,1,0,0)}A_{3.5}^0\oplus A_1,\, \langle e_1,e_4\rangle;\quad
\\[0ex plus .2ex]
\null\quad
\xrightarrow{I_{26}W(1,1,1,0)}A_{4.1},\, \langle e_2\rangle;\quad
\xrightarrow{a\ne0,\;W(1,1,1,0)}A_{4.6}^{2a,a},\, \langle e_4\rangle$.
\\[1.3ex]                
$A_{4.10}\colon\quad
\xrightarrow{I_{13}(0)W(1,0,1,1)}A_{3.1} \oplus A_1,\, \langle e_2\rangle;\quad
\xrightarrow{U_1}A_{3.2} \oplus A_1,\quad
\xrightarrow{W(0,0,0,1)}A_{3.3} \oplus A_1,\, \langle e_1,e_2,e_3\rangle;\quad
\\[0ex plus .2ex]
\null\quad
\xrightarrow{I_{13}W(0,0,0,1)}A_{3.5}^b \oplus A_1,\, \langle e_1,e_2,be_3+e_4\rangle\quad
\xrightarrow{U_2}A_{4.1},\quad
\xrightarrow{I_{13}(0)W(1,0,1,0)}A_{4.8}^0,\, \langle e_2,e_3\rangle $.

\vspace{2ex}

\noindent
The constant parts of matrices of generalized In\"on\"u--Wigner contractions have the form
{\small
\begin{alignat*}{3}
&I_1=
\left(\begin{array}{cccc}
0 & 0 & 1 & 0\\
0 & 0 & 0 & 1\\
-1 & 0 & 1 & 0\\
0 & 1 & 0 & 1\\
\end{array}\right),
&&
I_2=
\left(\begin{array}{cccc}
1 & 2 & 0 & 0\\
0 & 0 & 1 & 0\\
0 & 1 & 0 & 0\\
0 & 0 & 1 & 1\\
\end{array}\right),
&&
I_3=
\left(\begin{array}{cccc}
0 & -1& 0 & 0\\
0 & 0 & 0 & 1\\
-1 & -1 & 0 & 0\\
0 & 0 & 1 & 1\\
\end{array}\right),
\\[2ex]&
I_4=
\left(\begin{array}{cccc}
0 & 0 & 0 & 1\\
-1 & -1 & 0 & 0\\
0 & 0 & 1 & 0\\
0 & -1 & 0 & 0\\
\end{array}\right),
&&
I_5=
\left(\begin{array}{cccc}
-1 & 0 & 1 & 0\\
0 & 0 & 0 & 1\\
0 & 1 & 0 & 0\\
0 & 0 & 1 & 0\\
\end{array}\!\!\right),\;
&&
I_6=
\left(\begin{array}{cccc}
-\frac1a & \frac1{a(a-1)} & \frac1{a(a-1)} & 0\\
0 & a & 1 & 0\\
0 & 0 & 0 & 1\\
0 & 0 & 1 & 0\\
\end{array}\right),
\\[2ex]&
I_7=
\left(\begin{array}{cccc}
1 & 0 & 0 & 0\\
0 & 0 & 1 & 0\\
0 & -1 & 0 & 0\\
0 & 0 & 0 & 1\\
\end{array}\right),
&&
I_8=
\left(\begin{array}{cccc}
0 & 1 & 0 & 0\\
2 & 0 & 0 & 0\\
0 & 0 & 1 & 0\\
0 & 0 & 0 & 1\\
\end{array}\right),
&&
I_9=
\left(\begin{array}{cccc}
1 & 0 & \frac{-1}{b^2+1} & 0\\
0 & 1 & \frac{b}{b^2+1} & 0\\
0 & 0 & 0 & 1\\
0 & 0 & 1 & 0\\
\end{array}\right),
\\[2ex]&
I_{10}=
\left(\begin{array}{cccc}
0 & 0 & \frac12 & 0\\
0 & 1 & 0 & 0\\
1 & 0 & \frac12 & 0\\
0 & 0 & 0 & 1\\
\end{array}\right),
&&
I_{11}=
\left(\begin{array}{cccc}
1 & 0 & 0 & 1\\
0 & 1 & 0 & 0\\
0 & 0 & 1 & 0\\
0 & 0 & 0 & 1\\
\end{array}\right),
&&
I_{12}=
\left(\!\!\!\!\!\!
\begin{array}{cccc}
\frac{1}{b-1} & \frac{(a-b)^{-1}}{(b-1)} & \frac{(a-b)^{-1}}{(a-1)(b-1)}\!\!\!\!\! & 0\\
0 & b-1 & 1 & 0\\
0& 0 & 1 & 0\\
0 & 0 & 0 & 1\\
\end{array}
\!\!\!\right),
\\[2ex]&
I_{13}(b)=
\left(\begin{array}{cccc}
1 & 0 & 0 & 0\\
0 & 1 & 0 & 0\\
0 & 0 & b & 1\\
0 & 0 & 1 & 0\\
\end{array}\right),
&&
I_{14}=
\left(\begin{array}{cccc}
0 & 0 & 0 & 1\\
1 & 0 & 0 & 0\\
0 & 1 & 0 & 0\\
0 & 0 & 1 & 0\\
\end{array}\right),
&&
I_{15}=
\left(\begin{array}{cccc}
1 & \frac{-1}{b-1} & \frac{-1}{(b-1)^2} & 0\\
0 & 1 & 0 & 0\\
0 & 0 & 1 & 0\\
0 & 0 & 0 & 1\\
\end{array}\right),
\\[2ex]&
I_{16}=
\left(\begin{array}{cccc}
1 & 0 & 0 & 0\\
0 & 0 & 0 & 1\\
0 & 0 & 1 & 0\\
0 & 1 & 0 & 0\\
\end{array}\right),
&&
I_{17}=
\left(\begin{array}{cccc}
1 & 1 & 1 & 0\\
0 & 1 & 0 & 0\\
0 & 0 & 1 & 0\\
0 & 0 & 0 & 1\\
\end{array}\right),
&&
I_{18}=
\left(\begin{array}{cccc}
1 & \frac{1+b}{a} & 0 & 0\\
0 & 1 & 0 & 0\\
0 & 0 & 0 & 1\\
0 & 0 & 1 & 0\\
\end{array}\right),
\\[2ex]&
I_{19}=
\left(\begin{array}{cccc}
0 & 1 & 0 & 0\\
0 & 0 & 0 & 1\\
0 & 0 & 1 & 0\\
1 & 0 & 0 & -\tfrac 12\\
\end{array}\right),
&&
I_{20}=
\left(\!\begin{array}{cccc}
-1 & 0 & 0 & 0\\
0 & 1 & 0 & 0\\
0 & 0 & 0 & -1\\
0 & 0 & 1 & 0\\
\end{array}\!\!\right)\!,\;
&&
I_{21}=
\left(\!\!\!
\begin{array}{cccc}
1 &\!\frac{(a-b)(a-1)\!^{-1}}{a-b+1}\!&\!\frac{(a-1)\!^{-1}}{a-b+1}\!& 0\\
0 & 0 & 1 & 0\\
0 & -1 & 0 & 0\\
0 & 0 & 0 & 1\\
\end{array}
\!\!\!\right)\!,\;
\\[2ex]&
I_{22}=
\left(\begin{array}{cccc}
-\frac 12 & 0 & \frac 12 & \frac 12\\
0 & 1 & 0 & 0\\
-\frac 12 & 0 & -\frac 12 & \frac 12\\
0 & 0 & 0 & 1\\
\end{array}\right),\ \  
&&
I_{23}=
\left(\begin{array}{cccc}
0 & 0 & 0 & 1\\
0 & 1 & 0 & 0\\
0 & 0 & -\tfrac 12 & 0\\
1 & 0 & 0 & 1\\
\end{array}\right),\ \  
&&
I_{24}=
\left(\begin{array}{cccc}
1 & 0 & 0 & 0\\
0 & 1 & 0 & 0\\
0 & 0 & 1 & 0\\
0 & 0 & 1 & 1\\
\end{array}\right),
\\[2ex]&
I_{25}=
\left(\begin{array}{cccc}
1 & 0 & 0 & 0\\
0 & 1 & 0 & -1\\
0 & 0 & 0 & 1\\
0 & 0 & \frac{1}{b-1} & 0\\
\end{array}\right),
&&
I_{26}=
\left(\begin{array}{cccc}
-1 & 0 & 0 & 0\\
0 & 0 & 0 & 1\\
0 & 1 & 0 & 0\\
0 & 0 & 1 & 0\\
\end{array}\right),
&&
I_{27}=
\left(\begin{array}{cccc}
1 & 0 & 0 & 0\\
0 & 0 & 1 & 0\\
0 & 1 & 0 & 0\\
0 & 0 & a & -1\\
\end{array}\right),
\\[2ex]&
I_{28}=
\left(\begin{array}{cccc}
-1 & 0 & 0 & 0\\
0 & 0 & 0 & 1\\
0 & 1 & 0 & -1\\
0 & 0 & 1 & 0
\end{array}\right),
&&
I_{29}=
\left(\begin{array}{cccc}
1 & 0 & -1 & 0\\
0 & 1 & 1 & 0\\
0 & 0 & 0 & 1\\
0 & 0 & 1 & 0
\end{array}\right),
&&
I_{30}=
\left(\begin{array}{cccc}
1 & 0 & -1 & 0\\
0 & 1 & 0 & 0\\
0 & 0 & 1 & 0\\
0 & 0 & 0 & 1
\end{array}\right).
\end{alignat*}}

\begin{figure}[ht]
\centerline{\includegraphics[scale=0.88]{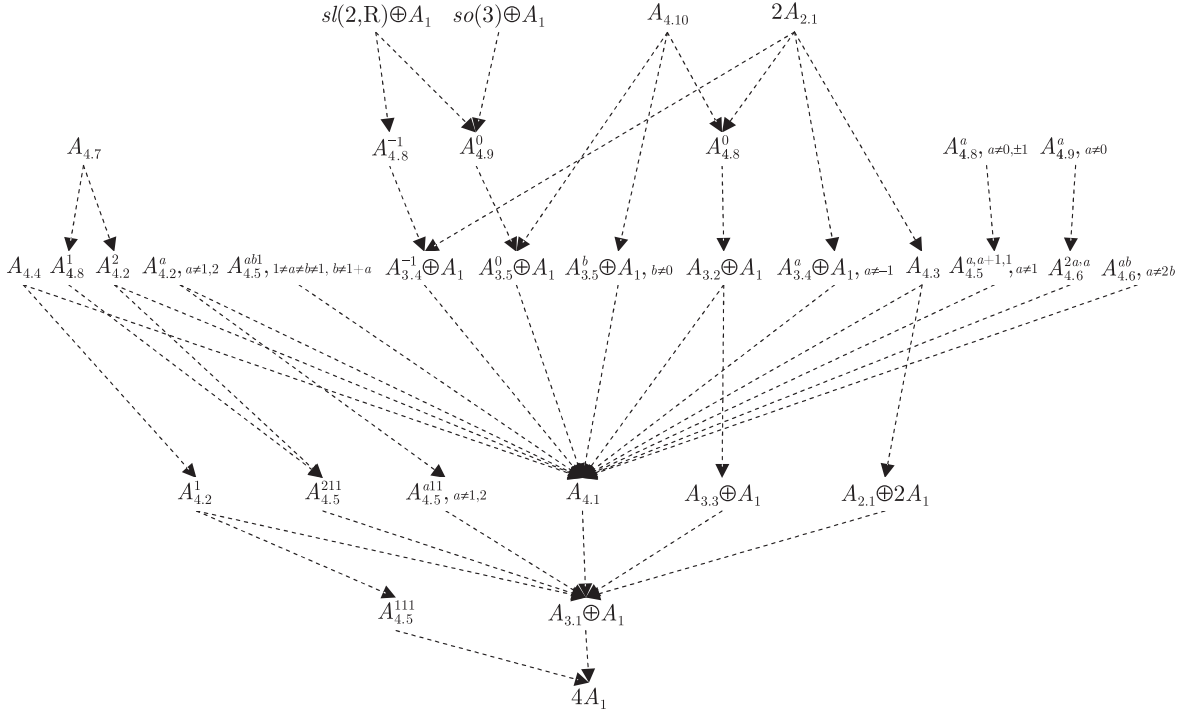}}
\caption{One-parametric contractions of real four-dimensional Lie algebras}\label{fig2}
\end{figure}

\begin{remark}\label{RemarkOnIWcontractionsOf4DimAlgeabras}
All the constructed contraction matrices include only nonnegative integer powers of~$\varepsilon$.
Therefore, they admit well-defined limit process under $\varepsilon\to+0$. 
Moreover, most contractions are equivalent to simple IW-contractions. 

All generalized IW-contractions of solvable real four-dimensional algebras (namely, 
$A_{3.2} \oplus A_1$, 
$A_{3.4}^a \oplus A_1$, 
$A_{3.5}^b \oplus A_1$, 
$A_{4.2}^b$, {\scriptsize $b{\ne}1$},
$A_{4.3}$, 
$A_{4.4}$, 
$A_{4.5}^{ab1}$, {\scriptsize $1{\ne}a{\ne}b{\ne}1$},
$A_{4.6}^{ab}$)
to $A_{4.1}$ are direct and, therefore, cannot be presented via composition of simple IW-contractions.
The same statement is true for the contractions of the unsolvable algebras  
($sl(2,{\mathbb R})\oplus A_1$ and $so(3)\oplus A_1$) to $A_{4.9}^{0}$.
Only three generalized IW-contractions ($so(3)\oplus A_1\to A_{3.1}\oplus A_1$, $so(3)\oplus A_1\to A_{4.1}$ and $A_{4.4}\to A_{4.5}^{111}$) 
are decomposed to sequences of simple IW-contractions.
The listed contractions exhaust a set of inequivalent `truly' generalized IW-contractions of the real four-dimensional algebras.

In contrast to three-dimensional Lie algebras,
there exist four contractions of four-dimensional Lie algebras,
which are inequivalent to generalized In\"on\"u--Wigner contractions,
namely
\begin{gather*}
A_{4.10}\xrightarrow{U_1}A_{3.2} \oplus A_1,\quad
2A_{2.1}\xrightarrow{U_2}A_{3.2}\oplus A_1,\\
A_{4.10}\xrightarrow{U_3}A_{4.1},\quad
2A_{2.1}\xrightarrow{U_4}A_{4.1},
\end{gather*}
They are provided by the `non-diagonalizable' matrices
\begin{gather*}
U_1=
\left(\begin{array}{cccc}
\varepsilon & 0 & 0 & 0\\
0 & 1 & 0 & 0\\
0 & 0 & 1 & \varepsilon\\
0 & 0 & \varepsilon & 0
\end{array}\right),
\quad
U_2=
\left(\begin{array}{cccc}
0 & -1 & 0 & 0\\
0 & 0 & 1 & \varepsilon\\
-\varepsilon & -1 & 0 & 0\\
0 & 0 & 1+\varepsilon & \varepsilon
\end{array}\right),
\\[2ex]
U_3=
\left(\begin{array}{cccc}
\varepsilon^2 & 0 & 0 & 0\\
0 & \varepsilon & 0 & -1\\
0 & 0 & \varepsilon & 0\\
0 & 0 & 0 & \varepsilon
\end{array}\right),\quad
U_4=
\left(\begin{array}{cccc}
-\varepsilon^2 & -\varepsilon & -1 & -1\\
0 & 0 &\varepsilon & 0\\
0 & -\varepsilon^2 & -\varepsilon & 0\\
0 & 0 & \varepsilon & \varepsilon
\end{array}\right).
\end{gather*}
The matrices $U_1$ and $U_2$ include only the zero and first powers of the contraction parameter. 
Therefore, the corresponding contractions are Saletan ones.
\end{remark}

\begin{remark}\label{RemarkOnReductionOfMaxContractionPower}
The maximal powers of contraction parameter, which are in components of contraction matrices, can be lowered 
if the restriction with the class of generalized IW-contractions in the case they exist will be neglected. 
For example, a generalized IW-contraction from the algebra $so(3)\oplus A_1$ to $A_{4.9}^0$ is generated by 
the matrix $I_{11}W(2,1,1,0)$ containing components with the second power of the contraction parameter. 
At the same time, it is known~\cite{Saletan1961} that there exist the Saletan contraction between these algebras 
which is provided by the matrix 
\[
\left(\begin{array}{cccc}
0 & \varepsilon & 0 & 0\\
0 & 0 & \varepsilon & 0\\
-\varepsilon & 0 & 0 & 1\\
-\varepsilon & 0 & 0 & 1-\varepsilon
\end{array}\right)
\]
obviously being of the first power with respect to~$\varepsilon$. 

Another example is given by the contraction $so(3)\oplus A_1\to A_{4.1}$. 
It is generated, as a generalized IW-contraction, with the matrix $I_5W(3,2,1,1)$ and has 
the essential contraction parameter power which is equal to 3 and is maximal among the generalized IW-contractions of the four-dimensional Lie algebras. 
All the other presented generalized IW-contractions contain at most the second power of the contraction parameter.
(The similar situation is in the three-dimensional case where the unique truly generalized IW-contraction is the contraction $so(3)\to A_{3.1}$ 
with the matrix $W(2,1,1)$ containing the second power of~$\varepsilon$.) 
The matrix $I_5W(3,2,1,1)$ can be replaced with the matrix
\[
\left(\begin{array}{cccc}
0 & 0 & \varepsilon & 0\\
0 & -\varepsilon^2 & 0 & 0\\
0 & 0 & 0 & \varepsilon\\
-\varepsilon^2 & 0 & -1 & 0
\end{array}\right)
\]
which has no `generalized IW-form' and contains at most the second power of the contraction parameter.
\end{remark}

\begin{remark}\label{RemarkOnRealCriteriaIn4Dim}
In each from the following pairs of Lie algebras
\begin{gather*}
(so(3)\oplus A_1,\, A_{4.8}^{-1}),\quad
(so(3)\oplus A_1,\,A_{3.4}^{-1}\oplus A_1),\quad
(A_{4.8}^{-1},\,A_{3.5}^{0}\oplus A_1),\quad
(A_{4.9}^{0},\,A_{3.4}^{-1}\oplus A_1),
\\
(A_{4.10},\,A_{4.3}),\quad
(A_{4.10},\,A_{2.1}\oplus 2A_1),\quad
(A_{4.10},\,A^a_{3.4}\oplus A_1),\quad
(2A_{2.1},\,A^b_{3.5}\oplus A_1),
\end{gather*}
the first algebra is contracted to the second one over the complex field.
See Section~\ref{SectionOnOne-parContractionsOfComplexLow-DimLieAlgebras} additionally. In particular,
\begin{gather*}
A_{4.10}\xrightarrow{I_{31}W(1,1,1,0)}A_{4.3},\quad
A_{4.10}\xrightarrow{I_{32}W(1,1,0,1)}A_{3.4}^a\oplus A_1,\quad
2A_{2.1}\xrightarrow{I_{33}W(0,0,0,1)}A^b_{3.5}\oplus A_1,
\end{gather*}
where
{\small\arraycolsep=1ex
\begin{gather*}
I_{31}=
\left(\begin{array}{cccc}
-i & i & 0 & -i\\[.5ex]
1 & 1 & 0 & -1\\[.5ex]
0 & 0 & \frac12 & \frac12\\[.5ex]
0 & 0 & \frac12 & -\frac i2\\[.5ex]
\end{array}\right),\;
I_{32}=
\left(\begin{array}{cccc}
i & -1 & 0 & 0\\[.5ex]
i & 1 & 0 & 0\\[.5ex]
0 & 0 & \frac{1+a}{2} & \frac{-i(1+a)}{2}\\[.5ex]
0 & 0 & -\frac 12 & -\frac i2\\[.5ex]
\end{array}\right),\;
I_{33}=
\left(\begin{array}{cccc}
-\frac i2 & -\frac 12 & 0 & 0\\[.5ex]
0 & 0 & b+i & 1\\[.5ex]
-\frac i2 & \frac 12 & 0 & 0\\[.5ex]
0 & 0 & b-i & 1
\end{array}\right).
\end{gather*}}
Therefore, almost all necessary criteria hold true since they do not discriminate between the real and complex fields.
At the same time, there are no real contractions in these pairs.
To prove it, we have to apply criteria specific for the real numbers,
e.g.,\ Criterion~\ref{criterion_Killing_formPNRanks} which is based 
on the law of inertia of quadratic forms over the real field.

For the first four pairs it is enough to consider only their Killing forms.
$\kappa_{so(3)\oplus A_1}=-2(u_1v_1+u_2v_2+u_3v_3)$, $\kappa_{A_{3.5}^{0}\oplus A_1}=-2u_3v_3$ and $\kappa_{A_{4.9}^0}=-2u_4v_4$
are nonpositively defined.
$\kappa_{A_{4.8}^{-1}}=2u_4v_4$ and $\kappa_{A_{3.4}^{-1}\oplus A_1}=2u_3v_3$
are nonnegatively defined.
All the above forms do not vanish identically.
Therefore, in each of these pairs an algebra has the nonpositively defined nonzero Killing form
and the other does the nonnegative defined nonzero one.
In view of necessary Criterion~\ref{criterion_Killing_formPNRanks}, there are no contractions in these pairs.

The criterion based on inertia of the Killing forms is powerless for the algebras from the other pairs.  
For them we consider the modified Killing forms with the specially chosen value $\alpha=-1/2$:
\begin{gather*}
\tilde\kappa_{A_{4.10}}^{-1/2}=2(1+2\alpha)u_3v_3-2u_4v_4\big|_{\alpha=-1/2}=-2u_4v_4,\\
\tilde\kappa_{A_{3.5}^b\oplus A_1}^{-1/2}=2((1+2\alpha)b^2-1)u_3v_3\big|_{\alpha=-1/2}=-2u_3v_3,\\
\tilde\kappa_{A_{4.3}}^{-1/2}=(1+\alpha)u_4v_4\big|_{\alpha=-1/2}=\frac12u_2v_2,\qquad
\tilde\kappa_{A_{2.1}\oplus 2A_1}^{-1/2}=(1+\alpha)u_2v_2\big|_{\alpha=-1/2}=\frac12u_4v_4,\\
\tilde\kappa_{A_{3.4}^a\oplus A_1}^{-1/2}=((1+a^2)+\alpha(1+a)^2)u_3v_3\big|_{\alpha=-1/2}=\frac12(1-a+a^2)u_3v_3,\\
\tilde\kappa_{2A_{2.1}}^{-1/2}=((1{+}\alpha)(u_2v_2{+}u_4v_4)+\alpha(u_2v_4{+}u_4v_2))\big|_{\alpha=-1/2}=\frac12((u_2v_2{+}u_4v_4)-(u_2v_4{+}u_4v_2)).
\end{gather*}
Two first forms are nonpositively defined and nonzero. The others are nonnegatively defined and also do not vanish identically.
In view of the second part of Criterion~\ref{criterion_Killing_formPNRanks}, there are no contractions in the pairs under consideration.
\end{remark}

\subsection{Levels and colevels of low-dimensional real Lie algebras}

Contractions assign the partial ordering relationship
on the variety $\mathcal L_n$ of $n$-dimensional Lie algebras.
Namely, we assume that $\mathfrak g\succ\mathfrak g_0$ if $g_0$ is a proper contraction of $g$.
The introduced strict order is well defined due to the transitivity property of contractions.
If improper contractions are allowed in the definition of ordering then the partial ordering becomes nonstrict.

The order $\succ$ generates separation of $\mathcal L_n$ to tuples of levels of different types.

\begin{definition}\label{def-level_of_alg}
The Lie algebra $\mathfrak g$ from $\mathcal L_n$ belongs to the \emph{zero level} of $\mathcal L_n$ if it has no proper contractions.
The other levels of~$\mathcal L_n$ are defined by induction.
The Lie algebra $\mathfrak g$ belongs to \emph{$k$-level} of $\mathcal L_n$
if it can be contracted to algebras from $(k-1)$-level and only to algebras from the previous levels.
\end{definition}

\begin{remark}
We have recently become aware due to~\cite{Lauret2003} that
the notion of level was introduced and investigated by Gorbatsevich~\cite{Gorbatsevich1991,Gorbatsevich1994,Gorbatsevich1998}.
He also proposed another notion of level based on interesting generalization of contractions to case of different dimensions
of initial and contracted algebras, which is reviewed in the Introduction.
\end{remark}

The zero level of $\mathcal L_n$ for any $n$ contains exactly one algebra, and it is the $n$-dimensional Abelian algebra
which is the unique minimal element in~$\mathcal L_n$.
The elements of the last level are maximal elements with respect to the ordering relationship
induced by contractions in $\mathcal L_n$ but do not generally exhaust the set of maximal elements of~$\mathcal L_n$.

Obtained exhaustive description of contractions of low-dimensional Lie algebras 
allows us to study completely levels of these algebras.

$\mathcal L_1$ consists of one element and has only one algebra level.
Analogously, $\mathcal L_2$ is formed by two elements and is separated
by contractions into exactly two levels.
The first level consists of the two-dimensional non-Abelian algebra $A_{2.1}$
and the zero level does the two-dimensional Abelian algebra $2A_1$.

The hierarchies of levels of real three- and four-dimensional Lie algebras are more complicated.
Actually, they are already represented in Figures~\ref{fig1} and~\ref{fig2}, where the level number grows upward.
It is the usage of the level ideology that makes the figures clear and elucidative. 
$\mathcal L_3$ and $\mathcal L_4$ have four and six levels, correspondingly.

\begin{remark}
Structure of Lie algebra is simplified under contraction.
The level number of an algebra can be assumed as a measure of complexity of its commutation structure,
i.e.,\ algebras with higher level numbers are more complicated than those with lower level numbers.
In particular, nilpotent algebras are in low levels. 
The simple algebras $sl(2,\R)$ and $so(3)$ having the most complicated structures among three-dimensional algebras 
form the highest 3-level of $\mathcal L_3$. 
The highest 6-level of $\mathcal L_4$ is formed by the unsolvable algebras $sl(2,\R)\oplus A_1$ and $so(3)\oplus A_1$ 
and the perfect (by Jacobson~\cite{Jacobson}) algebras $2A_{2.1}$ and $A_{4.10}$.
\end{remark}

\begin{remark}
There exists an inverse correlation of level numbers with dimensions of differentiation algebras
(or a direct correlation with dimensions of algebra orbits),
which is connected with necessary Criterion~\ref{criterion_dim_Der}.
As a rule, the algebras with the same dimension of differentiation algebras belong to the same level.
The dimensions of differentiation algebras of the algebras from $k$-level are not less and generally greater
than those of the algebras from $(k+1)$-level.

For the three-dimensional Lie algebras the correlation is complete.
Namely, the dimensions of differentiation algebras take the values of 9, 6, 4, 3 for the algebras from
0-, 1-, 2- and 3-level, correspondingly.

In $\mathcal L_4$ the correlation is partially broken.
Namely, for almost all algebras from $3$-level the dimensions of the differentiation algebras equal to six
and only the algebra $A^1_{4.8}$ which also belongs to this level has seven-dimensional differentiation algebra.
The same happens in $2$-level. Almost all algebras have
eight-dimensional differentiation algebras except the algebra $A_{4.1}$ with seven-dimensional differentiation algebra.
In other words, the four-dimensional Lie algebras with $\dim\Der=7$
are separated between the second and third levels, and the `simpler' nilpotent algebra~$A_{4.1}$ belongs to the lower level.
The algebras~$A_{4.5}^{111}$ ($\dim\Der=12$) and $A_{3.1}\oplus A_1$ ($\dim\Der=10$) form 1-level.
In all other cases the correlation is complete. 0-, 5- and 6-levels consist of the algebras having 16-, 5- and 4-dimensional
differentiation algebras correspondingly.
\end{remark}

Starting from the Lie algebras which are not proper contractions of any Lie algebras, 
we can introduce the related definition of colevel.

\begin{definition}\label{def-colevel_of_alg}
The Lie algebra $\mathfrak g$ from $\mathcal L_n$ belongs to the \emph{zero colevel} of $\mathcal L_n$
if it is not a proper contraction of any $n$-dimensional Lie algebra.
The other colevels of $\mathcal L_n$ are defined by induction.
The Lie algebra $\mathfrak g$ belongs to a \emph{colevel} of $\mathcal L_n$
if it is a proper contraction only of algebras from the previous colevels.
\end{definition}

0-colevel coincides with the set of maximal elements with respect to the order induced by contractions in~$\mathcal L_n$, 
i.e., it is formed by the algebras which are not proper contractions of the other algebras from~$\mathcal L_n$.
The last colevel of $\mathcal L_n$ for any $n$ contains exactly one algebra, and it is the $n$-dimensional Abelian algebra.

For the lowest dimensions structures of levels and colevels are analogous.
$\mathcal L_1$ has only  0-colevel which obviously coincides with 0-level.
$\mathcal L_2$ is separated by contractions into exactly two colevels.
The zero and first colevels coincide with the first and zero levels, correspondingly.

The hierarchies of colevels of real three- and four-dimensional Lie algebras differ from the hierarchies of levels
and adduced below.

\medskip

\noindent
Colevels of three-dimensional algebras:
\begin{enumerate}\itemsep=1ex
\renewcommand{\labelenumi}{{\rm \theenumi)}}
\setcounter{enumi}{-1}
\item\label{3level1}
$A_{2.1}\oplus A_1$, $A_{3.2}$, $A_{3.4}^a$,\,{\scriptsize$a\ne-1$}, $A_{3.5}^b$,\,{\scriptsize$b\ne0$}, $sl(2,\R)$, $so(3)$;
\item\label{3level2}
$A_{3.3}$, $A_{3.4}^{-1}$, $A_{3.5}^{0}$;
\item\label{3level3}
$A_{3.1}$;
\item\label{3level4}
$3A_{1}$.
\end{enumerate}

\medskip

\noindent
Colevels of four-dimensional algebras:
\begin{enumerate}\itemsep=0ex
\renewcommand{\labelenumi}{{\rm \theenumi)}}
\setcounter{enumi}{-1}
\item\label{level1}
$2A_{2.1}$, $sl(2,\R)\oplus A_1$, $so(3)\oplus A_1$,
$A_{4.2}^{b}$,\,{\scriptsize$b\ne1,2$}, $A_{4.4}$,
$A_{4.6}^{ab}$,\,{\scriptsize$a\ne2b$}, $A_{4.7}$,
$A_{4.8}^{b}$,\,{\scriptsize$b\ne0,\pm1$},
$A_{4.9}^{a}$,\,{\scriptsize$a\ne0$},
$A_{4.10}$,
$A_{4.5}^{abc}$,\,{\scriptsize$a\ne b\ne c\ne a,\, b\ne a+1$};

\item\label{level2}
$A_{3.4}^{a}\oplus A_1$,\,{\scriptsize$a\ne-1$}, $A_{3.5}^{b}\oplus A_1$,\,{\scriptsize$b\ne0$},
$A_{4.2}^1$, $A_{4.2}^2$, $A_{4.3}$, $A_{4.5}^{a,a+1,1}$,\,{\scriptsize$a\ne1$}, $A_{4.5}^{a11}$,\,{\scriptsize$a\ne1,2$}, $A_{4.6}^{2b,b}$,
$A_{4.8}^{-1}$, $A_{4.8}^{0}$, $A_{4.8}^1$, $A_{4.9}^{0}$;

\item\label{level3}
$A_{2.1}\oplus 2A_1$, $A_{3.2}\oplus A_1$, $A_{3.4}^{-1}\oplus A_1$,
$A_{3.5}^0\oplus A_1$, $A_{4.5}^{111}$, $A_{4.5}^{211}$;

\item\label{level4}
$A_{3.3}\oplus A_1$, $A_{4.1}$;

\item\label{level5}
$A_{3.1}\oplus A_1$;

\item\label{level6}
$4A_{1}$.
\end{enumerate}

\begin{remark}
The levels and colevels of $\mathcal L_n$ are related.
The numbers of levels and colevels of  $\mathcal L_n$ coincide and equal to the maximal length of chains of direct contractions.
If a fixed Lie algebra $\mathfrak g$ from $\mathcal L_n$ belongs to $k_1$-level and $k_2$-colevel then $k_1+k_2\leqslant n^2-n$.
\end{remark}

\begin{remark}\looseness=-1
Correlation of co-level numbers with dimensions of differentiation algebras (or orbit dimensions) is essentially weaker than for level numbers.
For each separated part of series of Lie algebras the orbit dimension of the whole part should be used here.
It equals the sum of orbit dimension of single algebras from this part and the number of essential parameters
parameterizing this part.
Even for the three-dimensional Lie algebras the correlation is broken in some cases.
For example, the orbit dimensions of the algebras $A_{3.1}$ and $A_{3.3}$ equal to three and they belong to different co-levels.
In spite of such weak correlation, the notion of co-level is useful in studying geometrical structure of~$\mathcal L_n$.
In particular, more regular (i.e.,\ having less constraints on parameters) parts of series of Lie algebras have less co-level numbers
than more singular ones.
\end{remark}

Analyzing obtained results for dimensions three and four, we induce a number of conjectures.
Testing and proof of them are out of the subject of this paper.
We have recently learned that some of them are already proved~\cite{Gorbatsevich1991,Lauret2003}.
We unite known statements in the following theorem.
Let $\mathfrak a_{E_{n-1}}$ be the almost Abelian algebra which contains an $(n-1)$-dimensional Abelian ideal 
and an element the adjoint action of which on the ideal is the identical operator~$E_{n-1}$.

\begin{theorem}
For any $n>2$ 1-level of $\mathcal L_n$ is formed by the algebras $A_{3.1}\oplus (n-3)A_1$ and~$\mathfrak a_{E_{n-1}}$.
\end{theorem}

\section{One-parametric contractions of complex low-dimensional\\ Lie algebras}\label{SectionOnOne-parContractionsOfComplexLow-DimLieAlgebras}

Some algebras which are inequivalent over the real field could be representatives of the
same class of algebras over the complex field.

Below we list pairs of real three- and four-dimensional Lie algebras
which are isomorphic or belong to the same series over the complex field.
For each of them we present the corresponding complex algebra (or series) together with the
appropriate basis transformation in the case it is non-identical.
The list is completed by the pairs of the direct sums
$(A_{3.4}^a\oplus A_1,A_{3.5}^b\oplus A_1)$ and $(sl(2,\R)\oplus A_1,so(3)\oplus A_1)$
isomorphisms of which become obvious.
Any complex indecomposable solvable algebra is denoted by $\mathfrak g_{n.k}$,
where $n$ is the dimension of the algebra and
$k$ is the number of the real algebra with the same form of canonical commutation relations.
\[\hspace*{-.5\arraycolsep}
\begin{array}{ll}
\lefteqn{\hspace*{-.5\arraycolsep}\begin{array}{l}\mathfrak g_{3.4}^\alpha,\\ \alpha\in \CC\end{array}}
\phantom{sl(2,{\CC})}
&\left\{\begin{array}{ll}
A_{3.5}^b,&
\tilde e_1 =e_1+ie_2,\;
\tilde e_2 =e_1-ie_2,\;
\tilde e_3 =\frac {1}{b+i}e_3,\;
\alpha=\frac{b-i}{b+i}\\[1ex]
A_{3.4}^a,&\alpha=a
\end{array}\right.
\end{array}
\]
\[\hspace*{-.5\arraycolsep}
\begin{array}{ll}
sl(2,{\CC})
&\left\{\begin{array}{ll}
so(3),&
\tilde e_1 =-ie_2+e_3,\;
\tilde e_2 =-ie_1,\;
\tilde e_3 =ie_2+e_3\\[1ex]
sl(2,{\mathbb R})&
\end{array}\right.
\end{array}
\]
\[\hspace*{-.5\arraycolsep}
\begin{array}{ll}
\lefteqn{\hspace*{-.5\arraycolsep}\begin{array}{l}\mathfrak g_{4.5}^{1,\alpha,\beta},\\ \alpha, \beta\in \CC\end{array}}
\phantom{sl(2,{\CC})}
&\left\{\begin{array}{ll}
A_{4.6}^{a,b}, &
\tilde e_1 =e_1,\;
\tilde e_2 =e_2-ie_3,\;
\tilde e_3 =e_2+ie_3,\;
\tilde e_4 =\frac {1}{a}e_4,\;
\alpha=\frac{b-i}{a},\;\beta=\frac{b+i}{a}\\[1ex]
A_{4.5}^{1,b,c},&\alpha=b,\;\beta=c
\end{array}\right.
\end{array}
\]
\[\hspace*{-.5\arraycolsep}
\begin{array}{ll}
\lefteqn{\hspace*{-.5\arraycolsep}\begin{array}{l}\mathfrak g_{4.8}^{\beta},\\ \beta\in \CC\end{array}}
\phantom{sl(2,{\CC})}
&\left\{\begin{array}{ll}
A_{4.9}^{a}, &
\tilde e_1 =-e_1,\;
\tilde e_2 =e_2+ie_3,\;
\tilde e_3 =-\frac{i}{2}e_2-\frac{1}{2}e_3,\;
\tilde e_4 =\frac {1}{a+i}e_4,\;
\beta=\frac{a-i}{a+i}\\[1ex]
A_{4.8}^{b},&\beta=b
\end{array}\right.
\end{array}
\]
\[\hspace*{-.5\arraycolsep}
\begin{array}{ll}
\lefteqn{2 \mathfrak g_{2.1}}
\phantom{sl(2,{\CC})}
&\left\{\begin{array}{ll}
A_{4.10}, &
\tilde e_1 =ie_1-e_2,\;
\tilde e_2 =\frac12e_3-\frac i2e_4,\;
\tilde e_3 = ie_1+e_2,\;
\tilde e_4 =\frac12e_3+\frac i2e_4\\[1ex]
2A_{2.1}&
\end{array}\right.
\end{array}
\]

\begin{figure}[ht]
\centerline{\includegraphics[scale=0.88]{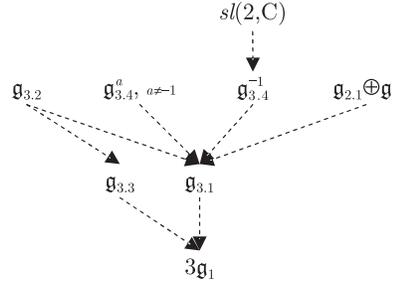}}
\caption{One-parametric contractions of three-dimensional complex Lie algebras}\label{fig3}
\end{figure}

\begin{figure}[ht]
\centerline{\includegraphics[scale=0.88]{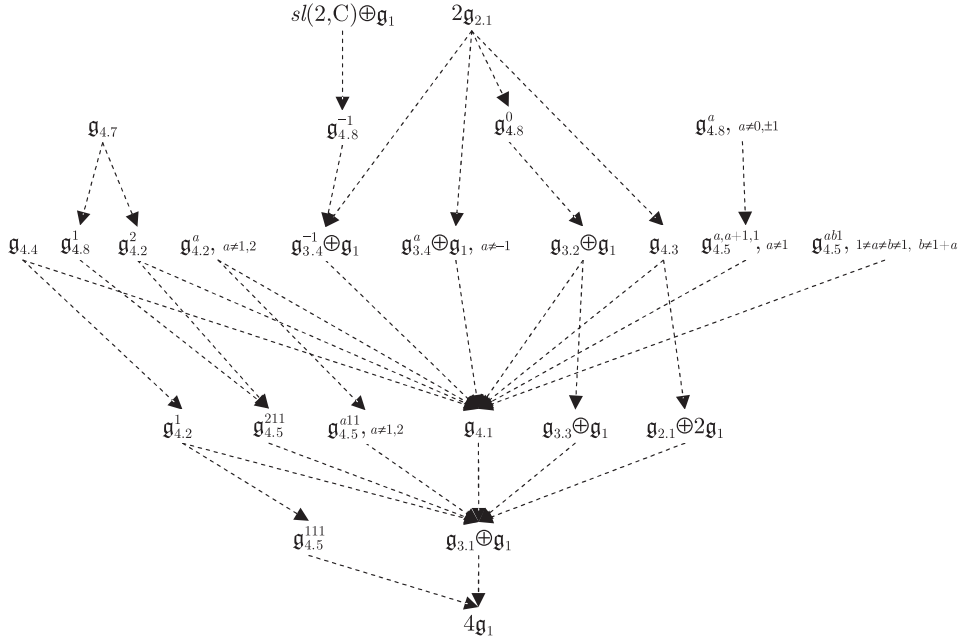}}
\caption{One-parametric contractions of four-dimensional complex Lie algebras}\label{fig4}
\end{figure}

Knowledge of the correspondences between real and complex Lie algebras
allows us to describe all continuous contractions of the complex low-dimensional Lie algebras.
The corresponding lists are produced from the analogous lists for the real low-dimensional Lie algebras
by accurate elimination of algebras which are equivalent to other forms over the complex field.
The contraction matrices are preserved.
The contractions of three- and four-dimensional complex algebras are visualized with Figure~\ref{fig3} and Figure~\ref{fig4}.
The one- and two-dimensional cases are trivial and are not considered.

\begin{theorem}
Any continuous contraction of complex three-dimensional Lie algebras is equivalent to
a simple In\"on\"u--Wigner contraction.
\end{theorem}

In four-dimensional case only the contractions
$2\mathfrak g_{2.1}\to \mathfrak g_{3.2}\oplus \mathfrak g_1$ and $2\mathfrak g_{2.1}\to \mathfrak g_{4.1}$
are not presented as generalized In\"on\"u--Wigner contractions.
All the constructed contraction matrices include only nonnegative integer powers of~$\varepsilon$. Therefore,
they admit well-defined limit process under $\varepsilon\to+0$.

A list of continuous contractions of the complex three-dimensional Lie algebras was adduced, e.g.,\  
in~\cite{Agaoka1999,Burde&Steinhoff1999,Steinhoff1997} in terms of orbit closures. 
It obviously coincides with that presented in Figure~\ref{fig3}.
In~\cite{Burde&Steinhoff1999,Steinhoff1997} and later in~\cite{Agaoka2002} 
contractions of the four-dimensional complex Lie algebras are also investigated.
Comparing these results with ours, at first
we determine correspondence between the used lists of algebras.
To avoid confusions, we add hats over the symbol $\mathfrak g$ denoting algebras from~\cite{Burde&Steinhoff1999}. 
Note also that the list used in~\cite{Burde&Steinhoff1999} is essentially based on classification obtained in~\cite{Patera&Zassenhaus1990b}.
\begin{gather*}\arraycolsep=0ex
\begin{array}[t]{ll}
4\mathfrak g_{1} \sim \CC^4;&
\mathfrak g_{4.1}\sim \mathfrak n_4;
\\[.4ex]
\mathfrak g_{2.1}\oplus 2\mathfrak g_1\sim \mathfrak r_2\oplus\CC^2;&
\mathfrak g_{4.2}^1\sim \hat{\mathfrak g}_5;\
\mathfrak g_{4.2}^{-2}\sim \hat{\mathfrak g}_3\Bigl(\frac{27}4\Bigr);\
\mathfrak g_{4.2}^{b\ne1,-2}\sim \hat{\mathfrak g}_2\Bigl(\frac b{(b+2)^3},\frac{2b+1}{(b+2)^2}\Bigr);
\\[1.1ex]
2\mathfrak g_{2.1} \sim \mathfrak r_2\oplus \mathfrak r_2;&
\mathfrak g_{4.3}\sim \hat{\mathfrak g}_2(0,0);
\\[1.1ex]
\mathfrak g_{3.1}\oplus \mathfrak g_1\sim \mathfrak n_3\oplus\CC;&
\mathfrak g_{4.4}\sim \hat{\mathfrak g}_2(\frac 1{27},\frac 1 3);
\\[1.1ex]
\mathfrak g_{3.2}\oplus \mathfrak g_1\sim \mathfrak r_3\oplus\CC;&
\mathfrak g_{4.5}^{a11}\sim \hat{\mathfrak g}_1(a);\
\\[1.1ex]
\mathfrak g_{3.3}\oplus \mathfrak g_1\sim \mathfrak r_{3,1}\oplus\CC;&
\mathfrak g_{4.5}^{ab1}\sim \hat{\mathfrak g}_2(\alpha,\beta),\hat{\mathfrak g}_3(\gamma),\hat{\mathfrak g}_4,\ 
\mbox{\scriptsize $1\ne a \ne b \ne 1,\ ab\ne0$};
\\[1.1ex]
\mathfrak g_{3.4}^a\oplus \mathfrak g_1\sim \mathfrak r_{3,a}\oplus\CC,\ \mbox{\scriptsize $a\ne1$};&
\mathfrak g_{4.7}\sim \hat{\mathfrak g}_8(\frac14);
\\[.8ex]
sl(2,\CC)\oplus \mathfrak g_1\sim sl_2(\CC)\oplus \CC;\qquad &
\mathfrak g_{4.8}^{1}\sim \hat{\mathfrak g}_6;\
\mathfrak g_{4.8}^{-1}\sim \hat{\mathfrak g}_7;\
\mathfrak g_{4.8}^{b\ne\pm1}\sim \hat{\mathfrak g}_8\Bigl(\frac b{(b+1)^2}\Bigr).
\end{array}
\end{gather*}
Let us give more details on the algebra series $\{\mathfrak g_{4.5}^{abc}, abc\ne0\}$. 
The parameter tuples $(a,b,c)$ and $(a',b',c')$ are associated with the same algebra if they are proportional up to a permutation. 
The algebra $\mathfrak g_{4.5}^{ab1}$, {\scriptsize $1\ne a \ne b \ne 1,\ ab\ne0$}, corresponds to 
\begin{gather*}
\hat{\mathfrak g}_2(\alpha,\beta), \quad\mbox{where}\quad \alpha=\frac{ab}{(a+b+1)^3},\ \beta=\frac{ab+a+b}{(a+b+1)^2}, \quad\mbox{if}\quad a+b+1\ne0;
\\[0.5ex] 
\hat{\mathfrak g}_3(\gamma), \quad\mbox{where}\quad \gamma=-\frac{(ab-1)^3}{a^2b^2}, \quad\mbox{if}\quad a+b+1=0,\ ab\ne1;
\\[1ex] 
\hat{\mathfrak g}_4 \quad\mbox{if}\quad a+b+1=0,\ ab=1.
\end{gather*}

In~\cite{Agaoka2002} a special classification of four-dimensional complex Lie algebras was used 
under study of contractions. 
Correspondences between ours and Agaoka's lists of inequivalent algebras are the following:
\begin{gather*}
L_0\sim 4\mathfrak g_1,\quad 
L_1\sim \mathfrak g_{3.1}\oplus\mathfrak g_1,\quad 
L_2\sim \mathfrak g_{4.1},\quad 
L_3\sim \mathfrak g_{4.5}^{111},\quad 
L_5\sim \mathfrak g_{4.8}^{1},\quad 
L_6\sim \mathfrak sl(2,\CC)\oplus\mathfrak g_1,
\\[1ex]
L_4(a)\sim \mathfrak g_{4.5}^{a11},\ \mbox{\scriptsize $a\ne0,1,$}\quad 
L_4(0)\sim L_7(0,1)\sim \mathfrak g_{3.3}\oplus\mathfrak g_1,\quad 
L_4(1)\sim \mathfrak g_{4.2}^1,\quad 
L_4(\infty)\sim \mathfrak g_{2.1}\oplus2\mathfrak g_1,
\\[1ex]
L_7(a,b)\sim \mathfrak g_{4.5}^{ab1},\ \mbox{\scriptsize $1\ne a\ne b\ne1,\ ab\ne0,$}\quad 
L_7(a,1)\sim \mathfrak g_{4.2}^a,\ \mbox{\scriptsize $a\ne0,1,$}\quad 
L_7(1,1)\sim \mathfrak g_{4.4}, 
\\[1ex]
L_7(a,0)\sim \mathfrak g_{3.4}^a\oplus\mathfrak g_1,\ \mbox{\scriptsize $a\ne0,1,$}\quad 
L_7(1,0)\sim \mathfrak g_{3.2}\oplus\mathfrak g_1,\quad 
L_7(0,0)\sim \mathfrak g_{4.3}, 
\\[1ex]
L_8(a)\sim \mathfrak g_{4.8}^a,\ \mbox{\scriptsize $a\ne1,$}\quad 
L_8(1)\sim \mathfrak g_{4.7},\quad
L_9\sim 2\mathfrak g_{2.1}.
\end{gather*}

It is easy to see from the performed comparisons that our list of contractions coincides with the ones adduced 
in~\cite{Steinhoff1997,Burde&Steinhoff1999,Agaoka2002}.

\section{Multi-parametric, decomposable and repeated contractions}\label{SectionOnMulti-parametricDecomposableAndRepeatedContractions}

One-parametric contractions exhaust the set of continuous contractions.
At the same time other types of contractions are also useful, in particular,
for finding one-parametric contractions.
Consider a class of continuous contractions which generalizes one-parametric ones, namely,
the class of multi-parametric contractions, following the notations and spirit of
Section~\ref{SectionOnDefsOfContractionsAndTheirEquvalence}.

Let $U\colon (0,1]^m\to GL(V)$, i.e.,\ $U_{\bar{\varepsilon}}=U(\varepsilon_1, \dots, \varepsilon_m)$
is a nonsingular linear operator on $V$ for any $\bar{\varepsilon}\in (0,1]^m$.
Here $m\in\N$, $\bar{\varepsilon}$ is the tuple of the parameters $\varepsilon_1$, \dots, $\varepsilon_m$.
We define a parameterized family of new Lie brackets on~$V$ via the old one by the following way:
\[
\forall\; \bar{\varepsilon} \in (0,1]^m,\ \forall \; x, y\in V\colon\quad
[x,y]_{\bar{\varepsilon}}=U_{\bar{\varepsilon}}^{-1}[U_{\bar{\varepsilon}} x,U_{\bar{\varepsilon}} y].
\]
For any $\bar{\varepsilon} \in (0,1]^m$ the Lie algebra
${\mathfrak g}_{\bar{\varepsilon}}=(V,[\cdot,\cdot]_{\bar{\varepsilon}})$
is isomorphic to ${\mathfrak g}$.

\begin{definition}\label{DefOfContractions_multi-param}
If the limit
$
\lim\limits_{\bar{\varepsilon} \to +\bar{0}}[x,y]_{\bar{\varepsilon}}=
\lim\limits_{\bar{\varepsilon} \to +\bar{0}}U_{\bar{\varepsilon}}^{-1}
[U_{\bar{\varepsilon}} x,U_{\bar{\varepsilon}} y]=:[x,y]_0
$
exists for any $x, y\in V$ then the Lie bracket~$[\cdot,\cdot]_0$ is well-defined.
The Lie algebra ${\mathfrak g}_0=(V,[\cdot,\cdot]_0)$ is called an
\emph{multi-parametric (continuous) contraction} of the Lie algebra~${\mathfrak g}$.
\end{definition}

The notation $\bar{\varepsilon} \to +\bar{0}$ means $\varepsilon_l \to +0$, $l=1,\dots,m$.

If a basis of~$V$ is fixed, the operator $U_{\bar{\varepsilon}}$ is defined by the corresponding matrix
(we will use the notation $U_{\bar{\varepsilon}}$ for the matrix also) and
Definition~\ref{DefOfContractions_multi-param} can be reformulated in terms of structure constants.

{\addtocounter{definition}{-1}\renewcommand{\thedefinition}{\arabic{definition}$'$}
\begin{definition}\label{Def-m-Contractions}
If the limit
$
\lim\limits_{\bar{\varepsilon} \to +\bar{0}}(U_{\bar{\varepsilon}})_{i'}^i
(U_{\bar{\varepsilon}})_{j'}^j(U_{\bar{\varepsilon}}^{-1})_k^{k'}c^{k}_{ij}=:\tilde c^{k'}_{i'\!j'}
$
exists for all values of $i'$, $j'$ and $k'$ then
$\tilde c^{k'}_{i'\!j'}$ are components of the well-defined structure constant tensor of a Lie algebra~${\mathfrak g}_0$.
In this case the Lie algebra~${\mathfrak g}_0$ is called
a \emph{$m$-parametric (continuous) contraction} of the Lie algebra~${\mathfrak g}$.
The parameters $\varepsilon_1, \dots, \varepsilon_m$ and the matrix-function $U_{\bar{\varepsilon}}$
are called \emph{contraction parameters} and a \emph{contraction matrix} correspondingly.
\end{definition}}

\begin{remark}
Any multi-parametric contraction generates a set of strongly equivalent
(in the sense of Definition~\ref{DefStrongEquivOfContractions}) one-parametric contractions
via replacement $\varepsilon_i=f_i(\varepsilon)$ of the parameters $\bar\varepsilon$
by functions of one parameter $\varepsilon$.
For the replacement to be correct, the functions $f_i\colon(0,1]\to(0,1]$
should be monotonic, continuous and
$f_i(\varepsilon)\to+0$, $\varepsilon\to+0$.
\end{remark}


It is obvious that the notion of orbit closure~\cite{Burde&Steinhoff1999} is \emph{transitive}.
The same statement is true for one-parametric contractions.
Due to the transitivity, we can easily construct new continuous contractions
from the ones adduced in Section~\ref{SectionOnOne-parContractionsOfRealLow-DimLieAlgebras}.
But simple multiplication of the matrices of successive contractions does not give the matrix of the resulting contraction.
Below we introduce necessary notions concerning successive contractions and discuss significant examples.

Let the algebra~${\mathfrak g}_2$ be contracted with the matrix $U_1(\varepsilon'_1,\dots,\varepsilon'_{m_1})$ to the algebra~${\mathfrak g}_1$
which is further contracted with the matrix $U_2(\varepsilon''_1,\dots,\varepsilon''_{m_2})$ to the algebra~${\mathfrak g}_0$.
If the matrix \[U_1(\varepsilon'_1,\dots,\varepsilon'_{m_1})U_2(\varepsilon''_1,\dots,\varepsilon''_{m_2})\] provides
an $(m_1+m_2)$-parametric contraction from ${\mathfrak g}_2$ to~${\mathfrak g}_0$
then this contraction is called \emph{composition} of two initial contractions.

\begin{definition}\label{DefDecomposibleContractions}
A multi-parametric contraction is called \emph{decomposable} if it can be presented as a composition of
two proper multi-parametric contractions.
\end{definition}

More precisely, an $m$-parametric contraction from the algebra ${\mathfrak g}$ to the algebra~${\mathfrak g}_0$ is decomposable iff
there exists an algebra~${\mathfrak g}_1$ (nonisomorphic to ${\mathfrak g}$ and~${\mathfrak g}_0$) such that
the contraction from ${\mathfrak g}$ to~${\mathfrak g}_0$ can be presented as a composition of
$m_1$-parametric contraction from ${\mathfrak g}$ to~${\mathfrak g}_1$ and
$m_2$-parametric contraction from ${\mathfrak g}_1$ to~${\mathfrak g}_0$, where $m_1+m_2=m$.

\begin{definition}\label{DefCompletelyDecomposibleContractions}
An $m$-parametric contraction is called \emph{completely decomposable} if it can be presented as a composition of
$m$ one-parametric contractions.
\end{definition}

Any two-parametric decomposable contraction is obviously completely decomposable.

\begin{definition}\label{DefRepeatedContractions}
If there exist two one-parametric contractions from ${\mathfrak g}$ to~${\mathfrak g}_1$ and
from ${\mathfrak g}_1$ to~${\mathfrak g}_0$
then ${\mathfrak g}_0$ is called a \emph{repeated contraction} of ${\mathfrak g}$.
\end{definition}

Analogously, any $l$-repeated contraction is a result of $l$ one-parametric successive contractions.
The notion of repeated multi-parametric contractions can also be introduced in a similar way.

The above definition can be justified in the following way.
Let $U_1(\varepsilon_1)$ and $U_2(\varepsilon_2)$ be the contraction matrices
of one-parametric contractions from ${\mathfrak g}$ to~${\mathfrak g}_1$ and
from ${\mathfrak g}_1$ to~${\mathfrak g}_0$ correspondingly and
$U_{\bar\varepsilon}=U_1(\varepsilon_1)U_2(\varepsilon_2)$, where $\bar\varepsilon=(\varepsilon_1,\varepsilon_2)$.
Then there exists the \emph{repeated} limit
\begin{gather*}
\lim\limits_{\varepsilon_2\to+0}\left(
\lim\limits_{\varepsilon_1\to+0}
(U_{\bar\varepsilon})_{i'}^i (U_{\bar\varepsilon})_{j'}^j(U_{\bar\varepsilon}^{-1})_k^{k'}c^k_{ij}
\right)=:\tilde c^{k'}_{i'\!j'}
\end{gather*}
for all values of $i'$, $j'$ and $k'$,
i.e.,\ $\tilde c^{k'}_{i'\!j'}$ are components of the well-defined structure constant tensor of
the Lie algebra~${\mathfrak g}_0$.

\begin{remark}\label{rem_repeated_to_decomposable}
If the repeated limit can be replaced by the
well-defined simultaneous limit
\begin{gather*}
\lim\limits_{\bar\varepsilon \to +\bar0}
(U_{\bar\varepsilon})_{i'}^i (U_{\bar\varepsilon})_{j'}^j(U_{\bar\varepsilon}^{-1})_k^{k'}c^k_{ij}=
\lim\limits_{\varepsilon_2\to+0}\left(
\lim\limits_{\varepsilon_1\to+0}
(U_{\bar\varepsilon})_{i'}^i (U_{\bar\varepsilon})_{j'}^j(U_{\bar\varepsilon}^{-1})_k^{k'}c^k_{ij}
\right)
\end{gather*}
with $\bar\varepsilon=(\varepsilon_1,\varepsilon_2)$, then the repeated contraction turns into
completely decomposable multi-parametric contraction.
\end{remark}

\begin{example}\label{ExampleSo3+A1ToA41}
Consider the  algebra pair $(so(3)\oplus A_1,A_{4.1})$. 
In view of the necessary contraction criteria, the algebra $so(3)\oplus A_1$ may be contracted to $A_{4.1}$.
It is difficult to construct a contraction matrix for this pair by the direct algorithm. 
Instead of this, we study repeated contractions from $so(3)\oplus A_1$ to $A_{4.1}$ in detail. 
It is easy to see due to the level structure of the real four-dimensional Lie algebras (Figure~\ref{fig2}) that
there are two different ways for the repeated contractions, 
which are associated with intermediate algebras $A_{3.5}^0\oplus A_1$ and $A_{4.9}^0$.

The generalized In\"on\"u--Wigner contractions
from $so(3)\oplus A_1$ to $A_{3.5}^0\oplus A_1$ and 
from $A_{3.5}^0\oplus A_1$ to $A_{4.1}$
are provided by the matrices
$U_1=\mathop{\rm diag}(\varepsilon_1, \varepsilon_1, 1, 1)$ and 
$U_2=I_9(b)\mathop{\rm diag}(\varepsilon_2^2, \varepsilon_2, 1, \varepsilon_2)$ correspondingly.
Their product
\begin{gather*}
U_{\bar\varepsilon}=U_1(\varepsilon_1)U_2(\varepsilon_2)=
\mathop{\rm diag}(\varepsilon_1, \varepsilon_1, 1, 1)
I_9(0)\mathop{\rm diag} (\varepsilon_2^2, \varepsilon_2, 1, \varepsilon_2),
\end{gather*}
gives a matrix-valued function of two variables $\bar\varepsilon=(\varepsilon_1,\varepsilon_2)$.
Let us investigate how the contraction generated by the matrix $U_{\bar\varepsilon}$
acts on the algebra $so(3)\oplus A_1$.

The matrix $U_{\bar\varepsilon}$ and its inverse matrix $U_{\bar\varepsilon}^{-1}$ have the explicit forms
\begin{gather*}
U_{\bar\varepsilon}=
\left(\begin{array}{cccc}
\varepsilon_1\varepsilon_2^{2} & 0 & -\varepsilon_1& 0\\
0 & \varepsilon_1\varepsilon_2 & 0 & 0\\
0 & 0 & 0 & \varepsilon_2\\
0 & 0 & 1 & 0\\
\end{array}\right),\qquad
U_{\bar\varepsilon}^{-1}=
\left(\begin{array}{cccc}
\varepsilon_1\varepsilon_2^{-2} & 0 & 0& \varepsilon_2^{-2}\\
0 & \varepsilon_1^{-1}\varepsilon_2^{-1} & 0 & 0\\
0 & 0 & 0 & 1\\
0 & 0 & \varepsilon_2^{-1} & 0\\
\end{array}\right).
\end{gather*}
We calculate all the different (up to antisymmetry) transformed commutators of the canonical basis elements of the algebra $so(3)\oplus A_1$
using the formula $[e_i, e_j]_{\bar\varepsilon}=U^{-1}_{\bar\varepsilon}[U_{\bar\varepsilon}e_i, U_{\bar\varepsilon}e_j]$
and the canonical commutation relations $[e_1,e_2]=e_3$, $[e_2,e_3]=e_1$, $[e_3,e_1]=e_2$:
\begin{gather*}\samepage
[e_1,e_2]_{\bar\varepsilon}=\varepsilon_1^2\varepsilon_2^2e_4,
\quad
[e_1,e_4]_{\bar\varepsilon}=-\varepsilon_2^2e_2,
\quad
[e_2,e_3]_{\bar\varepsilon}=\varepsilon_1^2e_4,
\\ 
[e_1,e_3]_{\bar\varepsilon}=0,\quad
[e_2,e_4]_{\bar\varepsilon}=e_1,\quad
[e_3,e_4]_{\bar\varepsilon}=e_2.
\end{gather*}

Under the repeated limit
$\varepsilon_1\to+0$ and then $\varepsilon_2\to+0$
these commutation relations go to the canonical ones of the Lie algebra $A_{4.1}$,
i.e.,\ composition of two successive one-parametric contractions results
in the repeated contraction from $so(3)\oplus A_1$ to $A_{4.1}$.
Moreover, the simultaneous limit $\bar\varepsilon=(\varepsilon_1,\varepsilon_2)\to\bar 0$
exists for the transformed structure constants. It implies in view of Remark~\ref{rem_repeated_to_decomposable} that
the matrix $U_{\bar\varepsilon}$ also gives the completely decomposable two-parametric contraction
from $so(3)\oplus A_1$ to $A_{4.1}$.
After putting $\varepsilon_1=\varepsilon_2=:\varepsilon$,
we construct a well-defined one-parametric contraction between the algebras under consideration.
Unfortunately, it is not a generalized IW-contraction.

Consider the way via the algebra~$A_{4.9}^0$.
The generalized In\"on\"u--Wigner contractions
from $so(3)\oplus A_1$ to $A_{4.9}^0$ and 
from $A_{4.9}^0$ to $A_{4.1}$ 
are provided by the matrices
$U_1=I_{11}\mathop{\rm diag}(\varepsilon_1^2, \varepsilon_1, \varepsilon_1, 1)$ and
$U_2=I_{26}\mathop{\rm diag}(\varepsilon_2, \varepsilon_2, \varepsilon_2, 1)$, correspondingly.
The repeated limit $\varepsilon_1\to+0$ and then $\varepsilon_2\to+0$ in the commutation relations 
\begin{gather*}
[e_1,e_2]_{\bar\varepsilon}=\varepsilon_1^2\varepsilon_2^2e_4,
\quad
[e_1,e_4]_{\bar\varepsilon}=-\varepsilon_1^2e_2,
\quad
[e_2,e_3]_{\bar\varepsilon}=\varepsilon_2^2e_4,
\\
[e_1,e_3]_{\bar\varepsilon}=0,\quad
[e_2,e_4]_{\bar\varepsilon}=e_1,\quad
[e_3,e_4]_{\bar\varepsilon}=e_2
\end{gather*}
obtained by transformation of the canonical relations of $so(3)\oplus A_1$ with the matrix
\begin{gather*}
U_{\bar\varepsilon}=U_1(\varepsilon_1)U_2(\varepsilon_2)=
\left(\begin{array}{cccc}
-\varepsilon_1^{2}\varepsilon_2 & 0 & \varepsilon_2& 0\\
0 & 0 & 0 & \varepsilon_1\\
0 & \varepsilon_1\varepsilon_2 & 0 & 0\\
0 & 0 & \varepsilon_2 & 0\\
\end{array}\right)
\end{gather*}
results in the canonical commutation relations of $A_{4.1}$.
The simultaneous limit $\bar\varepsilon=(\varepsilon_1,\varepsilon_2)\to\bar 0$ also exists. 
After putting $\varepsilon_1=\varepsilon_2=:\varepsilon$,
we construct the matrix 
\begin{gather*}
U_{\varepsilon,\varepsilon}=
\left(\begin{array}{cccc}
-\varepsilon^3 & 0 & \varepsilon& 0\\
0 & 0 & 0 & \varepsilon\\
0 & \varepsilon^2 & 0 & 0\\
0 & 0 & \varepsilon & 0\\
\end{array}\right)=I_5\diag(\varepsilon^3,\varepsilon^2,\varepsilon,\varepsilon)
\end{gather*} 
of a well-defined one-parametric generalized IW-contraction $so(3)\oplus A_1\to A_{4.1}$. 
Note that possibility of IW-decomposition of the matrix $U_{\varepsilon,\varepsilon}$ into the product 
of a constant matrix and a diagonal matrix with powers of~$\varepsilon$ on the diagonal is obvious since 
the elements of any column of $U_{\varepsilon,\varepsilon}$ contain the same power of~$\varepsilon$.
\end{example}

In fact, a regular procedure for construction of generalized IW-contractions via repeated contractions 
is described in Example~\ref{ExampleSo3+A1ToA41}.

The repeated contractions of Example~\ref{ExampleSo3+A1ToA41} lead to well-defined decomposable multi-parametric contraction. 
This fact is not true in the general case that is illustrated by the next example.

\begin{example}\label{ex_2a21-a41}
We failed to construct a generalized In\"on\"u--Wigner contraction between the algebras $2A_{2.1}$ and $A_{4.1}$.
At the same time, there exist the one-parametric generalized In\"on\"u--Wigner contractions
from $2A_{2.1}$ to $A_{4.3}$ and from $A_{4.3}$ to $A_{4.1}$ with the contraction matrices
$U_1=I_{28}\mathop{\rm diag}(0, \varepsilon_1, \varepsilon_1, 0)$ and
$U_2=I_{17}\mathop{\rm diag}(\varepsilon_2^2, \varepsilon_2, 1, \varepsilon_2)$, correspondingly.
Product of these matrices,
\begin{gather*}
U_{\bar\varepsilon}=U_1(\varepsilon_1)U_2(\varepsilon_2)=
I_{28}\mathop{\rm diag}(1, \varepsilon_1, \varepsilon_1, 1)
I_{17}\mathop{\rm diag}(\varepsilon_2^2, \varepsilon_2, 1, \varepsilon_2),
\end{gather*}
defines a matrix-valued function of two variables $\bar\varepsilon=(\varepsilon_1,\varepsilon_2)$.
The matrix $U_{\bar\varepsilon}$ and its inverse matrix $U_{\bar\varepsilon}^{-1}$ have the explicit forms
\begin{gather*}
U_{\bar\varepsilon}=
\left(\begin{array}{cccc}
-\varepsilon_2^2 & -\varepsilon_2 & -1 & 0\\
0 & 0 & 0 & \varepsilon_2\\
0 & \varepsilon_1\varepsilon_2 & 0 & -\varepsilon_2\\
0 & 0 & \varepsilon_1 & 0\\
\end{array}\right),\quad
U_{\bar\varepsilon}^{-1}=
\left(\begin{array}{cccc}
-\varepsilon_2^{-2} & -\varepsilon_1^{-1}\varepsilon_2^{-2} & -\varepsilon_1^{-1}\varepsilon_2^{-2} & -\varepsilon_1^{-1}\varepsilon_2^{-2}\\
0 & \varepsilon_1^{-1}\varepsilon_2^{-1} & \varepsilon_1^{-1}\varepsilon_2^{-1} & 0\\
0 & 0 & 0 & \varepsilon_1^{-1}\\
0 & \varepsilon_2^{-1} & 0 & 0\\
\end{array}\right).
\end{gather*}
The nonzero canonical commutation relations of the algebra $2A_{2.1}$ are $[e_1,e_2]=e_1$, $[e_3,e_4]=e_3$.
We calculate all different (up to antisymmetry) transformed commutators of the basis elements
using the formula $[e_1, e_2]_{\bar\varepsilon}=U_{\bar\varepsilon}^{-1}[U_{\bar\varepsilon}e_1, U_{\bar\varepsilon}e_2]$:
\begin{gather*}
[e_1,e_4]_{\bar\varepsilon}=\varepsilon_2e_1,
\quad
[e_2,e_3]_{\bar\varepsilon}=-\frac{\varepsilon_1}{\varepsilon_2}e_1+\varepsilon_1e_2,
\\
[e_1,e_2]_{\bar\varepsilon}=0,
\quad
[e_1,e_3]_{\bar\varepsilon}=0,\quad
[e_2,e_4]_{\bar\varepsilon}=e_1,\quad
[e_3,e_4]_{\bar\varepsilon}=e_2.
\end{gather*}
The transformed commutation relations go to the canonical ones of the Lie algebra~$A_{4.1}$
only under the repeated limit $\varepsilon_1\to+0$ and then $\varepsilon_2\to+0$.
The simultaneous limit $\bar\varepsilon\to\bar 0$ does not exist,
i.e.,\ the repeated contraction does not result in a multi-parametric one in this case.
Therefore, to derive a matrix of one-parametric contraction, we have to constrain
the parameters $\varepsilon_1$ and $\varepsilon_2$ in a special way.
Namely, the condition $\varepsilon_1=f(\varepsilon_2)=o(\varepsilon_2)$, $\varepsilon_2\to +0$ guarantees 
that the one-parametric contraction with the matrix $U_{f(\varepsilon),\varepsilon}$ exists 
and the resulting algebra has the same commutation relations as in the case of the repeated contraction. 
We put $\varepsilon_1=\varepsilon_2^2$.
The matrix $U_{\varepsilon^2\!,\varepsilon}$ gives a well-defined one-parametric contraction
between the algebras $2A_{2.1}$ and $A_{4.1}$ under $\varepsilon\to +0$:
\begin{gather*}
[e_1,e_4]_{\varepsilon}=\varepsilon e_1\to 0,
\quad
[e_2,e_3]_{\varepsilon}=-\frac{\varepsilon^2}{\varepsilon}e_1+\varepsilon e_2\to 0,
\\
[e_1,e_2]_{\varepsilon}=0,
\quad
[e_1,e_3]_{\varepsilon}=0,\quad
[e_2,e_4]_{\varepsilon}=e_1,\quad
[e_3,e_4]_{\varepsilon}=e_2.
\end{gather*}

The matrix $U_{\varepsilon,\varepsilon}$ also provides a well-defined one-parametric contraction. 
Although the obtained commutation relations differ from the ones in the case of the repeated contraction, 
the resulting algebra is isomorphic to $A_{4.1}$ via the matrix 
\[
I_{31}=
\left(\begin{array}{cccc}
1 & 0 & 0 & 0\\
0 & -1 & 0 & 0\\
0 & 0 & 1 & 1\\
0 & 0 & 1 & 0
\end{array}\right).
\]
Finally, the matrix $U_4=U_{\varepsilon,\varepsilon}I_{31}$ the explicit form of which is adduced in Remark~\ref{RemarkOnIWcontractionsOf4DimAlgeabras} 
generates the contraction $2A_{2.1}\to A_{4.1}$ in the canonical bases.
It is not a generalized In\"on\"u--Wigner contraction.

Note also that there are other possibilities for repeated contractions from $2A_{2.1}$ to $A_{4.1}$. 
The algebras $A_{4.8}^0$, $A_{3.2}\oplus A_1$ and $A_{3.4}^a\oplus A_1$ 
can be used as intermediate ones similar to $A_{4.3}$.
\end{example}

What is a condition for repeated contractions to produce well-defined decomposable multi-parametric contractions?
What is a way in order to obtain  corresponding one-parametric contractions otherwise?

Let Lie algebras in the pairs $(\mathfrak g,\hat{\mathfrak g})$ and $(\hat{\mathfrak g},\tilde{\mathfrak g})$
be connected by the one-parametric contractions with the matrices $\hat U_{\hat\varepsilon}$ and $\tilde U_{\tilde\varepsilon}$,
$c^k_{ij}$, $\hat c^{k''}_{i''\!j''}$ and $\tilde c^{k'}_{i'\!j'}$ be components of the structure constant tensors of
the algebras $\mathfrak g$, $\hat{\mathfrak g}$ and $\tilde{\mathfrak g}$ correspondingly,
$U_{\bar\varepsilon}=\hat U_{\hat\varepsilon}\tilde U_{\tilde\varepsilon}$, where $\bar\varepsilon=(\hat\varepsilon,\tilde\varepsilon)$.
In view of the contraction definition,
\[
\lim\limits_{\hat\varepsilon\to+0}
(\hat U_{\hat\varepsilon})_{i''}^i (\hat U_{\hat\varepsilon})_{j''}^j(\hat U_{\hat\varepsilon}{}^{-1})_k^{k''}c^k_{ij}
=\hat c^{k''}_{i''\!j''},
\quad
\lim\limits_{\tilde\varepsilon\to+0}
(\tilde U_{\tilde\varepsilon})_{i'}^{i''} (\tilde U_{\tilde\varepsilon})_{j'}^{j''}(\tilde U_{\tilde\varepsilon}{}^{-1})_{k''}^{k'}\hat c^{k''}_{i''\!j''}
=\tilde c^{k'}_{i'\!j'},
\quad
\]
and therefore we have the repeated contraction
\begin{gather*}
\lim\limits_{\tilde\varepsilon\to+0}
(\tilde U_{\tilde\varepsilon})_{i'}^{i''} (\tilde U_{\tilde\varepsilon})_{j'}^{j''}(\tilde U_{\tilde\varepsilon}{}^{-1})_{k''}^{k'}
\left(\lim\limits_{\hat\varepsilon\to+0}
(\hat U_{\hat\varepsilon})_{i''}^i (\hat U_{\hat\varepsilon})_{j''}^j(\hat U_{\hat\varepsilon}{}^{-1})_k^{k''}c^k_{ij}
\right)=
\\ \qquad\qquad\qquad
{}=\lim\limits_{\tilde\varepsilon\to+0}\left(\lim\limits_{\hat\varepsilon\to+0}
(U_{\bar\varepsilon})_{i'}^i (U_{\bar\varepsilon})_{j'}^j(U_{\bar\varepsilon}{}^{-1})_k^{k'}c^k_{ij}
\right)=\tilde c^{k'}_{i'\!j'}.
\end{gather*}
The latter condition is rewritten in the tautological form
\[
\lim\limits_{\tilde\varepsilon\to+0}\left(\lim\limits_{\hat\varepsilon\to+0} g^{k'}_{i'\!j'}(\bar\varepsilon)
\right)=0,
\]
where
$g^{k'}_{i'\!j'}(\bar\varepsilon)=\tilde g^{k'\!i''\!j''\!}_{i'\!j'\!k''\!}(\tilde\varepsilon)\,\hat g^{i''\!j''\!}_{\!k''\!}(\hat\varepsilon)$,
\[
\tilde g^{k'\!i''\!j''\!}_{i'\!j'\!k''\!}(\tilde\varepsilon)=
(\tilde U_{\tilde\varepsilon})_{i'}^{i''} (\tilde U_{\tilde\varepsilon})_{j'}^{j''}(\tilde U_{\tilde\varepsilon}{}^{-1})_{k''}^{k'},
\quad
\hat g^{i''\!j''}_{\!k''}(\hat\varepsilon)=
(\hat U_{\hat\varepsilon})_{i''}^i (\hat U_{\hat\varepsilon})_{j''}^j(\hat U_{\hat\varepsilon}{}^{-1})_k^{k''}c^k_{ij}-\hat c^{k''}_{i''\!j''}.
\]

If the repeated limit in the tautological equation can be correctly replaced by the simultaneous limit $\bar\varepsilon\to +\bar0$
or a simple limit with constrained values of $\varepsilon_1$ and $\varepsilon_2$ then
the matrix~$U_{\bar\varepsilon}$ results in a well-defined multi- or one-parametric contraction.
More precisely, the following statement is obviously true.

\begin{lemma}
The matrix $U_{\bar\varepsilon}=\hat U_{\hat\varepsilon}\tilde U_{\tilde\varepsilon}$
(or $U_{\bar f(\varepsilon)}=\hat U_{\hat f(\varepsilon)}\tilde U_{\tilde f(\varepsilon)}$,
where $\hat f,\tilde f\colon(0,1]\to(0,1]$ are continuous monotonic functions, $\hat f,\tilde f\to0$, $\varepsilon\to +0$)
gives a multi-parametric (one-parametric) contraction of the algebra $\mathfrak g$ to the algebra $\tilde{\mathfrak g}$ iff
\[
\lim\limits_{\bar\varepsilon\to+0}g^{k'}_{i'\!j'}(\bar\varepsilon)=0\qquad
\left(\lim\limits_{\varepsilon\to+0}g^{k'}_{i'\!j'}(\hat f(\varepsilon),\tilde f(\varepsilon))=0\right).
\]
\end{lemma}

\begin{corollary}
If the functions $\tilde g^{k'\!i''\!j''\!}_{i'\!j'\!k''\!}(\tilde\varepsilon)$ are bounded for any values of indices
under $\tilde\varepsilon\to +0$ then the matrix $U_{\bar\varepsilon}=\hat U_{\hat\varepsilon}\tilde U_{\tilde\varepsilon}$ generates
a two-parametric contraction of the algebra $\mathfrak g$ to the algebra $\tilde{\mathfrak g}$.
\end{corollary}

\begin{theorem}
Let the algebra $\mathfrak g$ be contracted to the algebra $\hat{\mathfrak g}$ with the matrix $\hat U_{\hat\varepsilon}$
and
the algebra $\hat{\mathfrak g}$ be contracted to the algebra $\tilde{\mathfrak g}$ with the matrix $\tilde U_{\hat\varepsilon}$.
Then there exists a continuous monotonic function $f\colon(0,1]\to(0,1]$, $f\to0$, $\varepsilon\to+0$, such that
the matrix $\check U_{\varepsilon}=\hat U_{f(\varepsilon)}\tilde U_{\varepsilon}$ results in a one-parametric continuous contraction
from the algebra $\mathfrak g$ to the algebra $\tilde{\mathfrak g}$.
\end{theorem}

\begin{proof}
Since $\tilde g^{k'\!i''\!j''\!}_{i'\!j'\!k''\!}$ are continuous functions for any values of indices then
for any $p\in\N$ there exists $\varkappa_p>0$ that
$\tilde g^{k'\!i''\!j''\!}_{i'\!j'\!k''\!}(\varepsilon)<\varkappa_p$ if $\varepsilon\in[\frac1{p+1},\frac1p]$.
Hereafter we assume that the indices $i$, $j$, $k$,~\dots\ run the whole range from 1 to $n$.
In view of $\hat g^{i''\!j''}_{\!k''}(\hat\varepsilon)\to0$ under $\hat\varepsilon\to+0$,
for any $p\in\N$ there exists $\delta_p\in(0,1]$ that
$|g^{i''\!j''}_{\!k''}(\hat\varepsilon)|<n^{-3}p^{-1}\min(1,\varkappa_p^{-1})$ if $\hat\varepsilon\in(0,\delta_p]$.
Without loss of generality we put $\delta_1>\delta_2>\cdots$. Then the desired function~$f$ can be defined by the formula
\[
f(\varepsilon)=p\delta_p((p+1)\varepsilon-1)-(p+1)\delta_{p+1}(p\varepsilon-1), \quad \varepsilon\in[\tfrac1{p+1},\tfrac1p], \quad p\in\N.
\tag*{\qed}
\]\renewcommand{\qed}{}
\end{proof}

\begin{theorem}
Let the algebra $\mathfrak g$ be contracted to the algebra $\hat{\mathfrak g}$ with the matrix $\hat U_{\hat\varepsilon}$,
the algebra $\hat{\mathfrak g}$ be contracted to the algebra $\tilde{\mathfrak g}$ with the matrix $\tilde U_{\hat\varepsilon}$
and the coefficients of $\hat U_{\hat\varepsilon}$ and $\tilde U_{\hat\varepsilon}$ be expanded in Laurent series in a neighborhood of 0.
Then there exists a positive integer $\nu$ such that
the matrix $\check U_{\varepsilon}=\hat U_{\varepsilon^\nu}\tilde U_{\varepsilon}$ generates a one-parametric continuous contraction
from $\mathfrak g$ to $\tilde{\mathfrak g}$.
\end{theorem}

\begin{proof}
In view of conditions of the theorem,
the functions $\tilde g^{k'\!i''\!j''\!}_{i'\!j'\!k''\!}$ and $\hat g^{i''\!j''}_{\!k''}$ are
also expanded in Laurent series in a neighborhood of 0.
Since $\hat g^{i''\!j''}_{\!k''}(\hat\varepsilon)\to0$ under $\hat\varepsilon\to+0$ then
$\hat g^{i''\!j''}_{\!k''}(\hat\varepsilon)=O(\hat\varepsilon)$ under $\hat\varepsilon\to+0$.
Let $\mu$ be the maximal module of powers in singular parts of $\tilde g^{k'\!i''\!j''\!}_{i'\!j'\!k''\!}$.
Then $\nu=\mu+1$ is the desired positive integer.
\end{proof}

\begin{corollary}
If the contractions $\mathfrak g\to\hat{\mathfrak g}$ and $\hat{\mathfrak g}\to\tilde{\mathfrak g}$
are generated by matrices with  coefficients being polynomial in the contraction parameters then
the corresponding contraction $\mathfrak g\to\tilde{\mathfrak g}$ can also be realized with a matrix of the same kind.
\end{corollary}

\section{Conclusion}\label{conclusions}

We study the contractions of low-dimensional Lie algebras using
inequalities between algebraic quantities of initial and contracted algebras.
These inequalities are necessary conditions for contraction existence 
and are collected as a list of criteria in Theorem~\ref{TheoremOnNecessaryContractionCriteria1}.
In addition to a number of previously known criteria, we formulate several new ones
which concern ranks of adjoint representation,
ranks and the inertia law of Killing and modified Killing forms,
dimensions of radicals and nilradicals, etc. 
Criterion~\ref{criterion_Killing_formPNRanks} is most important among the new criteria since 
it tells the real and complex cases apart.

Due to wide variety the adduced criteria allow us to work with contractions effectively.
As a result, complete sets of inequivalent continuous contractions of real and complex Lie algebras 
of dimensions not greater than four are constructed.
Obtained results are presented in Sections~\ref{SectionOnOne-parContractionsOfRealLow-DimLieAlgebras} 
and~\ref{SectionOnOne-parContractionsOfComplexLow-DimLieAlgebras}.
The lists of contractions of three- and four-dimensional Lie algebras
are supplied with the explicit forms of the contraction matrices and are illustrated by diagrams.
Since contractions assign the partial ordering relationship on the variety $\mathcal L_n$ of $n$-dimensional Lie algebras,
the levels and co-levels of low-dimensional Lie algebras are also discussed in detail.

Analysis of obtained results shows that 
any one-parametric contraction of a real or complex three-dimensional algebra is equivalent
to a generalized In\"on\"u--Wigner contraction.
Contractions of four-dimensional Lie algebras except only four and two cases over the real and complex fields correspondingly 
are also represented via generalized IW-contractions. 
Accurate proof on impossibility of such representation for the exceptional cases is based on exhaustive study of filtrations 
on the initial algebras and is not presented in our paper. 
A sketch of the proof for the contraction $2\mathfrak g_{2.1}\to \mathfrak g_{4.1}$ was adduced in~\cite{Burde&Steinhoff1999}.

All constructed contraction matrices include only nonnegative powers of contraction parameters, 
i.e.,\ there exist limits of the contraction matrices under $\varepsilon\to+0$. 
It seems that this phenomenon is broken for higher dimensions at least in the class of generalized IW-contractions. 
Thus, it is stated in~\cite{Weimar-Woods2000} that a generalized IW-contraction 
of the algebra $A_{5.38}$ ($[e_1,e_4]=e_1$, $[e_2,e_5]=e_2$, $[e_4,e_5]=e_3$) to the algebra $2A_{2.1}\oplus A_1$ 
necessarily contains negative powers of contraction parameters.

An important by-consequence of complete knowledge on limit processes between Lie algebras is creation of 
additional possibilities in studying the variety formed by these algebras.
Structure of the variety $\mathcal L_n(\CC)$ of $n$-dimensional complex Lie algebras is well known for any $n$ 
from 1 to~7
\cite{CarlesDiakite1984,GozeKhakimdjanov1996,Kirillov&Neretin1984-1987,OnishchikVinberg1994}. 
Over the real field we observe the natural effect of component bifurcation in comparison with the complex case.
Thus, $\mathcal L_3(\R)$ has four irreducible (over $\R$) components 
\begin{gather*}
\overline{\mathcal O(sl(2,\R))}, \quad 
\overline{\mathcal O(so(3))}, \quad 
\overline{\cup_a\mathcal O(A_{3.4}^a)}, \quad 
\overline{\cup_b\mathcal O(A_{3.5}^b)}
\end{gather*}
having same dimension 6. 
$\mathcal L_4(\R)$ consists of eight irreducible (over $\R$) components 
\begin{gather*}
\overline{\mathcal O(sl(2,\R\oplus A_1))}, \quad 
\overline{\mathcal O(so(3)\oplus A_1)}, \quad 
\overline{\mathcal O(2A_{2.1})}, \quad 
\overline{\mathcal O(A_{4.10})}, \quad 
\\
\overline{\cup_{a\ne 1}\mathcal O(A_{4.8}^a)}, \quad
\overline{\cup_a\mathcal O(A_{4.9}^a)}, \quad
\overline{\cup_{1\ne a\ne b\ne 1}\mathcal O(A_{4.5}^{ab1})}, \quad
\overline{\cup_{a,b}\mathcal O(A_{4.6}^{ab})}
\end{gather*}
each of which is 12-dimensional. 
Here the series parameters are assumed to satisfy usual normalization conditions of this Lie algebra classification 
\cite{Mubarakzyanov1963a,Popovych&Boyko&Nesterenko&Lutfullin2003a,Popovych&Boyko&Nesterenko&Lutfullin2003b}.
Precise description of structure of the variety $\mathcal L_n(\R)$ for small~$n$'s will be a subject of a further paper.  

Fulfilled investigation does not only solve previously posed problems but also generates new ones. We remark only on some of them.

A family of new problems concerns necessary contraction criteria. 
Is the adduced list of criteria sufficient in the case of higher dimensions 
for separating all the pairs of Lie algebras the first of which is contracted to the other?  
What is a sufficient list to do it for a fixed dimension~$n$, e.g., $n=5$? 
Are there sufficient lists which are suitable for an arbitrary dimension? 
And if such lists exist, what list is minimal? 
Will Criterion~\ref{criterion_Killing_formPNRanks} based on the inertia law of Killing and modified Killing forms
tell the real and complex cases apart for an arbitrary dimension similar to the low dimensions or 
should additional criteria of such type be found?

At the moment we know only one independent criterion with strict inequality for proper contractions, 
namely, Criterion~\ref{criterion_dim_Der} on dimensions of differentiation algebras.
The other well-known criterion with strict inequality between the orbit dimensions
is equivalent to Criterion~\ref{criterion_dim_Der} 
since $\dim O(\mathfrak g)=(\dim\mathfrak g)^2-\dim\Der(\mathfrak g)$.
It seems true that there are no other such criteria
which are inequivalent to Criterion~\ref{criterion_dim_Der} but it is not known certainly. 

Consideration of three- and four-dimensional algebras allows us to conjecture 
that the applied algorithm will also be effective in dimensions five and six. 
Classifications of five- and six-dimen\-sional Lie algebras are known. 
(See~\cite{Popovych&Boyko&Nesterenko&Lutfullin2003a,Popovych&Boyko&Nesterenko&Lutfullin2003b} for review of these results.) 
Thus, the five-dimen\-sional algebras were classified in~\cite{Mubarakzyanov1963b} over both the real and complex fields.
Unfortunately, the classifications should be tested and enhanced before application. 
A way in what the classifications should be modified to be more suitable for investigation of contractions and deformations 
was pointed out, e.g., in~\cite{Agaoka2002,Fialowski&Penkava2005}.  

In the future we also plan to extend our investigations by studying
deformations and contractions of realizations of low-dimensional Lie algebras.
The necessary background to this
is given by~\cite{Popovych&Boyko&Nesterenko&Lutfullin2003a}
in the form of classification of such realizations.

\subsection*{Acknowledgements}

The authors are grateful to Profs.\ V.~Boyko, J.~Niederle, and A.~Nikitin 
for productive and helpful discussions and useful censorious remarks.
M.\,N.\ also thanks for the hospitality to the Institute of Physics of the Academy of Sciences
of the Czech Republic.
The research of M.\,N.\ was supported by INTAS
in the form of PhD Fellowship (INTAS Ref.\ Nr 04-83-3217) and
partially by the Grant of the President of Ukraine
for young scientists.
The research of R.\,P.\ was supported by the Austrian Science Fund (FWF), Lise Meitner
project M923-N13.

\end{document}